%% file: main.tex
\documentclass[twocolumn]{svjour3}
\usepackage{times,amsmath,amssymb,epsfig}
\usepackage[T1]{fontenc}
\usepackage[latin1]{inputenc}
\usepackage{framed}
\usepackage{stmaryrd}
\usepackage{listings}
\usepackage{cite}
\usepackage{nalgo}
\usepackage{xcolor}
\usepackage{url}
\newcommand{\KIM}[1]{{\color{orange}\bf Kim: #1 }}

\newcommand{\no}[1]{}

\spnewtheorem{algorithm}{Algorithm}{\bf}{\rm}
\newcommand{\FN}[1]{\textbf{#1}}
\newcommand{\TAG}[1]{\mbox{\texttt{#1}}}

\title{Fast In-Memory XPath Search using Compressed Indexes}
\author{%
Diego Arroyuelo{\small $~^{\#1}$},
 Francisco Claude{\small $~^{*2}$},
 Sebastian Maneth{\small $~^{+\%3}$},
 Veli M\"akinen{\small $~^{\dag 4}$},\\
 Gonzalo Navarro{\small $~^{\$5}$},
 Kim Nguy$\tilde{\hat{\textrm{e}}}$n{\small $~^{+6}$},
 Jouni Sir\'en{\small $~^{\dag 7}$},
 Niko V\"alim\"aki{\small $~^{\dag 8}$}%
\vspace{1.6mm}\\
\ \\
\centerline{
\begin{minipage}{9cm}
\begin{center}
\fontsize{10}{10}\selectfont\itshape
$~^{\#}$Yahoo! Research Latin America, Chile\\
\fontsize{9}{9}\selectfont\ttfamily\upshape
$~^{1}$darroyue@dcc.uchile.cl
\vspace{1.2mm}\\
\fontsize{10}{10}\selectfont\rmfamily\itshape
$~^{*}$David R. Cheriton School of Computer Science,\\ University of Waterloo, Canada\\
\fontsize{9}{9}\selectfont\ttfamily\upshape
$~^{2}$fclaude@cs.uwaterloo.ca
\vspace{1.2mm}\\
\fontsize{10}{10}\selectfont\rmfamily\itshape
$~^{+}$NICTA, Australia \\
\fontsize{9}{9}\selectfont\ttfamily\upshape
$~^{3}$sebastian.maneth@nicta.com.au,\\
$~^{6}$kim.nguyen@nicta.com.au
\end{center}
\end{minipage}
\begin{minipage}{9cm}
\begin{center}
\fontsize{10}{10}\selectfont\rmfamily\itshape
$~^{\dag}$Dept. of Computer Science, University of Helsinki, Finland\\
\fontsize{9}{9}\selectfont\ttfamily\upshape
$~^{4}$vmakinen@cs.helsinki.fi\\
$~^{7}$jltsiren@cs.helsinki.fi\\
$~^{8}$nvalimak@cs.helsinki.fi
\vspace{1.2mm}\\
\fontsize{10}{10}\selectfont\rmfamily\itshape
$~^{\$}$Dept. of Computer Science, University of Chile, Chile\\
\fontsize{9}{9}\selectfont\ttfamily\upshape
$~^{5}$gnavarro@dcc.uchile.cl
\vspace{1.2mm}\\
\fontsize{10}{10}\selectfont\rmfamily\itshape
$~^{\%}$School of Computer Science and Engineering,\\
University of New South Wales, Australia\\
{\large \ \\ }
\end{center}
\end{minipage}}
}
\begin{document}
\maketitle

\begin{abstract}
A large fraction of an XML document typically consists of text data.
The XPath query language allows to search such text
via its equal, contains, and starts-with predicates.
These predicates can be efficiently implemented using
a compressed self-index of the document's text data.
Most queries, however, are hybrid: they contain parts which
query the text of the document, and contain other parts
which query the tree structure of the document.
It is therefore a challenge to appropriately choose a
query evaluation order which optimally leverages
the execution speeds of the text and tree indexes.
Here the SXSI system is introduced. It stores the tree
structure of an XML document using a bit array of
opening and closing brackets plus a sequence of labels,
and stores the text nodes of the
document using a global compressed self-index.
On top of these indexes sits an XPath query engine that is
based on tree automata. The engine uses fast counting queries
on the text index in order to determine whether to
evaluate top-down or bottom-up with respect to
the tree structure. The resulting system has several
advantages over existing systems:
(1) on pure tree queries (without text search) such
as the XPathMark queries, the SXSI system performs on par or
better than the fastest known systems MonetDB and Qizx,
and
(2) on queries that use text search, SXSI outperforms
the existing systems by 1--3 orders of magnitude
(depending on the size of the result set), and
(3) for all tested data and queries, SXSI's memory
consumption consistently stays below two times the document size.
\end{abstract}

\input{introduction}

\input{textcollection}

\input{tree}

\input{automata}

\input{experiments}


\input{conclusions}

\section*{Acknowledgements}

We would like to thank Schloss Dagstuhl for the very pleasant and
stimulating research environment it provides;
the work of this paper was initiated during the Dagstuhl seminar
``Structure-Based Compression of Complex Massive Data'' (Number 08261).
We are grateful to Kunihiko Sadakane for making available to us
his implementation of parentheses structure for succinct trees,
and to Juha Karjalainen for composing the BioXML data.
Diego Arroyuelo and Francisco Claude were partially funded by NICTA,
Australia.
Francisco Claude was partially funded by NSERC of Canada and the Go-Bell
Scholarships Program. Francisco Claude and Gonzalo Navarro
were partially funded by Fondecyt Grant 1-080019, Chile.
Gonzalo Navarro
was partially funded by Millennium Institute for Cell Dynamics and
Biotechnology (ICDB), Grant ICM P05-001-F, Mideplan, Chile.
Veli M\"akinen and Jouni Sir\'en
were funded by the Academy of Finland under grant 1140727.
Niko V\"alim\"aki
was funded by the Helsinki Graduate School in Computer Science and
Engineering.

\bibliographystyle{plain}
\bibliography{main}

\end{document}

%% file: introduction.tex
\section{Introduction}
\label{sec:intro}

As more and more data is stored, transmitted, queried, and manipulated in
XML, the popularity of XPath and XQuery as query languages 
for semi-structured data is increasing. 
Evaluating such XML queries efficiently is challenging,
and has triggered much research. Today there is a wealth of public and 
commercial XPath/XQuery engines, apart from several theoretical proposals.

In this paper we focus on XPath, which is simpler and forms the basis of
XQuery. XPath query engines can be roughly divided into two categories:
{\em sequential} and {\em indexed}. In the former, which follows a {\em
streaming} approach, no preprocessing of the XML data is performed. Each
query sequentially reads the whole document,
and the goal is to be as close as possible to making just one pass 
over the data, while using as little main memory as possible to hold 
intermediate results and data structures. 
Instead, the indexed approach preprocesses the XML document
to build a data structure on it, so that later queries can be evaluated
without traversing the whole document.  A serious shortcoming of the
indexed approach is that the index can use much more space than the
original data, and thus may have to be manipulated on disk.
There are two approaches for dealing with this problem:
(1) to load the index only partially (by using clever clustering
techniques), or
(2) to use less powerful indexes which require less space. 
Examples of systems using these approaches are Qizx/DB~\cite{quizx09},
MonetDB/XQuery~\cite{bonea06}, and Tauro~\cite{tauro}.

In this work we aim at an index for XML that uses little space compared to
the size of the data, so that the indexed document can fit in main memory
for moderate-sized data, thereby solving XPath queries without any need of
resorting to disk. An in-memory index should outperform streaming
approaches, even when the data fits in RAM.  
This is confirmed when comparing our indexed approach against
two well-known streaming XPath engines (over data coming from a RAM-disk):
GCX~\cite{DBLP:conf/vldb/KochSS07} 
and SPEX~\cite{DBLP:journals/tkde/Olteanu07}
run about $50$ and $350$ times, respectively, 
slower than our system. Of course such a comparison is hardly fair
because the streaming engines need to \emph{parse} the 
entire XML input document at each run.

Note that usually main memory
XML query systems (such as Saxon~\cite{kay08}, Galax~\cite{ferea03},
Qizx/Open~\cite{quizx09}, etc.) use machine pointers to represent XML data.
We observe that on various well-estab\-lished DOM implementations, this
representation blows up memory consumption to about 5--10 times the size of
the original XML document.

An XML document can be regarded essentially as a {\em text collection}
(that is, a set of strings) organized into a {\em tree structure}, so that
the strings correspond to the text data and the tree structure corresponds
to the nesting of tags.  The problem of manipulating text collections
within compressed space is now well understood~\cite{NM07,CHLS07,MN08}, and
also much work has been carried out on compact data structures for trees
\cite{Jacobson89,MunRam01,GearyRRR04,GRR04,BDMRRR05,DRR06,HMR07,FarMun08,Arr08,SN08}.
In this paper we show how both types of compact data structures can be
integrated into a compressed index representation for XML data, which is
able to efficiently solve XPath queries.

A feature inherited from its components is that the compressed index {\em
replaces} the XML document, in the sense that the data (or any part of
it) can be efficiently reproduced from the index (and thus the data itself
can be discarded). The result is called a {\em self-index}, as the data is
inextricably tied to its index. A self-index for XML data was recently
proposed~\cite{FLMM05,FLMM06}, yet its support for XPath is reduced to a
very limited class of queries that are handled particularly well.
Namely, they handle ``simple paths'', that is, 
queries of the form
//$t_1$/$t_2$/ $\dots$ /$t_k$,
where each $t_i$ is a tagname. For such queries they can count in 
time $O(k)$, and can report in time $O(\log^{1+\epsilon} n)$ per result.
See~\cite{DBLP:journals/corr/abs-1012-5696} 
where a comparison is presented which allows to conclude that,
for result counting
over XMark documents, their self-index can be between
one and two orders of magnitude faster than our system.

The main value of our work is to provide the first practical and public
tool for compressed indexing of XML data, dubbed {\em Succinct XML
Self-Index} (SXSI), which takes little space, solves a significant portion
of XPath, and largely outperforms the best public software supporting
XPath we are aware of, namely MonetDB/XQuery~\cite{bonea06} and
Qizx/DB~\cite{quizx09}.  Currently we support at
least \emph{forward Core XPath}~\cite{gotkocpic05}, i.e., all forward navigational axes,
plus, additionally text() and the attribute axis, and 
the three text predicates $=$ (equality), \emph{contains}, and
\emph{starts-with}.  The main challenges in achieving our results have been
to obtain practical implementations of compact data structures (for texts,
trees, and others) that are at a theoretical stage, to develop new compact
schemes tailored to this particular problem, and to develop query
processing strategies tuned for the specific cost model that emerges from
the use of these compact data structures. The limitations of our scheme are
that it is in-memory, that it is static (i.e., the index must be rebuilt
when the XML data changes), and that it does not handle XQuery. The first
limitation is a design decision; the last two are subject of future work.

This paper introduces the three main ingredients of SXSI: (i)~the text index,
(ii)~the tree index, and (iii)~the query evaluator. While technical details on
the first two components can be found elsewhere in the literature, 
we here only mention the main aspects of these components and focus on 
how they are integrated inside of the SXSI system. 
A new aspect of executing automata to solve XPath queries 
is the notion of \emph{relevant nodes}~\cite{xpathwhole};
intuitively, a node is relevant if the automaton must
necessarily visit it to solve the query. 
A large speed-up is obtained by using the new ``jump'' primitives of our 
tree index in order to visit only relevant nodes.
We present an algorithm for ``true bottom-up runs'', which 
are beneficial if the query contains highly selective text predicates. 
In our experimental section we 
first test the core speeds of our indexes;
for instance, timing global pattern counting over the text
index against a naive string buffer, or,
full traversals over the tree index against traversing
a pointer-based tree store. 
The main part of the experimental section is about comparing
SXSI against the state-of-the-art XPath
engines MonetDB/XQuery and Qizx/DB.
We use two batches of experiments: the ``tree oriented'' queries
of the XPathMark benchmark~\cite{DBLP:conf/xsym/Franceschet05} 
(over XMark data~\cite{DBLP:conf/vldb/SchmidtWKCMB02}) and
our own ``text oriented'' queries (over Medline documents).
Our results show that SXSI outperforms the other systems for virtually 
all tested queries, and moreover that 
the running times of SXSI are more predictable
and ``robust'' than those of other systems.

\section{Basic Concepts and Model}

We regard an XML document as $(i)$~an ordered set of strings and 
$(ii)$~a labeled tree. The latter is the natural XML parse tree defined by the hierarchical
tags, where the (normalized) tag name labels the corresponding node. 
We add an extra root node (labeled ``\&'') on top of the document's root node;
this node is needed for XPath semantics, but could also be used
to hold additional information such as the document name.
Each text node is represented as a leaf labeled {\tt \#}. 
Attributes are handled as follows in this model. 
Each node with attributes gets an additional single child
labeled {\tt @} (at the first child position), 
and for each attribute {\tt @attr=value} of the node,
a child labeled {\tt attr} is added to its {\tt @}-node, and a leaf child
labeled {\tt \%} to the {\tt attr}-node.  The text content {\tt value} is
then associated to that leaf.  Therefore, there is exactly one string
content associated to each tree leaf labeled {\tt \#} or {\tt \%}.
We refer to those strings as \emph{texts}. 
Note that we do not store empty texts; for instance, 
the XML document \verb!<a></a>! is stored as 
a single {\tt a}-labeled leaf node 
(which is the unique child of the $\&$-labeled root node).

\begin{table}[t]
\caption{Notation.}
\begin{center}
\begin{tabular}{c|p{6.2cm}}
Term & Meaning \\
\hline
$T$ & Concatenation of all the texts in the collection \\
$u$ & Length of $T$ in symbols \\
$\Sigma$ & Alphabet of the distinct text symbols \\
$\sigma$ & Size of $\Sigma$ \\
\$ & Character that terminates each text in the collection \\
$n$ & Number of nodes in the XML tree \\
$t$ & Number of different tag and attribute names in the document \\
$d$ & Number of texts in the XML tree (in our model, tree leaves) \\
$H_k(S)$ & $k$-th order empirical entropy of string $S$
\end{tabular}
\end{center}
\label{tab:not}
\end{table}

Let us call $T$ the set of all the texts and $u$ its total length measured
in symbols, $n$ the total number of tree nodes, $\Sigma$ the alphabet of
the strings and $\sigma=|\Sigma|$, $t$ the total number of different tag
and attribute names, and $d$ the number of texts (or tree leaves).  These
receive {\em text identifiers} which are consecutive numbers assigned in a
left-to-right parsing of the data.  In our implementation $\Sigma$ is
simply the set of byte values 1 to 255, and 0 will act as a special
terminator called $\$$. This symbol occurs exactly once at the end of each
text in $T$.  
Note that our implementation can easily support also
UTF-8 encoding and hence adheres to the XML standard.
Table~\ref{tab:not} summarizes the notation.

\begin{figure*}[tb]
\centerline{\includegraphics[width=0.95\textwidth]{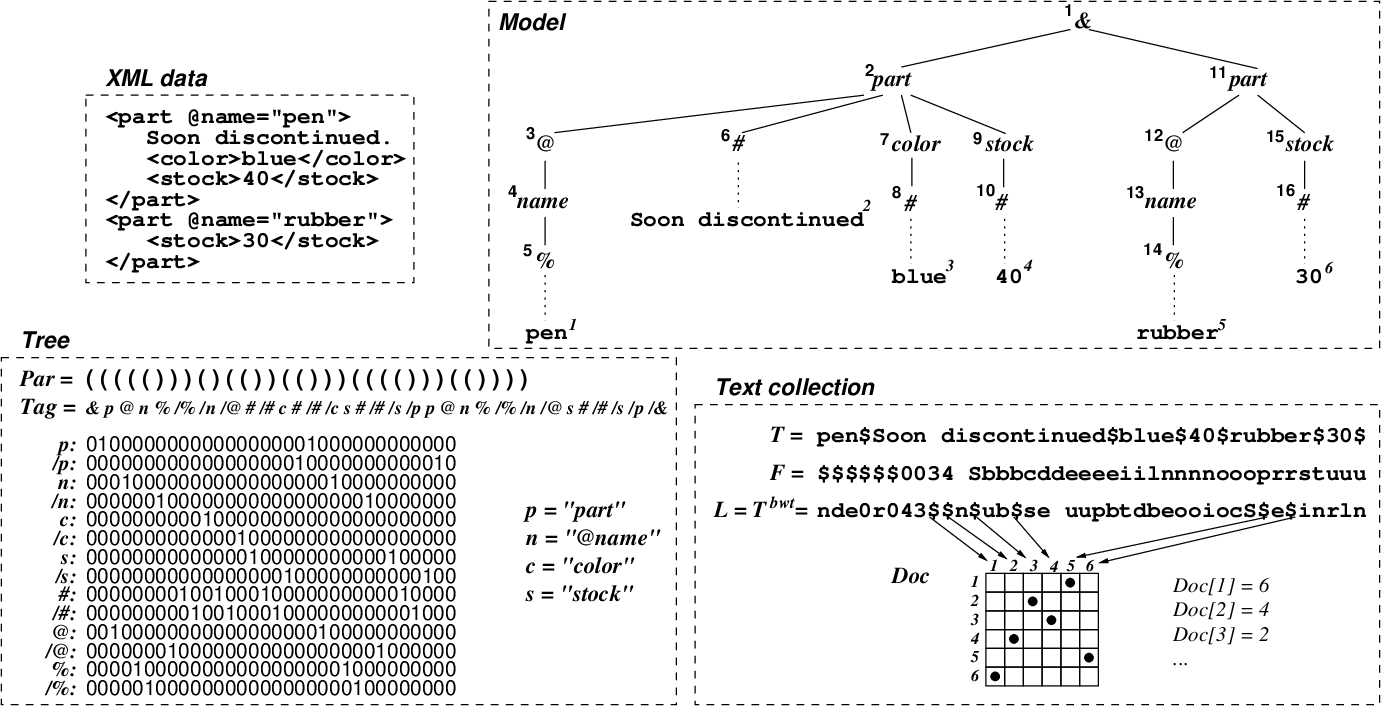}}
\caption{Our running example on representing an XML document.}
\label{fig:example}
\end{figure*}

To connect tree nodes and texts, we define {\em global identifiers}, which
give unique numbers to both internal and leaf nodes, in depth-first
preorder.  Figure~\ref{fig:example} shows a toy document (top left) and our
model of it (top right), as well as its representation using our data
structures (bottom), which serves as a running example for the rest of the
paper. In the model, the tree is formed by the solid edges, whereas dotted
edges display the connection with the set of texts.
The tree contains the extra root node (labeled {\tt \&}), 
as well as extra internal nodes (labeled {\tt \#}, {\tt @}, and
{\tt \%}). Note how the attributes are handled. There are six texts, which are
associated to the tree leaves and receive consecutive text numbers (marked
in italics at their right). Global identifiers are associated to each node
and leaf (drawn at their left). The conversion between tag names and
symbols, drawn within the bottom-left component, is used to translate
queries and to recreate the XML data.
Note that if the return and space (indentation) characters are present
precisely as shown in the ``XML data'' box of the figure,
then there are indeed several additional {\tt \#}-leaves in the tree:
for instance, the whitespace (return and space
characters) after the initial \texttt{<parts>} and
before the final \texttt{</parts>} give rise to two extra texts
(and therefore the \texttt{parts}-node in the tree has
additional first and last children labeled {\tt \#}).
In total there are seven such whitespace texts which 
have been omitted in our figure
for reasons of readability.

Some notation and measures of compressibility follow, preceding a rough
description of our space complexities.  The {\em empirical $k$-th order
entropy}~\cite{Man01} of a sequence $S$ over alphabet $\Sigma$, $H_k(S) \le
\log \sigma$, is a lower bound to the output size per symbol of any $k$-th
order compressor applied to $S$. The formula of the zero-order entropy is
as follows: \[ H_0(S) = \sum_{c\in\Sigma} \frac{s_c}{s}\log\frac{s}{s_c},
\] where $s_c$ is the number of occurrences of $c$ in $S$ and $s=|S|$.  We
assume $\log = \log_2$ and $0\log 0 = 0$ henceforth. 
Let $\Sigma^k$ denote the set of words over $\Sigma$ of length $k$.
Now let $S_W$ be the
set of characters preceding the occurrences of $W \in \Sigma^k$ in $S$,
then for $k>0$, \[ H_k(S) = \frac{1}{s}\sum_{W\in\Sigma^k} |S_W|H_0(S_W).
\] 
Note $0 \le H_k(S) \le H_{k-1}(S) \le \ldots \le H_0(S) \le \log\sigma$.

We will build on self-indexes able of handling text collections $T$ of
total length $u$ within $uH_k(T) + o(u\log\sigma)$ bits
\cite{NM07,FMMN07,MN08}. On the other hand, representing an unlabeled tree
of $n$ nodes requires $2n-O(\log n)$ bits, and several representations
using $2n+o(n)$ bits support many tree query and navigation operations in
constant time (e.g., \cite{SN08}). The labels require in principle other
$n\log t$ bits.  Sequences $S$ can be stored within $|S|\log\sigma
(1+o(1))$ bits (and even $|S|H_0(S)+o(|S|\log\sigma)$), so that any element
$S[i]$ can be accessed, and they can also efficiently answer the following
queries \cite{GGV03,GMR06,FMMN07}: 
\begin{description}
\item[$rank_c(S,i)$] is the number of $c$'s in $S[1,i]$; 
\item[$select_c(S,j)$] is the position of the $j$-th $c$ in $S$.
\end{description}
These are essential building blocks for more complex functionalities, as seen 
later.

The final space requirement of our index will include:
\begin{enumerate}
\item $uH_k(T)+o(u\log\sigma)$ bits for representing the text collection $T$ 
in self-indexed form. This supports the string searches of XPath and can
(slowly) reproduce any text.
\item $d\log d + o(d\log d)$ bits for the mapping between the self-index and
the text identifiers, e.g., to determine to which text identifier a self-index 
position belongs, or restricting self-index searches to some texts.
\item $2n+o(n)$ bits for representing the tree structure. This supports many
navigational operations in constant time.
\item $4n\log t + 2n + o(n)$ bits to represent the tags in a way that they 
support very fast XPath searches.
\item $2n+o(n)$ bits for mapping between tree nodes and text identifiers.
\item Optionally, $u\log\sigma$ or $uH_k(T)+o(u\log\sigma)$ bits, plus
$O(d\log\frac{u}{d})$, to achieve faster text extraction than in 1).
\end{enumerate}

As a practical yardstick, without the extra storage of texts (Item 6) the
memory consumption of our system is about the size of the original XML file
(and, being a self-index, includes it!), and with the extra text store the
memory consumption is 1--2 times the size of the original XML file.

In Section~\ref{sec:text} we describe our representation of the set of
strings, including how to obtain text identifiers from text positions. This
explains items 1, 2, and 6 above. Section~\ref{sec:tree} describes our
representation for the tree and the labels, and the way the correspondence
between tree nodes and text identifiers works. This explains items 3, 4,
and 5. Section~\ref{sec:queries} describes how we process XPath queries on
top of these compact data structures. In Sections~\ref{sec:imple} and
\ref{sec:exper} we give some implementation details and empirically compare
our SXSI engine with the most relevant public engines we are aware of.  We
conclude in Section~\ref{sec:concl}.

%% file: textcollection.tex
\section{Text Representation}
\label{sec:text}

Text data is represented as a succinct full-text self-index \cite{NM07}
that is generally known as the {\em FM-index} \cite{FM05}. The index
supports efficient pattern matching operations that can be easily extended
to support different XPath predicates.

\subsection{FM-Index and Backward Searching}

Given a string $T$ of total length $u$, from an alphabet of size $\sigma$,
the {\em alphabet-friendly FM-index} \cite{FMMN07} requires $uH_k(T) + o(u
\log \sigma)$ bits of space.  The index supports counting the number of
occurrences of a pattern $P$ in $O(|P|\log\sigma)$ time.  Locating the
occurrences takes extra $O(\log^{1+\epsilon} u)$ time per answer, for any
constant $\epsilon > 1$.

\begin{figure}[t]
\includegraphics[width=0.5\textwidth]{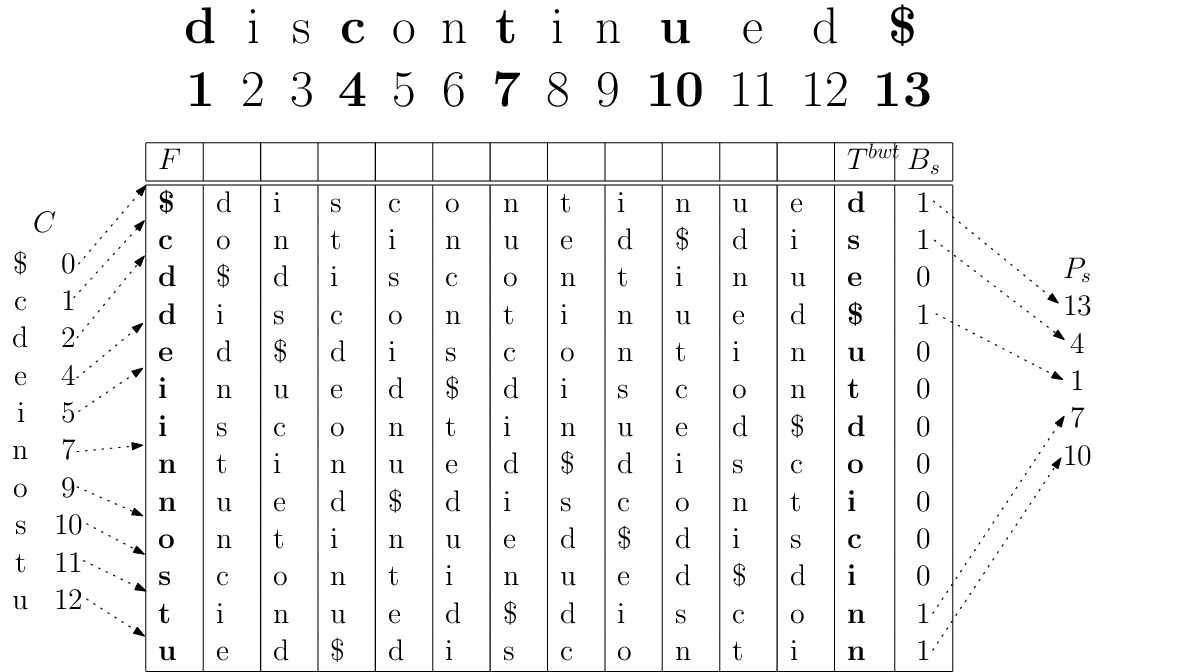}
\caption{Example of the FM-index for text $T=\mathtt{``discontinued"}$ sampled
each $l=3$ positions.}
\label{fig:exbwt}
\end{figure}

The FM-index is based on the Burrows--Wheeler transform (BWT) of string $T$
\cite{BW94}.  Assume $T$ ends with the special end-marker $\$$.  Let
$\mathcal{M}$ be a matrix whose rows are all the cyclic rotations of $T$ in
lexicographic order.  The first column of $\mathcal{M}$, denoted $F$,
contains all symbols of $T$ in lexicographic order. The last column $L$ of
$\mathcal{M}$ forms a permutation of $T$ which is the BWT string $L =
T^{\text{\it bwt\/}}$.  The matrix is only conceptual; the FM-index uses only on the
$T^{\text{\it bwt\/}}$ string.  Figure~\ref{fig:exbwt} illustrates the matrix
$\mathcal{M}$ with its first and last rows ($F$ and $L=T^{\text{\it bwt\/}}$) in bold.
Figure~\ref{fig:example} (bottom right) shows how this fits in our overall
scheme. 

The resulting permutation from $T$ to $T^{\text{\it bwt\/}}$ is reversible.  There
exists a simple last-to-first mapping from symbols in $L$ to $F$
\cite{FM05}: Let $C[c]$ be the total number of symbols in $T$ that are
lexicographically less than $c$.  Then the {\em LF-mapping} is defined as
\[ \text{\it LF\/}(i) = C[L[i]] + \text{\it rank}_{L[i]}(L, i). \] Note that $L[i]$ is the symbol
preceding the $i$-th lexicographically smallest row of $\mathcal{M}$. Thus,
if $T^{\text{\it bwt\/}}[i] = T[j]$, then $T^{\text{\it bwt\/}}[\text{\it LF\/}(i)] = T[j-1]$.  The symbols of $T$
can therefore be read in reverse order by starting from the location $i$
such that $T^{\text{\it bwt\/}}[i]=\$$, and applying $\text{\it LF\/}$ recursively:

\smallskip\noindent
$T[u] = \$ = T^{\text{\it bwt\/}}[i]$, \\
$T[u-1] = T^{\text{\it bwt\/}}[\text{\it LF\/}(i)]$, \\
$T[u-2] = T^{\text{\it bwt\/}}[\text{\it LF\/}(\text{\it LF\/}(i))]$, 

\smallskip\noindent
and so on until, after $u$ steps, we get the first symbol $T[1]$.  The
values $C[c]$ can be stored in a small array of $\sigma \log u$ bits.
Function $\text{\it rank}_c(L, i)$ can be computed in $O(\log \sigma)$ time with a
data structure called {\em wavelet tree} which, when built on $T^{\text{\it bwt\/}}$,
uses only $uH_k(T) + o(u \log \sigma)$ bits \cite{GGV03,FMMN07}.  In
practice we opt for a Huffman-shaped wavelet tree using uncompressed
bitmaps inside \cite{CNspire08.1}. Despite this achieves space
$u(H_0(T)+1)(1+o(1))$, it is much faster than the other implementations. In
particular, operations cost $O(H_0(T))$ time on average, an improvement
that applies to all the $O(\log\sigma)$ worst-case complexities that
follow.

Pattern matching is supported via {\em backward searching} on the BWT
\cite{FM05}.  Given a pattern $P[1,m]$, the backward search starts with the
range $[sp, ep] = [1,u]$ of rows in $\mathcal{M}$.  At each step $i \in \{
m, m-1, \ldots, 1 \}$ of the backward search, the range $[sp, ep]$ is
updated to match all rows of $\mathcal{M}$ that have $P[i, m]$ as a prefix.
New range $[sp', ep']$ is given by $sp' = C[P[i]] + \text{\it rank}_{P[i]}(L, sp-1)+1$
and $ep' = C[P[i]] + \text{\it rank}_{P[i]}(L, ep)$.  Each step takes $O(\log \sigma)$
time using the wavelet tree, and finally $ep - sp + 1$ gives the number of
times $P$ occurs in $T$.  Figure~\ref{fig:fmcount} gives the pseudocode.

\begin{figure}[t]
\hrulefill
\vspace*{-5mm}
\begin{algorithme}
{\bf FM-Count($p_1p_2\ldots p_m$)}\\
\lign $i := m$ \\
\lign $sp := 1$ \\
\lign $ep := n$ \\
\lign \While{$sp \leq ep$ and $i\geq 1$}\\
\lign \> $c := p_i$\\
\lign \> $sp := C[c]+\text{\it rank}_c(T^{\text{\it bwt\/}},sp-1)+1$\\
\lign \> $ep := C[c]+\text{\it rank}_c(T^{\text{\it bwt\/}},ep)$\\
\lign \> $i := i-1$\\
\lign \If{$ep<sp$}\\
\lign \> \Return $0$\\
\lign \Else \\
\lign \> \Return $ep-sp+1$
\end{algorithme}

\vspace*{-4mm}
\hrulefill
\caption{Counting on the FM-index.}
\label{fig:fmcount}
\end{figure}

To find out the location of each occurrence, the text is traversed
backwards from each $sp \le i \le ep$ (virtually, using $\text{\it LF\/}$ on $T^{\text{\it bwt\/}}$)
until a {\em sampled} position is found. This is a sampling carried out at
regular text positions, so that the corresponding positions in $T^{\text{\it bwt\/}}$
are marked in a bitmap $B_s[1,u]$, and the text position corresponding to
$T^{\text{\it bwt\/}}[i]$, if $B_s[i]=1$, is stored in a samples array
$P_s[\text{\it rank}_1(B_s,i)]$.  If every $l$-th position of $T$ is sampled, the
extra space is $O((n/l)\log n)$ (including the compressed $B_s$
\cite{RRR02}) and the locating takes $O(l \log \sigma)$ time per
occurrence. Using $l = \Theta(\log^{1+\epsilon} u / \log\sigma)$ yields
$o(u\log\sigma)$ extra space and locating time $O(\log^{1+\epsilon} u)$.

Figure~\ref{fig:exbwt} illustrates a sampling of $T$ each $l=3$ symbols.
Assume we look for $P=\mathtt{``n"}$; then backward search finds
$[sp,ep]=[8,9]$. Now to locate the occurrence at $8$ we see that
$B_s[8]=0$, $B_s[\text{\it LF\/}(8)]=B_s[10]=0$, and finally $B_s[\text{\it LF\/}(10)]=B_s[2]=1$.
This corresponds to position $P_s[\text{\it rank}_1(B_s,2)] = P_s[2] = 4$. Since we
applied $\text{\it LF\/}$ twice, the answer is $4+2=6$. We have found the occurrence
$T[6..]=\mathtt{``n.."}$.

\subsection{Text Collection and Queries}

The textual content of the XML data is stored as $\$$-terminated strings so
that each text corresponds to one string.  Let $T$ be the concatenated
sequence of the $d$ texts.  Array $P_s$ is extended to record both the text
identifier and the offset inside it.  Since there are several $\$$'s in
$T$, we fix a special ordering such that the end-marker of the $i$-th text
appears at $F[i]$ in $\mathcal{M}$ (see Figure~\ref{fig:example}, bottom
right).  This generates a valid $T^{\text{\it bwt\/}}$ of all the texts and makes it
easy to extract the $i$-th text starting from its $\$$-terminator.

Now $T^{\text{\it bwt\/}}$ contains all end-markers in some permuted order.  This
permutation is represented with a data structure $\text{\it Doc}$, that maps from
positions of \$s in $T^{\text{\it bwt\/}}$ to text identifiers.  Let $T^{\text{\it bwt\/}}[j]$
correspond to the first symbol of the text with identifier $x$, thus if
$i=\text{\it LF\/}(j)$ it holds $T^{\text{\it bwt\/}}[i]=\$$. Then we store $\text{\it Doc\/}[\text{\it rank}_{\$}(T^{\text{\it bwt\/}},
i)]=x$. Furthermore, $\text{\it Doc\/}$ can be stored in a format that allows for range
searching (as illustrated in Figure~\ref{fig:example} (right)): Given a range
$[sp, ep]$ of $T^{\text{\it bwt\/}}$ and a range of text identifiers $[x, y]$, $\text{\it Doc\/}$ can
be used to output identifiers of all $\$$-terminators within the range $[sp, ep]
\times [x, y]$, in $O(\log d)$ time per answer. In practice, because
we only use the simpler functionality in the current system, $\text{\it Doc\/}$ is
implemented as a plain array using $d\log d$ bits.

Note $\text{\it Doc\/}$ allows us to never switch from one text to another while
looking for the preceding sampled value: If we reach a \$ before finding
any $B_s[i]=1$, array $\text{\it Doc\/}$ can be used to determine that we are at the
first position of some text with identifier $x$. 

The basic pattern matching feature of the FM-index can be extended to
support XPath functions such as {\em starts-with}, {\em ends-with}, {\em
contains}, and operators $=$, $\le$, $<$, $>$, $\ge$ for lexicographic
ordering.  Given a pattern and a range of text identifiers to be searched,
these functions return all text identifiers that match the query within the
range.  In addition, existential (is there a match in the range?) and
counting (how many matches in the range?) queries are supported.  Time
complexities are $O(|P| \log \sigma)$ for the search phase, plus an extra
for reporting. While we describe the operators in their general form, which
need the range reporting functionality from $\text{\it Doc}$, our current prototype
implements only the simple case $[x,y]=[1,d]$, where $\text{\it Doc\/}$ can be an array.

\paragraph*{starts-with$(P, [x, y])$}: \ 
The goal is to find texts in $[x, y]$ range prefixed by the given pattern
$P$.  After the normal backward search, the range $[sp, ep]$ in $T^{\text{\it bwt\/}}$
contains the end-markers of all the texts prefixed by $P$.  Now $[sp, ep]
\times [x, y]$ can be mapped to $\text{\it Doc}$, and existential and counting queries
can be answered in $O(\log d)$ time.  Matching text identifiers can be
reported in $O(\log d)$ time per identifier.

\paragraph*{ends-with$(P, [x, y])$}: \ 
Backward searching is localized to texts in $[x, y]$ by choosing $[sp,ep] =
[x,y]$ as the starting interval, since we have forced the ordering of
$F[1,d]$ so that $F[z]=\$$ is the terminator of text with identifier $z$.
After the backward search, the resulting range $[sp, ep]$ contains all
possible matches, thus existential and counting queries are answered in
constant time after the search.  To find out text identifiers for each
occurrence, the text must be traversed backwards to find a sampled position
(or a \$).  The cost is $O(l\log\sigma)$ per answer, where $l$ is the
sampling step.

\paragraph*{operator $=(P, [x, y])$}: \
Whole texts which are equal to $P$, and with identifiers in the range
$[x,y]$, can be found as follows. Start with a backward search as in {\em
ends-with}, and then map to the $\$$-terminators as in {\em starts-with}.
The time complexities are same as in {\em starts-with}.

\paragraph*{contains$(P, [x, y])$}: To find texts that contain $P$, 
we start with the normal backward search and finish like in {\em
ends-with}.  In this case there might be several occurrences inside one
text, which have to be filtered.  Thus, the time complexity is proportional
to the total number of occurrences, $O(l \log \sigma)$ for each.
Existential and counting queries are as slow as reporting queries. The
basic $O(|P|\log\sigma)$-time counting of all the occurrences of $P$ can
still be useful for query optimization.

\paragraph*{operators $\le$, $<$, $>$, $\ge$}:
Operator $\le$ matches texts that are lexicographically smaller than or
equal to the given pattern. It can be solved like the {\em starts-with}
query, but updating only the $ep$ of each backward search step, while $sp =
1$ stays constant. While $[sp,ep]$ delimits the rows of $\mathcal{M}$ that
start with $P[i,m]$, $[1,ep]$ delimits the rows that start with a prefix
lexicographically smaller than or equal to $P[i,m]$.  If at some point
there are no occurrences of $P[i]=c$ within the prefix $L[1,ep]$, this
means that $P[i,m]$ does not appear in $T$. To continue the search we
replace $ep = C[c]$ and continue for $P[1,i-1]$.  Other operators can be
supported analogously, and costs are as for {\em starts-with}.

\bigskip

The new XPath extension, \textit{XPath Full Text
1.0}~\cite{xqueryfulltext}, suggests a wider selection of functionality for
text searching. Implementation of these extensions requires regular
expression and approximate searching functionalities, which can be
supported within our index using the general {\em backtracking framework}
\cite{Lametal08}: The idea is to alter the backward search to branch
recursively to different ranges $[sp', ep']$ representing the suffixes of
the text prefixes (i.e., substrings).  This is done by computing $sp_c' =
C[c] + \text{\it rank}_{c}(L, sp-1)+1$ and $ep_c' = C[c] + \text{\it rank}_{c}(L, ep)$ for all
$c\in \Sigma$ at each step and recursing on each $[sp_c', ep_c']$. Then the
pattern (or regular expression) can be compared with all substrings of the
texts, allowing us to search for approximate occurrences \cite{Lametal08}.
The running time becomes exponential in the number of errors allowed, but
different branch-and-bound techniques can be used to obtain practical
running times \cite{LTPS09,LD09}.  We omit further details, as these
extensions are out of the scope of this paper.

\subsection{Construction and Text Extraction}

The FM-index can be built by adapting any BWT construction algorithm.
Linear time algorithms exist for the task, but their practical bottleneck
is the peak memory consumption. Although there exist general time- and
space-efficient construction algorithms, it turned out that our special
case of text collection admits a tailored incremental BWT construction
algorithm \cite{Sir09} (see the references and experimental comparison
therein for previous work on BWT construction): The text collection is
split into several smaller collections, and a temporary index is built for
each of them separately. The temporary indexes are then merged, and finally
converted into a static FM-index.
The BWT allows extracting the $i$-th text by successively 
applying $\text{\it LF\/}$
from $T^{\text{\it bwt\/}}[i]$, at $O(\log\sigma)$ cost per extracted symbol.  


\subsection{Additional Text Collections}

To enable faster text extraction, we allow storing the texts in plain 
format in $n \log \sigma$ bits, 
or in an enhanced LZ78-compressed format (derived from
the LZ-index \cite{ANS06}) using $u H_k(T) + o(u \log \sigma)$ bits.  These
secondary text representations are coupled with a delta-encoded bit vector
storing starting positions of each text in $T$.  This bitmap requires
$O(d\log\frac{u}{d})$ more bits.

In fact, keeping next to the FM-index an additional copy of all
texts in plain format has more advantages.
As mentioned before, the time complexity of contains-queries
is proportional to the total number of occurrences.
This implies that for large occurrence numbers, it becomes
faster to search over the plain texts than over the
FM-index. The precise cut-off point depends on the sampling
factor $l$, see Section~\ref{RawText} for mor details.
Since a global count over the FM-index is fast
($O(|P|\log\sigma)$ time), we use it to determine whether
to search over the plain text or over the FM-index.

Since search over our implementaiton of the LZ-index
is problematic, we only consider plain and FM-index 
from now on, and do not mention the LZ-index anymore.

%% file: tree.tex
\section{Tree Representation}
\label{sec:tree}

\subsection{Data Representation}

The tree structure of an XML collection is represented by the
following compact data structures, which provide navigation and
indexed access to it.
See also the bottom left of Figure~\ref{fig:example}.

\subsubsection{$\mbox{\it Par\/}$}
This is the \emph{balanced parentheses}
representation of the tree structure (see, e.g.,~\cite{MR97}).
It is obtained by traversing the tree in \emph{depth-first-search}
(DFS) order, writing a {\tt "("} whenever we arrive at a node, and a
{\tt ")"} when we leave it (thus it is easily produced during the XML
parsing).  In this way, every node is represented by a pair of
matching opening and closing parentheses. A tree node is identified by
the position of its opening parenthesis
in $\mbox{\it Par\/}$ (that is, a node is just an integer index within
$\mbox{\it Par\/}$).  In particular, we use the balanced parentheses
implementation of~\cite{SN08}, which supports a very complete set
of operations, including finding the $i$-th child of a node, in constant
time; for more information concerning implementation details and
performance, see~\cite{DBLP:conf/alenex/ArroyueloCNS10}.
Overall $\mbox{\it Par\/}$ uses $2n + o(n)$ bits. This includes the
space needed for constant-time binary {\em rank} on $\mbox{\it Par}$,
which is very fast in practice.

\newcommand{\Tag}{\textit{Tag\/}}

\subsubsection{$\Tag$}
\label{ss:treejump}
This is the sequence of the tag identifiers
of each tree node, including an opening and a closing version of each
tag, to mark the beginning and ending point of each node.  These tags
are numbers in $[1,2t]$ and are aligned with $\mbox{\it Par\/}$ so
that the tag of node $i$ is simply $\Tag[i]$.

We also need $\text{\it rank\/}$ and $\text{\it select\/}$ queries on $\Tag$.
They allow to realize special operations such as ``TaggedDesc'' which
``jumps'' to the first descendant of the given node, such that the
descendant has a given label (see Section~\ref{sect:TagOps}).
Several sequence
representations supporting access and these operations are known
\cite{GGV03,GMR06,CNspire08.1}. Given that $\Tag$ is not too critical in
the overall space, but it is in time, we opt for a practical representation
that favors speed over space.  First, we store the tags in an array using
$\lceil \log 2t \rceil$ bits per field, which gives constant time access to
$\text{\it Tag}[i]$. The rank and select queries over the sequence of tags are
answered by a second structure.  Consider the binary matrix
$R[1..2t][1..2n]$ such that $R[i,j]=1$ if $\Tag[j]=i$. We represent each
row of the matrix using Okanohara and Sadakane's structure {\tt sarray}
\cite{os07}. Its space requirement for each row $i$ is
$n_i\log\frac{2n}{n_i} +n_i(2+o(1))$ bits, where $n_i$ is the number of
times symbol $i$ appears in $\text{\it Tag}$. The total space of both structures adds
up to $2n \log(2t)+2nH_0(\Tag)+n(2+o(1)) \le 4n\log t + 2n+o(n)$ bits.
They support access and {\em select} in $O(1)$ time, and {\em rank} in
$O(\log n)$ time.\footnote{They report higher complexities, but these are
  easily improved by using a representation for dense arrays that supports
  {\em select} in constant time.}


\subsection{Tree Navigation} \label{navi}

We define the following operations over the tree structure, which are
useful to support XPath queries over the tree.  Most of these operations
are supported in constant time, except when a $\text{\it rank\/}$ over $\Tag$ is
involved.  Let $tag$ be a tag identifier.

\subsubsection{Basic Tree Operations} These are direcly inherited from
Sadakane's implementation \cite{SN08}. We mention only the most important
ones for this paper; $x$ is a node (a position in $\text{\it Par\/}$).

\begin{itemize}
\item Close$(x)$: The closing parenthesis matching $\text{\it Par\/}[x]$. If $x$ is a
  small subtree this takes a few local accesses to $\text{\it Par}$, otherwise a few
  non-local table accesses.
\item Preorder$(x) = \text{\it rank\/}_((\text{\it Par},i)$: Preorder number of $x$.
\item SubtreeSize$(x) = (\textrm{Close}(x)-x+1)/2$: Number of nodes in the
  subtree rooted at $x$.
\item IsAncestor$(x,y) = x \le y \le \textrm{Close}(x)$: Whether $x$ is an
  ancestor of $y$.
\item FirstChild$(x)=x+1$: First child of $x$, if any.
\item NextSibling$(x)=\textrm{Close}(x)+1$: Next sibling of $x$, if any.
\item Parent$(x)$:  Parent of $x$. Somewhat costlier than Close$(x)$ in
  practice, because the answer is less likely to be near $x$ in $\text{\it Par}$.
\end{itemize}

\subsubsection{Connecting to Tags}
\label{sect:TagOps}

The following operations are essential for our fast XPath evaluation.

\begin{itemize}
\item SubtreeTags$(x,\text{\it tag\/})$: Returns the number of occurrences of $\text{\it tag\/}$
  within the subtree rooted at node $x$. This is \linebreak
  $\text{\it rank\/}_{\text{\it tag\/}}(\text{\it Tag},\textrm{Close}(x))-\text{\it rank\/}_{\text{\it tag\/}}(\text{\it Tag},x-1)$.
\item Tag$(x)$: Gives the tag identifier of node $x$.  In our
  representation this is just $\Tag[x]$.
\item TaggedDesc$(x,\text{\it tag\/})$:  The first node (in pre-order) labeled $\text{\it tag\/}$ strictly within
  the subtree rooted at $x$.  It is obtained as $select_{\text{\it tag\/}}(\text{\it Tag},$
  $\text{\it rank\/}_{\text{\it tag\/}}(\text{\it Tag},x)+1)$ if it is $\le \textrm{Close}(x)$, and undefined
  otherwise.
\item TaggedPrec$(x,\text{\it tag\/})$:  The last node labeled $\text{\it tag\/}$ with preorder
  smaller than that of node $x$, and not an ancestor of $x$.  Let
  $r=\text{\it rank\/}_{\text{\it tag\/}}(\text{\it Tag}, x-1)$. If $select_{\text{\it tag\/}}(\text{\it Tag}, r)$ is not an ancestor of
  node $x$, we stop. Otherwise, we set $r=r-1$ and iterate.
\item TaggedFoll$(x,\text{\it tag\/})$:  The first node labeled $\text{\it tag\/}$ with preorder
  larger than that of $x$, and not in the subtree of $x$.  This is
  $select_{\text{\it tag\/}}(\text{\it Tag},\text{\it rank\/}_{\text{\it tag\/}}(\text{\it Tag},\textrm{Close}(x))+1)$.
\end{itemize}

\subsubsection{Connecting the Text and the Tree}
\label{sssec:connect}

Conversion between text numbers, tree nodes, and global identifiers, is
easily carried out by using $\text{\it Par\/}$ and a bitmap $B$ of $2n$ bits that marks
the opening parentheses of tree leaves containing text, plus $o(n)$ extra
bits to support rank/select queries; the latter uses an implementation
of~\cite{RRR02} which is described in~\cite{CNspire08.1}.
The bitmap $B$ enables the computation of
the following operations:

\begin{itemize}
\item LeafNumber$(x)$: Gives the number of leaves up to $x$ in $\text{\it Par}$. This is
  $\text{\it rank\/}_1(B,x)$.
\item TextIds$(x)$: Gives the range of text identifiers that descend from
  node $x$. This is simply
  $[\textrm{LeafNumber}(x-1)+1,\textrm{LeafNumber}(\textrm{Close}(x))]$.
\item XMLIdText$(d)$: Gives the global identifier for the text with
  identifier $d$.  This is Preorder$(select_1(B,d))$.
\item XMLIdNode$(x)$: Gives the global identifier for a tree node $x$. This
  is just Preorder$(x)$.
\end{itemize}

\subsection{Displaying Contents}

Given a node $x$, we want to recreate its XML serialization, that is,
return (a portion of) the original XML string.
We traverse the structure starting from $\mbox{\it
  Par\/}[x]$, retrieving the tag names and the text contents, from the text
identifiers. The time is $O(\log\sigma)$ per text symbol (or $O(1)$ if we
use the redundant text storage described in Section~\ref{sec:text}) and
$O(1)$ per tag.


\begin{itemize}
\item GetText$(d)$: Generates the text with identifier $d$.
\item GetSubtree$(x)$: Generates the subtree at node $x$.
\end{itemize}

%% file: automata.tex
\section{XPath Queries}
\label{sec:queries}

Our goal is to support a practical subset of XPath, while being able
to guarantee efficient evaluation based on the data structures
described in the previous sections.  As a first shot we target the
forward fragment of ``Core XPath'' \cite{gotkocpic05}. We focus our
presentation on the \texttt{descendant} and \texttt{child} axes, but
\texttt{self}, \texttt{attribute} and \texttt{following-sibling} are
also part of our implementation.  Thus, the non-terminal ``Axis'' in
the following EBNF can here be thought of as generating only the
terminals ``\texttt{descendant}'' and ``\texttt{child}''.  A node test (non-terminal
``NodeTest'' below) is either the wildcard (``*''), a tag name, or a
node type test, i.e., one of ``\texttt{text()}'' or ``\texttt{node()}''.
Note that our current prototype does not support 
 the node type
tests ``\texttt{comment()}'' and ``\texttt{processing\-instruction()}''.  
Of course, additional to Core XPath, we
support all text predicates of XPath~1.0, i.e., the $=$, contains, and
starts-with predicates.  Here is an EBNF for Core XPath.

\begin{small}
\begin{center}
\begin{tabular}{lcl}
Core&::=& LocationPath $|$ `/' LocationPath\\
LocationPath&::=&LocationStep (`/' LocationStep)*\\
LocationStep&::=&Axis `::' NodeTest\\
&&$|$ Axis `::' NodeTest `[' Pred `]'\\
Pred&::=& Pred `and' Pred $|$ Pred `or' Pred\\
&&$|$ `not' `(' Pred `)' $|$ Core $|$ `(' Pred `)'
\end{tabular}
\end{center}
\end{small}

A {\em data value} is the value of an attribute or
the content of a text node. Here, all data values are
considered as strings.
If an XPath expression selects only data values
then we call it {\em value expression}.
In our fragment, $p$ is a value expression if
its last axis is the attribute axis
or the text() test. Inside of a filter
we call ``self'' (and ``.'')
a value expression if the last axis to the left of
the filter is a value expression.
Our XPath fragment (``Core+''), consists of Core XPath
plus the following data value comparisons which may
appear inside filters (that is, may be generated
by the nonterminal Pred of above).
Let $w$ be a string and $p$ a value expression.
\begin{itemize}
\item $p$ = $w$ (equality): tests if the string $w$ is
equal to a string selected by $p$.
\item contains$(p,w)$: tests if the string $w$ is
contained in a string selected by $p$.
\item starts-with$(p,w)$: tests if the string $w$ is
a prefix of a string selected by $p$.
\end{itemize}

\subsection{Tree Automata Representation}

Tree automata are a well-known and popular tool for reasoning about
XML, see, e.g.,~\cite{DBLP:journals/sigmod/Neven02,
  DBLP:journals/jcss/Schwentick07, DBLP:conf/icde/GenevesL10,
  DBLP:journals/japll/LibkinS10}.  Only seldomly have they been used
as a tool for query evaluation. In~\cite{greenEA04} automata are used
to evaluate, on an XML stream, many (very simple) XPath queries in
parallel.  It is well-known that Core XPath can be evaluated using
tree automata; see, e.g.,~\cite{kochvldb03} and \cite{bjorklund09}.
Here we use alternating tree automata (as in \cite{tata07} and
\cite{haruo:book}).  Such automata work with Boolean formulas over
states, which must become satisfied for a transition to fire.  This
allows a much more compact representation of queries through automata,
than ordinary tree automata (without formulas).  Our tree automata
are defined over a binary tree view of the XML tree where the left child is
the first child of the XML node and the right child is the next
sibling of the XML node.

\begin{definition}[Non-deterministic marking automaton]
  An automaton $\mathcal{A}$ is a tuple
  $(\mathcal{L},\mathcal{Q},\mathcal{T},\mathcal{B}, \delta)$, where:
  \begin{itemize}
  \item $\mathcal{L}$ is a countable (possibly infinite) set of tree
    labels;
  \item $\mathcal{Q}$ is the finite set of states;
  \item $\mathcal{T}\subseteq \mathcal{Q}$ is the set of top states
    (that is, states that must be satisfied at the root node);
  \item $\mathcal{B}\subseteq \mathcal{Q}$ is the set of bottom states
    (that is, states that must be satisfied at the leaves);
  \item $\displaystyle\delta : \mathcal{Q}\times 2_f^\mathcal{L}\cup
    2_{\textit{cof}}^\mathcal{L} \rightarrow F$ is the transition
    function, where $F$ is the set of Boolean formulas
    \footnote{We denote by $2_f^\mathcal{L}$ the set of finite subsets
      of $\mathcal{L}$ and by $2_{\textit{cof}}^\mathcal{L}$ the set
      of co-finite subsets of $\mathcal{L}$.}.  A
    \emph{Boolean formula} $\phi$ is a finite production of the
    grammar:
\begin{displaymath}
\begin{array}{lcll}
    \phi & ::= & \top ~|~ \bot ~|~ \phi\lor\phi ~|~ \phi\land\phi ~|~
    \lnot \phi ~|~    a ~|~ p &
    \textrm{(formula)}\\
    a & ::= & \downarrow_1\! q ~|~ \downarrow_2\! q & \textrm{(atom)}\\
  \end{array}
\end{displaymath}
where $p\in P$ is a built-in \emph{predicate} and $q$ is a state.
\end{itemize}
\end{definition}

Before explaining in details the use of formulas, we motivate our use
of finite or co-finite sets as guards for transitions. While
traditionally automata transitions are guarded by a state and a single
label, this would make the encoding of XPath into automata very
tedious and needlessly complicate the algorithms. Indeed, one of the
features of XPath is a wildcard node test, namely ``\texttt{*}''. One
solution could be to suppose that for a given automaton the set of
labels of the input document is known in advance and that this set is
used as alphabet for the automaton. Unfortunately, this does not 
accurately reflect the semantics of XPath in which a query can be defined
independently of any document and can even be executed on \emph{any}
document (it might not yield any result but its application is
nonetheless valid). Another solution (as in
\cite{greenEA04}) is to equip automata with a special ``default''
transition, labelled for instance ``\texttt{\_}'', which is taken if
in the current state no other transition can be evaluated. This has
two drawbacks. Firstly, it is only well-defined for deterministic tree
automata (our encoding makes heavy use of
nondeterminism). Secondly, the evaluation function is
polluted by the special cases which handle this default transition.
Our solution is more blunt. We guard transitions by finite or
co-finite sets of labels, and a transition is taken if the label of the
current node is a member of that set. For instance, 
the ``\texttt{*}'' XPath test is encoded as a transition guarded by the
set $\mathcal{L} - \{ \text{@}, \text{\#}\}$, where ``@'' and ``\#''
represent labels of subtrees containing attribute nodes and text nodes
in our encoding. This allows us to give a very straightforward
evaluation function for tree automata, which relies on the evaluation
of Boolean formulas,  presented next.

\begin{definition}[Evaluation of a formula]
  Given an automaton $\mathcal{A}$ and an input tree $t$, the
  evaluation of a formula is given by the judgment
  $\mathcal{R}_1,\mathcal{R}_2, n\vdash_{\mathcal{A}} \phi = (b,R)$
  where $\mathcal{R}_1$ and $\mathcal{R}_2$ are mappings from states
  to sets of nodes of $t$, $n$ is a node of $t$, $\phi$ is a
  formula, $b\in\{\top,\bot\}$, and $R$ is a set of nodes of
  $t$. We define the semantics of this judgment by the means of the
  inference rules given in Figure~\ref{fig:formsem}.
\end{definition}

\newcommand{\smallfrac}[2]{\frac{\textrm{\footnotesize
      $#1$}}{\textrm{\footnotesize $#2$}}}
\begin{figure}[ht!]
\begin{displaymath}
\begin{array}{c}
  \smallfrac{}{
    \mathcal{R}_1,\mathcal{R}_2,t'\vdash_{\mathcal{A}} \top
    = (\top,\emptyset)}\textbf{\small(true)} \\[10pt]

  \smallfrac{\mathcal{R}_1,\mathcal{R}_2,t'\vdash_{\mathcal{A}} \phi
    = (b,R)}
  {\mathcal{R}_1,\mathcal{R}_2,t'\vdash_{\mathcal{A}} \lnot\phi
    = (\overline{b},\emptyset)}\textbf{\small(not)} \\[10pt]

  \smallfrac{\begin{array}{c}
      \mathcal{R}_1,\mathcal{R}_2,t'\vdash_{\mathcal{A}}
      \phi_1=(b_1,R_1)\\
      \mathcal{R}_1,\mathcal{R}_2,t'\vdash_{\mathcal{A}}
      \phi_2=(b_2,R_2)\\
      \end{array}
  }
  {\mathcal{R}_1,\mathcal{R}_2,t'\vdash_{\mathcal{A}} \phi_1\lor\phi_2
    = (b_1,R_1)\varovee (b_2,R_2)}\textbf{\small(or)} \\[10pt]
    \smallfrac{\begin{array}{c}
      \mathcal{R}_1,\mathcal{R}_2,t'\vdash_{\mathcal{A}}
      \phi_1=(b_1,R_1)\\
      \mathcal{R}_1,\mathcal{R}_2,t'\vdash_{\mathcal{A}}
      \phi_2=(b_2,R_2)\\
      \end{array}
  }
  {\mathcal{R}_1,\mathcal{R}_2,t'\vdash_{\mathcal{A}} \phi_1\land\phi_2
    = (b_1,R_1)\varowedge (b_2,R_2)}\textbf{\small(and)} \\[10pt]
    \smallfrac{q\in\textrm{dom}(\mathcal{R}_i)}
  {\mathcal{R}_1,\mathcal{R}_2,t'\vdash_{\mathcal{A}}
    \downarrow_i\! q = (\top,\mathcal{R}(q))}\text{\small for~$i\in\{1,2\}$}~\textbf{\small(left,right)}\\[10pt]
  \smallfrac{}
  {\mathcal{R}_1,\mathcal{R}_2,t'\vdash_{\mathcal{A}}
    \texttt{mark} = (\top,\{t'\})}\textbf{\small(mark)}\\[5mm]
  \smallfrac{EvalPred($p$, $t'$ )=$b, R$}
  {\mathcal{R}_1,\mathcal{R}_2,t'\vdash_{\mathcal{A}}
    p = (b, R)}\textbf{\small(pred)}\\[10pt]

\smallfrac{\textrm{when no
      other rule applies}}{\mathcal{R}_1,\mathcal{R}_2,t'\vdash_{\mathcal{A}}
    \phi = (\bot,\emptyset)}
\end{array}
\end{displaymath}
{\small
where:
\begin{displaymath}
\begin{array}{l}
  \overline{\top} =  \bot\quad\text{ and }\quad\overline{\bot}  =  \top\\
  (b_1,R_1)\ovee(b_2,R_2) =  \left\{\small\begin{array}{cr}
      \top,R_1 & \textrm{if $b_1 = \top$, $b_2 = \bot$}\\
      \top,R_2 & \textrm{if $b_2 = \top$, $b_1 = \bot$}\\
      \top,R_1\cup R_2 & \textrm{if $b_1 = \top$, $b_2 = \top$}\\
      \bot,\emptyset & \textrm{otherwise}\\
    \end{array}\right.\\[10pt]
  (b_1,R_1)\owedge(b_2,R_2) =  \left\{\small\begin{array}{cr}
      \top,R_1\cup R_2 & \textrm{if $b_1 = \top$, $b_2 = \top$}\\
      \bot,\emptyset & \textrm{otherwise}\\
    \end{array}\right.
\end{array}
\end{displaymath}}
\caption{Inference rules defining the evaluation of a formula}
\label{fig:formsem}
\end{figure}

These rules are straightforward and combine the rules for a
classical alternating automaton, with the rules of a marking
automaton.
Rules~\textbf{(or)} and \textbf{(and)}
implements the Boolean connective of the formula and
collect the marking found in their true sub-formulas.
Rules~\textbf{(left)} and \textbf{(right)} (written as a rule
schema for
concision) evaluate to true if the state $q$ is in the corresponding
set. Intuitively, $\mathcal{R}_1$ (resp. $\mathcal{R}_2$) is the set
of states recognizing the left (resp. right) subtree of the input
tree. Rule \textbf{(pred)}
supposes the existence of an evaluation function for built-in
predicates. Among the latter, we suppose the existence of a special
predicate \texttt{mark}, which evaluates to $\top$ and returns the
singleton set containing the current node.

We now give the semantics of an automaton by means of the \emph{run
  function} \FN{TopDownRun} (see Figure~\ref{fig:algtd}).
\begin{figure}[ht!]
\hrulefill
\vspace*{-5mm}
\begin{algorithme}
\FN{TopDownRun}($\mathcal{A}$, $t$, $r$)\\
\lign \If{$t$ is the empty tree}\\
\lign \> \Return $\{q\rightarrow\varnothing\mid q\in \mathcal{B} \cap
r\}$\\
\lign \Else\\
\lign \> $\textit{trans} := \{ q,\ell\,\rightarrow \phi\mid
q\in r \textrm{~and~} \textrm{Tag}(t) \in \ell\}$\\[1mm]
\lign \> $r_i := \{ q\mid\downarrow_i\! q \in \phi, \forall
\phi \in \textit{trans}\}$,  ~for $i\in\{1, 2\}$\\[1mm]
\lign \> $\mathcal{R}_1 := \FN{TopDownRun}(\mathcal{A},
\textrm{FirstChild}(t), r_1)$\\[1mm]
\lign \> $\mathcal{R}_2 := \FN{TopDownRun}(\mathcal{A},
\textrm{NextSibling}(t), r_2)$\\[1mm]
\lign \> \Return $\displaystyle\hspace*{-4mm}\bigcup_{~~~~(q,\ell\rightarrow \phi) \in \textit{trans}}\hspace*{-5mm}\{ q\mapsto R\mid\mathcal{R}_1,\mathcal{R}_2,t\vdash_{\mathcal{A}},\phi= (\top,R)\}$\\
\end{algorithme}
\vspace*{-5mm}
\hrulefill
\caption{Evaluation function for tree automata}
\label{fig:algtd}
\end{figure}
This algorithm is based on the text book algorithm for recursive
bottom-up evaluation of tree automata (see e.g. \cite{haruo:book}).
The algorithm performs a recursive first child/next sibling
traversal of the tree until a leaf is reached (base case for the
recursion). When returning form the recursive evaluation on the left
and right subtrees (Lines~6 and 7, Figure\ref{fig:algtd}) the function
evaluate the set of transitions for the current node, based on the set
of states recognizing the left and right subtree. However,
instead of blindly doing a recursive descent from the root to the
leaves and evaluating when returning from the recursive calls, the
transitions are restricted by the set of states $r$ (Line~4). This
technique is dubbed ``bottom-up evaluation with top-down
preprocessing'' in \cite{haruo:book}. We therefore named the run
function \FN{TopDownRun} to differentiate it from a real bottom-up
run (starting from the leaves of the tree) that we present in
Section~\ref{sss:bu}. The novelty is our use of maps from states to
nodes instead of only sets of states, to efficiently implement the
marking of selected nodes.

\subsection{From XPath to Automata}
\label{aut}

The translation of an XPath query to an alternating automaton is a
simple syntax-directed translation which can be carried out in one
pass through the parse tree of the query.  Roughly
speaking, the resulting automaton is ``isomorphic'' to the original
query (and has essentially the same size). 
%
We illustrate the translation by an example. Consider the query
\begin{multline*}
\texttt{/descendant::listitem/descendant::keyword[}\\
\texttt{emph]}
\end{multline*}
for which the automaton is
\[
\mathcal{A}=(\mathcal{L}, \{q_0,q_1,q_2\}, \{q_0\}, \{
  q_1, q_2\}, \delta)
\]
where $\delta$ contains the following transitions (recall that \&, @,
and \# denote the special tags for the document node, attribute nodes,
and text nodes, respectively):
\begin{displaymath}
\begin{array}{l|@{\hspace{3pt}}lcl}
1 & q_0, \{\text{\&}\} & \rightarrow & \downarrow_1\! q_1\\[1mm]
2 & q_1, \{\TAG{listitem}\} & \rightarrow & \downarrow_1\!
q_2 ~\land~ \downarrow_1\! q_1 ~\land~ \downarrow_2\! q_1 \\
3 & q_1, \mathcal{L}-\{\text{@},\text{\#}\} & \rightarrow &
\downarrow_1\! q_1 ~\land~ \downarrow_2\! q_1\\[1mm]
4 & q_2, \{\TAG{keyword}\} & \rightarrow & \texttt{mark}~\land~
\downarrow_1\! q_3 ~\land~ \downarrow_1\! q_2 ~\land~ \downarrow_2\! q_2 \\
5 & q_2, \mathcal{L}-\{\text{@},\text{\#}\} & \rightarrow &
\downarrow_1\! q_2 ~\land~ \downarrow_2\! q_2\\[1mm]
6 & q_3, \{ \TAG{emph} \} & \rightarrow & \top\\
7 & q_3, \mathcal{L}-\{\text{@},\text{\#}\} & \rightarrow &
\downarrow_2\! q_3\\
\end{array}
\end{displaymath}
It is clear that this encoding is linear in the size of the query. For
each step, we create one state and two transitions. The first
transition (Transitions 2, 4, and 6 above) represents the action the
automaton performs when the current node matches the step at
issue. For instance if Transition 2 is taken, it means that the
current node has label \TAG{listitem} and that:
\begin{itemize}
\item the state encoding the rest of the query, namely $q_2$,
  recognizes the left subtree (hence the $\downarrow_1\! q_2$)
\item the current step also holds recursively for the descendant
  and following nodes of the current node (because of
  \mbox{$\downarrow_1\! q_1\land\downarrow_2\! q_1$}).
\end{itemize}
The second transition associated with a step handles the default case
and the recursion. For instance, in Transition~3, if the current node
is not a \TAG{listitem}~ \emph{or} if it is a \TAG{listitem} for
which the continuation path does not hold, then we just stay in the
same states for descendant and following nodes. This use of
non-determinism (since $\{\texttt{listitem}\}\subseteq
\mathcal{L}-\{\text{@}, \text{\#}\}$) is crucial to keep the automaton
linear in the size of the query. Lastly, note how ``top-level steps''
(\TAG{listitem} and \TAG{keyword} in our example) are encoded by
universal transitions, in which the use of the ``$\land$'' connectives
forces the formula to be recursively checked down to the leaves (and
therefore explores the tree to find all occurrences of such steps) while
``filter steps'' (here \TAG{emph}) are encoded as existential
transitions, which become satisfied as soon as one node verifies them.

\subsection{Leveraging the Speed of the Low-Level Interface}
\label{subsec:leverage}
We have seen how to evaluate an XPath query by compiling it into a
tree automaton and running the latter on the input document. We present
now several techniques that make use of the tree and text index
presented in Sections~\ref{sec:tree} and \ref{sec:text}.
It is these techniques that make our SXSI prototype
competitive in speed with state-of-the-art XML databases.

\subsubsection{Jumping to relevant nodes}

Conventionally, the run of a tree automaton
visits every node of the input tree. This is for instance the
behaviour of the tree automata presented in \cite{kochvldb03}, which
perform two scans of the whole XML document (the latter being stored
on disk in a particular format). However, for typical queries, most of
the nodes are ``useless'' in the sense that the automaton only loops
through them staying in the same set of states.
In other words, the
automaton \emph{ignores} most of the nodes.
To restrict the run to interesting nodes, we use the notion of
\emph{relevant nodes} introduced in~\cite{xpathwhole}.  While the full
characterization is out of the scope of this paper, we give a flavor
of relevant nodes, using an example.  Consider the query
\[
\texttt{/descendant::listitem/descendant::keyword}
\]
Clearly, we only care about \TAG{listitem} and \TAG{keyword} and how
they are positioned with respect to each other.  This is precisely the
information that is provided through the TaggedDesc and TaggedFoll
functions of the tree representation. These functions allow us to have
a ``contracted'' view of the tree, restricted to nodes with certain
labels of interest (but preserving the overall tree structure).  For
instance, to solve the above query we can call
TaggedDesc(Root,~\TAG{listitem}) which selects the first
\TAG{listitem} node $x$. Now, simply traverse recursively the subtree
rooted at $x$, using TaggedDesc(\_, \TAG{keyword})  and
TaggedFoll(\_, \TAG{keyword}) instead of FirstChild and
NextSibling.  After this, we determine the next \TAG{listitem}
node using TaggedFoll($x$, \TAG{listitem}).
We do this optimization of ``jumping run'' based on the automaton: for
a given set of states of the automaton we compute the set of relevant
transitions which cause a state change. The labels of those
transitions are the relevant labels to which we jump, using
TaggedDesc and TaggedFoll.  For instance, in the automaton for the
above query (which is the same as the one given in Section~\ref{aut},
minus state $q_3$ and corresponding transitions) only Transitions~2
and 4 are relevant (that is, these transitions are valid only when the
automaton is on a relevant node). Thus, in state $q_0$ the
automaton can use TaggedDesc to jump to \TAG{listitem} nodes, and in state
$q_1$ it can jump to \TAG{keyword} nodes. It should be noted that for
such a query, our ``jumping run'' is optimal: the automaton only
visits the top-most \TAG{listitem} nodes and all the \TAG{keyword}
nodes below them.
This behaviour is similar to the idea of
``partitioning and pruning'' in the staircase
join~\cite{grukeuteu03}, but here achieved by means of automata.

\subsubsection{Bottom-Up Runs}
\label{sss:bu}

While the previous technique works well for
tree-based queries it still remains slow for very selective
value-based queries. For instance, consider the query
\[
\texttt{//listitem//keyword[contains(.,"Unique")]}
\]
The text interface described in Section~\ref{sec:text} can answer
the text predicate very efficiently returning the set of text nodes
matching this \emph{contains} query.
If the number of occurrences is low, and in particular
smaller than the number of \texttt{listitem} or \texttt{keyword} tags
in the document
(which can also be determined efficiently through the tree
structure interface),
then it would be faster to take these text nodes as starting point for
query evaluation and test if their path upward to the root matches the
XPath expression \texttt{//listitem//keyword}. This scheme is
particularly useful for text oriented queries with low selectivity
text predicates.  However, it also applies for tree only queries:
imagine the query \texttt{//listitem//keyword} on a tree with many
listitem nodes but only a few keyword nodes. We can start bottom-up by
jumping to the keyword nodes and then check their ancestors for
listitem nodes. (Note that with the tree index described here, we
cannot directly jump to all bottom-most keyword nodes. We would need
to iterate through all keyword nodes. Direct access could be provided
through additional {\tt sarray}s storing for each label its
bottom-most nodes.)

We now devise a real bottom-up evaluation algorithm of our
automata. The algorithm takes an automaton and a sequence of potential
match nodes (in our example, the text nodes containing the string
\texttt{"Unique"}). It then moves up to the root, using the
Parent function and checks that the automaton arrives at the
root node in its initial state $q_i$.  Note that, if naively done,
such a bottom-up run will visit many nodes repeatedly: if a node is
the common-ancestor of $m$ potential match nodes, then it would be
visited $m$ times.

\begin{figure*}[t]
\begin{framed}
\definecolor{tdrun}{HTML}{1851C4}
\definecolor{matchabove}{HTML}{B7001F}
\definecolor{pausenext}{HTML}{00A300}
\begin{minipage}[t]{4.2cm}
\vspace{0pt}
\def\svgwidth{4.2cm}
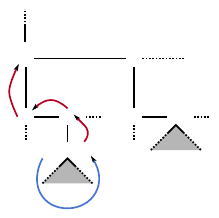\\[-2mm]
\color{tdrun}{\textit{(a)} Initial \FN{TopDownRun}}\\
\color{matchabove}{\textit{(b)} \FN{MatchAbove} walks upwards and
ignores non-matching subtrees}\\
\end{minipage}\hfill
\begin{minipage}[t]{5.3cm}
  \vspace{0pt} \def\svgwidth{4.2cm}
  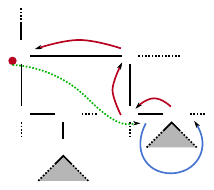\\
  \color{matchabove}{\textit{(a)} \FN{MatchAbove} stops at the common
    ancestor of   $t_1$ and $t_2$}\\
  \color{pausenext}{\textit{(b)} Recursive call on $t_2$, climbs up to
    $t'_1$}
\end{minipage}
\hfill
\begin{minipage}[t]{4.5cm}
  \vspace{0pt}
  \def\svgwidth{4.2cm}
 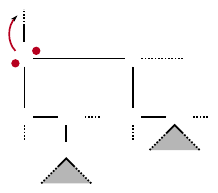\\
\color{matchabove}{Results of left
  and right subtrees are merged at $t'_1$ and the initial
  \FN{MatchAbove} restarts}
\end{minipage}
\caption{Illustration of the bottom-up run}
\label{fig:bottom_up_run}
\end{framed}
\end{figure*}

Instead, we move bottom-up left-to-right, and only move upwards from
the left-most potential match until we reach its lowest common
ancestor with the next potential match.  This technique is
similar in spirit to shift-reduce parsing (see \cite{compilers}).  This scheme is illustrated in Figure~\ref{fig:bottom_up_run}.
Consider a sequence \texttt{[}$t_1$,\ldots,$t_n$\texttt{]} (ordered in
pre-order)  of potentially  matching nodes.   The algorithm  starts on
node  $t_1$.  First, if the node is  not  a  leaf,  we  call  the
\FN{TopDownRun} function  on $t_1$ with  $r=\mathcal{Q}$. This
returns the mapping $\mathcal{R}_1$ of all states accepting $t_1$.  We
move from  $t_1$ upwards  to the document  root, starting  with states
$\textrm{dom}(\mathcal{R}_1)$.  Once we arrive at a node $t'_1$ which
is an ancestor  of the next potential matching subtree $t_2$, we stop
at  $t'_1$  and  start  the   algorithm  on  $t_2$  until  it  reaches
$t'_1$.  Upon reaching  $t'_1$, we  merge both  mappings  and continue
upwards until  we reach the  root or a  common ancestor of  $t'_1$ and
$t_3$, and so  on.  The idea of ``stopping'' at a  node to explore the
the  next potential matching  subtree is  similar to  the \emph{shift}
action of a bottom-up parser: the  stopped node is pushed onto a stack
(here  the recursive call  stack) while  the rest  of the  symbols are
processed.  Similarly, the  ``merging'' of  two nodes  which  are then
replaced  by   their  lowest  common   ancestor  is  similar   to  the
\emph{reduce} action  of a bottom-up  parser: two symbols  are removed
from  the  parsing stack  and  replaced  by  the non-terminal  of  the
corresponding  grammar rule.  \emph{Merging}  the runs  at the  lowest
common  ancestor guaranties  that we  never touch  any node  more than
once,  during a  bottom-up run.  The bottom-up run algorithm is given
in Figure~\ref{fig:bualg}.
\begin{figure}
\hrulefill
\vspace*{-5mm}
\begin{algorithme}
\FN{BottomUpRun}($\mathcal{A}$, $s$)\\
\lign \If{ $s$ is empty}\\
\lign \> \Return $\varnothing$\\
\lign \Else\\
\lign \> $t_1, s' := \text{Head(s)}, \text{Tail(s)}$\\
\lign \> $\mathcal{R} := \FN{TopDownRun}(\mathcal{A}, t_1,
  \mathcal{Q})$\\
\lign \> $\mathcal{R}',s'' := \FN{MatchAbove}(\mathcal{A}, t,
  s',  \mathcal{R}, \#)$\\
\lign \> \Return $\mathcal{R}'$\\[2mm]
\FN{MatchAbove}($\mathcal{A}$, $t$, $s$, $\mathcal{R}_1$, \textit{stop})\\
\lign \If{ $t$ = \textit{stop}}\\
\lign \> \Return $\mathcal{R}_1, s$\\
\lign \Else\\
\lign \> $p := \textrm{Parent}(t)$\\
\lign \> \If { $s$ is empty or not(\textrm{IsAncestor}($p$, $t_2$)) }\\
\lign \> \> $\mathcal{R}_2, s'' := \varnothing, s$\\
\lign \> \Else\\
\lign \> \> $t_2, s' := \text{Head(s)}, \text{Tail(s)}$\\
\lign \> \> $\mathcal{R} := \FN{TopDownRun}(\mathcal{A}, t_2, \mathcal{Q})$\\
\lign \> \> $\mathcal{R}_2, s'' := \FN{MatchAbove}(\mathcal{A},
  t_2, s', \mathcal{R}, p)$\\[2mm]
\lign \> $\textit{trans} := \{ q,\ell\rightarrow \phi \mid \begin{array}{l}
     \exists q'\in \textrm{dom}(\mathcal{R}_i) s.t. \downarrow_i\! q'\in \phi\\
     \textrm{Tag}(p)\in \ell
     \end{array} \}$\\[2mm]
\lign \> $\displaystyle\mathcal{R}' :=
\hspace{-5mm}\bigcup_{~~~(q,\ell\rightarrow \phi) \in \textit{trans}}\hspace{-5mm}\{ q\mapsto R \mid
     \mathcal{R}_1,\mathcal{R}_2,t\vdash_{\mathcal{A}},\phi= (\top,R)\}$\\
\lign \> \Return \FN{MatchAbove}($\mathcal{A}$, $p$, $s'$,
  $\mathcal{R}'$, \textit{stop})\\
\end{algorithme}
\vspace*{-4mm}
\hrulefill
\caption{Bottom-up evaluation function}
\label{fig:bualg}
\end{figure}

The first function takes an automaton and a sequence of potential
matches in pre-order, and proceed to run the automaton bottom-up from
the left-most potential match ($t_1$, Line~6, Figure \ref{fig:bualg}).
The \FN{MatchAbove} function is the one ``climbing-up'' the
tree. We assume that the Parent$(\_)$ function returns the empty tree
when applied to the root node. If the input node is not equal to the
sentinel \emph{stop} node (which is initially the empty tree
\texttt{\#}, allowing to stop only after the root node has been
processed) then we first check whether the next potential match
 is a descendant of our parent (Line
12). If so, then we pause for the current branch and recursively
call \FN{MatchAbove} with our parent as \emph{stop} tree. Once it
returns, we compute all the possible transitions that the automata can
take from the parent node to arrive on the left and right subtree with
the correct configuration (Line~18). We then
\emph{merge} both configurations using the same computation as in the
top-down algorithm (Line~19). Finally, we recursively call
\FN{MatchAbove} on the parent node, with the new configuration
and sequence of potential matching nodes (Line~20).

\subsection{General Optimizations, On-the-fly Determinization}
\label{sect:opt}
While the optimizations presented in the previous sections give the
most important speed-up we describe hereafter a series of
implementation techniques used for the efficient evaluation of
automata.

\subsubsection{Hash consing of data-structures}
\label{ssec:hashcon}
We use hash consing for all critical data-structures: sets of states,
formulas, sets of transitions, sets of labels and so on. Hash consed
values have the following two properties. First, structurally equal
values are shared in memory. Therefore testing for equality of such
values (for instance testing that two sets of transitions are equal)
consists in comparing their memory address (which is cheap). Second,
to each such value we can associate a unique integer id (this can be
its memory address for instance but more interestingly a small integer
assigned at the creation of the value). These two properties
---especially the second one--- are instrumental to the other
optimizations. Indeed, as described in \cite{ConchonFilliatre06}, we
can memoize (or cache) the results of expensive computations and reuse
them when needed instead of recomputing them. In particular, we can
associate to each function a table, indexed by the the argument's
id. While the first computation might be expensive, its result is
stored once and for all in the table and can be retrieved with one
pointer indirection later on, when the same computation is
requested. We explain now how this generic technique comes into play
for automata evaluation.

\subsubsection{Just-in-time compilation of automata}
\label{ssec:jit}

In the \FN{TopDownRun} algorithm (Figure~\ref{fig:algtd}) the most
expensive operations are in Lines~4, 5, and 8. By expensive we mean
that they take time $O(|Q|)$ where $|Q|$ is the number of steps in the
original query. At Line~4, we gather all the transitions that can be
selected from the current label $\ell$ and set of states $r$. From
these we compute, at Line~5, the new set of states $r_1$ and $r_2$
onto which we will launch the recursive call. As explained in
Section~\ref{subsec:leverage}, from the set of states $r_1$
(resp. $r_2$) we compute the ``jump'' moves that the automaton will do
to reach the next node in the left (resp. right) subtree. If none of
the formulas requires the evaluation of a value predicate (such as
\texttt{contains} for instance) then we can see that this whole
computation of Lines~4 and 5 can be cached in a 2-dimensional array,
using only $\ell$ (the current label, identified by a small integer)
and $r$ (a hash consed set of states with a unique small id) as
key. In practice we store in this table a small sequence of
instructions that are computed \emph{at run time} and which represent
the behaviour of the automaton for the next step (for instance ``jump
to the next \texttt{keyword} label in state $\{q_0, q_1\}$). This
just-in-time compilation scheme absorbs in practice most of the
overhead caused by the automaton machinery and makes running an
automaton almost as fast as executing a hand-written, precompiled
function. In the same fashion, the computation of the judgment
$\vdash_{\mathcal{A}}$ can be memoized, this time in two parts. First,
the sets of states (that is the domain of the resulting mapping) is
simply stored once and for all, and second, a sequence of instructions
telling how to propagate the results from the left and right subtrees
is stored and evaluated for each node.

\subsubsection{Handling of result sets}
\label{ssec:resultsets}
Maintaining sets of (result) nodes can be expensive. Our efficient
management of sets of nodes relies on the following two observations.
First, note that only the states outside of filters actually
accumulate nodes.  All other states always yield empty bindings.  Thus
we can split the set of states into marking and regular states.  This
reduces the number of $\ovee$ and $\owedge$ operations on result sets.
Note also that given a transition $q_i,\ell\rightarrow \downarrow_1\!
q_j \land \downarrow_2\!  q_k$ where $q_i$, $q_j$, and $q_k$ are
marking states, all nodes accumulated in $q_j$ are in the left subtree
of the current node. Likewise, all the nodes accumulated in $q_k$ are
subtrees of the right subtree of the current node. Thus both sets of
nodes are disjoint and we do not need to keep sorted sets of nodes but
only need sequences which support $O(1)$ concatenation. Computing the
union of two result sets $R_j$ and $R_k$ can therefore be done in
constant time and consequently $\ovee$ and $\owedge$ can be
implemented in constant time.

\subsubsection{Lazy result sets}
\label{ssec:lazyres}
Another way to leverage the speed
and jumping capabilities of our tree index is by making use of a lazy
result set. Consider the query
\texttt{//listitem//keyword}. When reaching a
\TAG{listitem} node, the automaton is in a state which encodes the
following behaviour: ``accumulate all \TAG{keyword} nodes below
this node''. Therefore instead of having the automaton jump through
the subtree to individually put each \TAG{keyword} node in the
result set, we only store the \TAG{listitem} node (\emph{i.e.} the
current node during evaluation) and a flag to remember that during
serialization, it is not the \TAG{listitem} node which should be
printed but rather all its \TAG{keyword} descendants. Since our
tree index allows us to reach each such node using a constant time
jump operation, we delay the process of getting all the final result
nodes until serialization, therefore speeding up the materialization
process. This not only saves time but also memory since the full set
of nodes do not have to by materialized in memory.


\subsubsection{Early evaluation of formulas}
\label{ssec:earlyeval}
Another optimization consists in evaluating the Boolean formulas of
the automaton as early as possible. First, remark that in the
\FN{TopDownRun} algorithm, a node is ``visited'' three times. Once
when the automaton enters the node, during the top-down phase
(Line~1). Here, we only know that at most all states in $r$ yield a
successful run. Then when returning from the left subtree (Line~6), we
know $\mathcal{R}_1$ that is, the states which yield an accepting run
for the left subtree. The idea now is to perform a partial evaluation
of formulas using only $\mathcal{R}_1$. If this happens to be
sufficient to prove or disprove the states in $r$, then the right
subtree can be skipped altogether. This optimization is very important
for filters as it insures that for instance in a query such as
\TAG{//listitem[.//keyword]} the run function only tests for the
presence of the left-most \TAG{keyword} node below a \TAG{listitem}
node.

\subsubsection{Relative Tag position tables}
\label{ssec:relativetag}
As explained earlier, the transitions for the query
\[
\mbox{\TAG{\ldots/descendant::keyword/\ldots}}
\]
 would be (just-in-time)
compiled into a piece of code performing a subtree traversal using
TaggedDesc(\_ , \TAG{keyword}) and TaggedFoll(\_ ,
\TAG{keyword}) instead of FirstChild and Next\-Sibling.  This is
already optimal for documents where \TAG{key}\-\TAG{word} nodes may appear
arbitrarily. However, it is often the case that labels are not
recursive (that is, nodes with a label $l$ do not occur below other
$l$-labelled nodes). To further optimize the compilation of the
automaton, we build --while indexing the document-- four relative
position tables, telling for each label $l$ in the document the sets
of labels that occur respectively in child position, descendant
position, following-sibling position and following position. When
compiling at runtime the automaton and generating a call to
TaggedDescendant for a label $l$, we check that this $l$ label can
indeed appear as descendant of the label of the current node (and
similarly for other jumping functions). If the label does not occur,
then the TaggedDescendant call is replaced by a constant function
returning directly the correct sets of states for the left subtree as
well as an empty result set, as if the automaton had made a full-run
on this subtree and found nothing.

%% file: bu_1.pdf_tex
\begingroup%
  \makeatletter%
  \providecommand\color[2][]{%
    \errmessage{(Inkscape) Color is used for the text in Inkscape, but the package 'color.sty' is not loaded}%
    \renewcommand\color[2][]{}%
  }%
  \providecommand\transparent[1]{%
    \errmessage{(Inkscape) Transparency is used (non-zero) for the text in Inkscape, but the package 'transparent.sty' is not loaded}%
    \renewcommand\transparent[1]{}%
  }%
  \providecommand\rotatebox[2]{#2}%
  \ifx\svgwidth\undefined%
    \setlength{\unitlength}{104.8bp}%
    \ifx\svgscale\undefined%
      \relax%
    \else%
      \setlength{\unitlength}{\unitlength * \real{\svgscale}}%
    \fi%
  \else%
    \setlength{\unitlength}{\svgwidth}%
  \fi%
  \global\let\svgwidth\undefined%
  \global\let\svgscale\undefined%
  \makeatother%
  \begin{picture}(1,0.99618321)%
    \put(0,0){\includegraphics[width=\unitlength]{bu_1.pdf}}%
    \put(0.27099237,0.30916029){\color[rgb]{0,0,0}\makebox(0,0)[lb]{\smash{$t_1$}}}%
    \put(0.30916031,0.46183204){\color[rgb]{0,0,0}\makebox(0,0)[b]{\smash{x}}}%
    \put(0.61450382,0.46183204){\color[rgb]{0,0,0}\makebox(0,0)[b]{\smash{x}}}%
    \put(0.61450382,0.72900761){\color[rgb]{0,0,0}\makebox(0,0)[b]{\smash{x}}}%
    \put(0.11832061,0.46183204){\color[rgb]{0,0,0}\makebox(0,0)[b]{\smash{x}}}%
    \put(0.08015267,0.72900761){\color[rgb]{0,0,0}\makebox(0,0)[lb]{\smash{$t'_1$}}}%
    \put(0.80534351,0.46183204){\color[rgb]{0,0,0}\makebox(0,0)[b]{\smash{$t_2$}}}%
    \put(0.49922832,0.19288367){\color[rgb]{0,0,0}\makebox(0,0)[lb]{\smash{\textit{(a)}}}}%
    \put(0.15267176,0.57251906){\color[rgb]{0,0,0}\makebox(0,0)[lb]{\smash{\textit{(b)}}}}%
  \end{picture}%
\endgroup%

%% file: bu_2.pdf_tex
\begingroup%
  \makeatletter%
  \providecommand\color[2][]{%
    \errmessage{(Inkscape) Color is used for the text in Inkscape, but the package 'color.sty' is not loaded}%
    \renewcommand\color[2][]{}%
  }%
  \providecommand\transparent[1]{%
    \errmessage{(Inkscape) Transparency is used (non-zero) for the text in Inkscape, but the package 'transparent.sty' is not loaded}%
    \renewcommand\transparent[1]{}%
  }%
  \providecommand\rotatebox[2]{#2}%
  \ifx\svgwidth\undefined%
    \setlength{\unitlength}{102.4bp}%
    \ifx\svgscale\undefined%
      \relax%
    \else%
      \setlength{\unitlength}{\unitlength * \real{\svgscale}}%
    \fi%
  \else%
    \setlength{\unitlength}{\svgwidth}%
  \fi%
  \global\let\svgwidth\undefined%
  \global\let\svgscale\undefined%
  \makeatother%
  \begin{picture}(1,0.8894043)%
    \put(0,0){\includegraphics[width=\unitlength]{bu_2.pdf}}%
    \put(0.2578125,0.19809818){\color[rgb]{0,0,0}\makebox(0,0)[lb]{\smash{$t_1$}}}%
    \put(0.296875,0.35434818){\color[rgb]{0,0,0}\makebox(0,0)[b]{\smash{x}}}%
    \put(0.609375,0.35434818){\color[rgb]{0,0,0}\makebox(0,0)[b]{\smash{x}}}%
    \put(0.609375,0.62778568){\color[rgb]{0,0,0}\makebox(0,0)[b]{\smash{x}}}%
    \put(0.1015625,0.35434818){\color[rgb]{0,0,0}\makebox(0,0)[b]{\smash{x}}}%
    \put(0.0625,0.62778568){\color[rgb]{0,0,0}\makebox(0,0)[lb]{\smash{$t'_1$}}}%
    \put(0.8046875,0.35434818){\color[rgb]{0,0,0}\makebox(0,0)[b]{\smash{$t_2$}}}%
    \put(0,0.5078125){\color[rgb]{0,0,0}\makebox(0,0)[lb]{\smash{\textit{(a)}}}}%
    \put(0.79348814,0.03483611){\color[rgb]{0,0,0}\makebox(0,0)[lb]{\smash{\textit{(b)}}}}%
  \end{picture}%
\endgroup%

%% file: bu_3.pdf_tex
\begingroup%
  \makeatletter%
  \providecommand\color[2][]{%
    \errmessage{(Inkscape) Color is used for the text in Inkscape, but the package 'color.sty' is not loaded}%
    \renewcommand\color[2][]{}%
  }%
  \providecommand\transparent[1]{%
    \errmessage{(Inkscape) Transparency is used (non-zero) for the text in Inkscape, but the package 'transparent.sty' is not loaded}%
    \renewcommand\transparent[1]{}%
  }%
  \providecommand\rotatebox[2]{#2}%
  \ifx\svgwidth\undefined%
    \setlength{\unitlength}{104.2bp}%
    \ifx\svgscale\undefined%
      \relax%
    \else%
      \setlength{\unitlength}{\unitlength * \real{\svgscale}}%
    \fi%
  \else%
    \setlength{\unitlength}{\svgwidth}%
  \fi%
  \global\let\svgwidth\undefined%
  \global\let\svgscale\undefined%
  \makeatother%
  \begin{picture}(1,0.88675624)%
    \put(0,0){\includegraphics[width=\unitlength]{bu_3.pdf}}%
    \put(0.26679463,0.19577733){\color[rgb]{0,0,0}\makebox(0,0)[lb]{\smash{$t_1$}}}%
    \put(0.30518234,0.34932819){\color[rgb]{0,0,0}\makebox(0,0)[b]{\smash{x}}}%
    \put(0.61228407,0.34932819){\color[rgb]{0,0,0}\makebox(0,0)[b]{\smash{x}}}%
    \put(0.61228407,0.61804221){\color[rgb]{0,0,0}\makebox(0,0)[b]{\smash{x}}}%
    \put(0.11324376,0.34932819){\color[rgb]{0,0,0}\makebox(0,0)[b]{\smash{x}}}%
    \put(0.07485605,0.61804221){\color[rgb]{0,0,0}\makebox(0,0)[lb]{\smash{$t'_1$}}}%
    \put(0.80422265,0.34932819){\color[rgb]{0,0,0}\makebox(0,0)[b]{\smash{$t_2$}}}%
  \end{picture}%
\endgroup%

%% file: experiments.tex

%
%
%

\section{Experimental Results}
\label{sec:exper}
This section presents our experimental results and is organized as
follows. We first describe our experimental settings, test machines,
and benchmark data. We then provide a first round of experiments
illustrating the raw performances of the tree and text index:
indexing time and resulting index size, performing full pre-order
traversal using FirstChild and NextSibling moves, and direct querying of
the text index. A third subsection illustrates how the tree index and
automata-based engine work together to achieve very fast tree-oriented
query evaluation (in particular using the jumping moves described in
Section~\ref{ss:treejump}). We then show how the automaton machinery
can leverage the speed of both the text and tree index by evaluating
queries containing both text and tree predicates. Lastly we illustrate
the versatility of our approach: our engine is easily extended to
support querying of XML document storing bio-genetic data.

We have implemented a prototype XPath evaluator based on the data
structures and algorithms presented in the previous sections. Both the
tree structure and the FM-index were developed in C++, while the
XPath engine was written using the OCaml language.

\subsection{Protocol}
To validate our approach we benchmark our implementation against
two well established XQuery implementations,
MonetDB/XQuery and Qizx/DB. We describe our experimental settings
hereafter.
\paragraph*{Test machine}\quad Our test machine features an Intel Core
i5 platform featuring 3.33Ghz cores, 3.8 GB of RAM and a S-ATA hard drive. The OS is
a 64-bit version of Ubuntu Linux (11.04). The kernel version is 2.6.38 and the
file system used to store the various files is ext4, with default
settings. All tests are run on a minimal environment where only the
tested program and essential services were running. We use the standard
compiler and libraries available on this distribution (namely g++
4.6.1, libxml2 2.7.8 for document parsing and OCaml 3.11.2).

\paragraph*{Qizx/DB}\quad We use version 3.0 of Qizx/DB engine (free
edition), running on top of the 64-bit version of the JVM (with the
\texttt{-server} flag set as recommended in the Qizx user manual).
The maximal amount of memory of the JVM is set to the maximal amount of
physical memory (using the \texttt{-Xmx} flag). We also use the flag
\texttt{-r} of the Qizx/DB command line interface, which allows us to re-run
the same query without restarting the whole program (this ensures
that the JVM's garbage collector and thread machinery do not impact
the performances). We use the timing provided by Qizx debugging flags,
and report the \emph{serialization time} (which actually includes
the materialization of the results in memory and the
serialization). \no{Indeed, since Qizx/DB seems to interleave query
computation and serialization, it is only possible to report a
``running time'' which includes both serialization and query
computation.}

\paragraph*{MonetDB/XQuery}\quad We use version Oct2010-SP1 of MonetDB,
and in particular, version 4.40.3 of MonetDB4 server and version
0.40.3 of the XQuery module (\emph{pathfinder}). We use the timing
reported by the ``\texttt{-t}'' flag of MonetDB client program,
\texttt{mclient}. We keep the materialization time and the
serialization time separate.

\paragraph*{Running times and memory reporting}\quad For each query, we keep
the best of five runs. For Qizx/DB, each individual run consists of two
repeated runs (``\texttt{-r 2}''), the second one being always
faster. For MonetDB, before each batch of five runs, the server is exited
properly and restarted. For all systems, we exclude from the running times the time
used for loading the index into main memory (based on the engines'
timing reports).
\no{We monitored the memory usage by reading, every 200 ms
during the duration of the tested program, the
\texttt{/proc/pid/statm} pseudo-file provided by Linux.
More specifically,
we monitored the so-called \emph{resident set size}, which corresponds
to the amount of process memory actually mapped in physical
memory.}
We monitor the \emph{resident set size} of each process
which corresponds to the amount of process memory actually mapped in
physical memory.
\no{For MonetDB, we kept track of the memory usage of both server
and client. }

For the tests in which serialization is involved we serialize to the
\texttt{/dev/null} device (that is, all the results are discarded
without causing any output operation).

\no{
However, we assured that for
all tested queries the amount of serialized data was roughly the same
(the handling of ignorable white spaces, order of attributes or
indentation of the output file caused the different outputs to be
slightly different), such that for each implementation the
serialization work was the same. This was necessary since, for
instance, Qizx/DB seems to interleave query execution and query
serialization: the serialization time of two different queries returning the
\emph{same results} would often differ by orders of magnitude, thus
indicating that some computation occurs during the
serialization of the results.
}

\paragraph*{Remarks}\quad We also compared with Tauro~\cite{tauro}.
Yet, as it uses a tailored query language, we could not
produce comparable results.

\subsection{Indexing}
Our implementation features a
versatile index. It is divided into three parts. First, the tree
representation composed of the parenthesis structure, as well as the
tag structure. Second, the FM-index encoding the
text collection. Third, the auxiliary text representation allowing fast
extraction of text content.

It is easy to determine from the query
which parts of the index are needed in order to solve it, and thus
load only those into main memory. For
instance, if a query only involves tree navigation,
then having the FM-index in memory is unnecessary.
\no{
Furthermore, if
we are only interested in count queries, then the auxiliary text
representation is not needed. This allows us to achieve a very low
memory footprint for tree-oriented count queries. If we wish to
serialize the result of a tree query, then only the tree structure and
the auxiliary text are needed. Again, we keep a very low memory
footprint (less than the size of the original document), while remaining
efficient to serialize the results.
}
On the other hand, if we are interested in
very selective text-oriented queries, then only the tree part and
FM-index are needed (both for counting and serializing the
results). In this case, serialization is a bit slower (due to the cost
of text extraction from the FM-index) but remains acceptable since the
number of results is low; see Table~\ref{textIndexPerf1} for a comparison
of the serialization speed of the FM-index versus serializing from plain
string buffers in memory.
\no{
We can trade memory for efficient
serialization by loading the
auxiliary text representation.
}

\begin{figure}
\begin{framed}
\vspace{-2mm}
\centerline{\includegraphics[width=6.5cm]{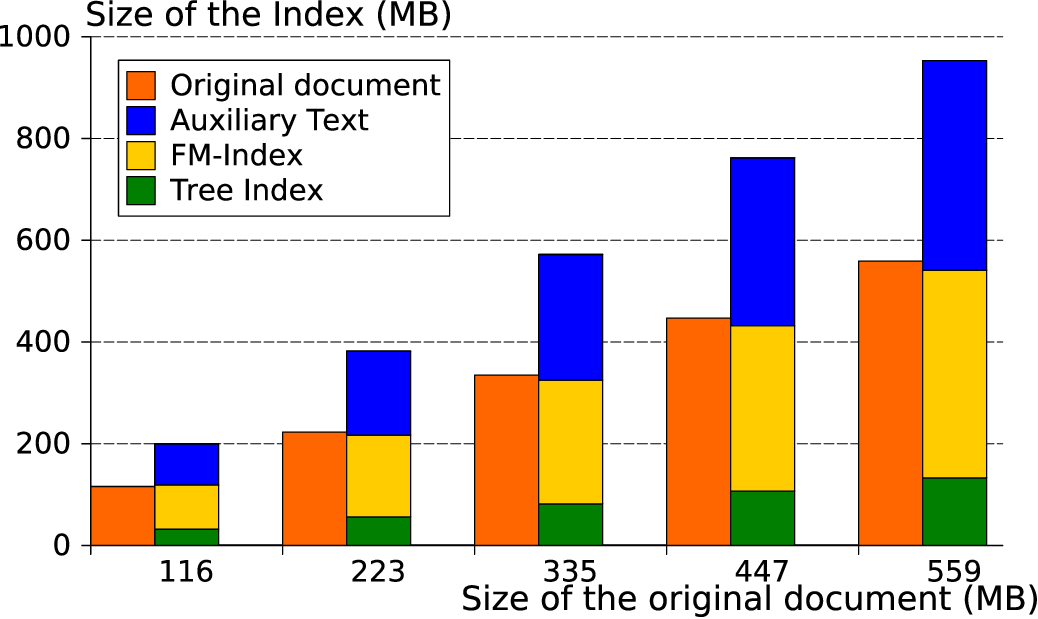}}
\vspace{5pt}
{\footnotesize
\centerline{\begin{tabular}{|l|*{5}{@{\hspace{2pt}}c@{\hspace{2pt}}|}}
\hline
Document Size (MB) & 116 & 223 & 335 & 447 & 559\\
\hline
Index construction time (min) & 5'1 & 10'40 & 17' & 23' & 29'40 \\
\hline
Index construction mem. use (MB)& 296 & 568 & 844 & 1085 & 1387 \\
\hline
\hline
Index loading time (s) & 0.5 & 1.5 & 2.0 & 2.4 & 2.5\\
\hline
\end{tabular}}
}
\caption{Indexing of XMark documents}
\label{fig:index}
\end{framed}
\end{figure}

Figure~\ref{fig:index} reports the construction time and memory
consumption of the indexing process, the loading time from disk
into main memory of a constructed index, and a comparison
between the size of the original document and the size of our
in-memory structures.
For these indexes, a sampling factor $l=64$
(cf. Section~\ref{sec:text}) was chosen. It should be noted that the
size of the tree index plus the size of the FM-index is always less
than the size of the original document.
\no{
As we see, the \emph{whole}
index, including tree structure, FM-index and auxiliary text representation,
is always smaller than twice the size of the original document. Since
it is always possible to choose which text representation to
use, the actual main-memory footprint of the index is close
to the original document size.}

It should further be noted that although loading time is acceptable, it
dominates query answering time. This is however not a problem for the
use case we have targeted: a main memory query engine where the same large
document is queried many times.
As mentioned in the Introduction, systems such as
MonetDB load their indexes only partially; this gives superior
performance in a cold-cache scenario when compared with our system.

\subsection{Raw Performance of Text Index}
\label{RawText}
\begin{table}[th]
{\small
\begin{center}
\begin{tabular}{|r|cc|cc|c|c|}
\hline
&\multicolumn{2}{|c|}{\it GlobalCount}&\multicolumn{2}{|c|}{\it ContainsCount}&
{\it Report-}&mem\\
q & number & time & number & time &
{\it Contains} & (MB)\\\hline
1& 1 & .004 & 1 & 0.04 & 0.012 & 61 \\
2& 22 & .009 & 19 & 2.281 & 1.588 &  61\\
3& 392 & .009 & 144 & 29.924 & 32.668 &  61\\
4& 438 & .009 & 438 & 4.616 & 4.457 &  61\\
5& 1472 & .008 & 966 & 128.28 & 122.014 &  61\\
6& 2685 & .005 & 1493 & 218.462 & 215.196 &  61\\
7& 6897 & .005 & 4690 & 553.496 & 548.009 &  62\\
8& 10402 & .005 & 8534 & 401.214 & 399.674 &  62\\
9& 20859 & .004 & 12073 & 1722.95 & 1717.83 &  62\\\hline
10& 63332 & .004 & 22974 & 5084.14 & 5083.77 &  63\\
11& 238638 & .003 & 42586 & 19641.8 & 19630.3 &  64\\
12& 2932251 & .001 & 595716 & 189299 & 188377 &  93\\
13& 9730750 & .001 & 5870474 & 132780 & 132241 & 86\\\hline
\end{tabular}
\end{center}
}
(q1, \dots, q13) = (
``Bakst'', ``ruminants'', ``morphine'', ``AUSTRALIA'', ``molecule'',
``brain'', ``human'', ``blood'', ``from'', ``with'', ``in'', ``a'',
``$\backslash$n'')
\caption{Search times of FM-index (in ms), sampling factor $l=64$}
\label{textIndexPerf1}
\end{table}

\begin{table}[t]
{\small
\begin{center}
\begin{tabular}{|r|cc|cc|c|c|}
\hline
&\multicolumn{2}{|c|}{\it GlobalCount}&\multicolumn{2}{|c|}{\it ContainsCount}&
{\it Contains-}&mem\\
q & number & time & number & time &
{\it Report} & (MB)\\\hline
1&1 & .005 & 1 & 0.049 & 0.013 & 100\\
2&22 & .01 & 19 & 0.156 & 0.086 & 100\\
3&392 & ..009 & 144 & 1.718 & 1.357 & 100\\
4&438 & .009 & 438 & 4.145 & 3.942 & 100\\
5&1472 & .009 & 966 & 6.247 & 5.853 & 101\\
6&2685 & .006 & 1493 & 12.24 & 11.588 & 101\\
7&6897 & .005 & 4690 & 25.403 & 27.344 & 101\\
8&10402 & .005 & 8534 & 77.175 & 73.613 & 101\\
9&20859 & .003 & 12073 & 84.012 & 78.663 & 101\\
10&63332 & .004 & 22974 & 242.834 & 235.043 & 102\\
11&238638 & .002 & 42586 & 1105.6 & 1091.43 & 103\\
12&411409 & .001 & 135307 & 1779.27 & 1762.62 & 108\\\hline
13&748326 & .001 & 320440 & 3411.65 & 3378.85 & 119\\
14&2932251 & 001 & 595716 & 13183.4 & 13173.4 & 133\\
15&9730750 & .001 & 5870474 & 87770.9 & 88230.4 & 126\\\hline
\end{tabular}
\end{center}
}
(q1, \dots, q15) = (
``Bakst'', ``ruminants'', ``morphine'', ``AUSTRALIA'', ``molecule'',
``brain'', ``human'', ``blood'', ``from'', ``with'', ``in'', ``b'',
``g'', ``a'', ``$\backslash$n'')
\caption{Search times of FM-index (in ms), sampling factor $l=4$}
\label{textIndexPerf2}
\end{table}
Here we give a short overview of the performance of
our implementation of the FM-index.
We present the search times for different versions of
\emph{contains}-queries:
\begin{enumerate}
\item $\text{\it GlobalCount\/}(P)$: returns the global
number of occurrences of the pattern $P$ in all texts.
\item $\text{\it ContainsCount\/}(P)$: returns the number
of texts that contain $P$,
\item $\text{\it ContainsReport\/}(P)$: returns the positions
of all occurrences of $P$ in the texts.
\end{enumerate}

Our experiments are over the text collection obtained from a
116MB XMark XML document~\cite{DBLP:conf/vldb/SchmidtWKCMB02}.
The size of this text is around 82MB (if stored in one-byte
per character ASCII format).
Our ``plain'' alternative to the FM-index is a naive
(byte-wise) string buffer (using precisely 82MB of memory).
To search over the plain buffer, we use
OCaml's regular string expression library.
The naive search time is constant for all our queries
at around $2700$ms.
For both the naive and the FM-index, the result positions
($32$ bit integers) for $\text{\it ContainsReport\/}$ queries
are materialized in an array.
Consider now the performance of our FM-index in comparison.
First at sampling factor $l=64$, shown in
Table~\ref{textIndexPerf1}.
As can be seen, the times for \textit{ContainsCount} and \textit{ContainsReport}
for the word ``from'' are at around $1720$ms.
Thus, in this case it is still faster to search over the FM-index.
On the other hand, for the word ``with'' the search time is over
$5000$ms, thus, here the plain search becomes faster.
Hence, somewhere between $20859$ and $63332$ occurrences lies the
cut-off point from which on searching over the plain text is faster
than over the FM-index.
Table~\ref{textIndexPerf2} shows timings obtained with
sampling factor $l=4$. As can be seen
the cut-off point is now much later, at a global
count somewhere between $411409$ and $748326$.
The last columns of Tables~\ref{textIndexPerf1} and~\ref{textIndexPerf2}
show the maximal memory consumption for these queries
over the FM-index. As mentioned in the beginning of this section,
we measure the maximum memory used by the process, as report by the
operating system( this is a slight over-approximation of the actual
memory).
The memory overhead for queries with large cardinality, such as the
last queries (q13 and q15),
is explained by the size of the result array: for both sampling factors
this is around 25MB. This query has around 6 million results
(\textit{ContainsCount}-number),
each result is stored as a 4 Byte integer. Thus, 23MB are needed.
However, additional memory
overhead occurs when results are removed from the \textit{GlobalCount} (because
they occur in the same XML text node). For instance, in the second to
last query (q12/q14) the ratio of \textit{GlobalCount}-number to
\textit{ContainsCount}-number is much larger than for the last query ($4.9$ versus
$1.7$); it means that on average there are around $5$ ``a''-characters
per text node, while there are only around $1.7$ return-characters per
text node. Correspondingly, the maximum memory consumption is much higher too.

%

\subsection{Raw Performance of Tree Index}
\begin{table}[th]
{\small
\begin{center}
\begin{tabular}{|r|c|c|c|c|c|c|}
\hline
file&parse&pointer&parent&tag&tag-tabs\\\hline
XMark116M&89446&373&504&4682&1324\\
XMark223M&220143&716&976&9051&2544\\
XMark559M&620479&7923&2415&22857&6283\\
Treebank83M&67412&465&615&14067&18867\\
medline122M&67935&537&760&6933&2036\\\hline
\end{tabular}
\end{center}
}
\caption{Construction times in ms for pointer versus SXSI tree store}
\label{tab:tree construction}
\end{table}
The performance of some low-level features of our tree index is
compared with the corresponding performance of a standard
pointer-based implementation of a tree.
The latter provides for each tree node two 64-bit pointers to
its first child and next sibling nodes (and does not store labels).
We first compare construction times.
Then we compare times for
a full depth-first left-to-right tree traversal on the different
structures.
Finally, we test the speed
of the taggedDesc and taggedFoll functions. We compare different
traversals through all nodes with a given label: (i) using
a pure C++ function, (ii) using our automata in counting mode, and
(iii) using our automata in materialization mode.

\paragraph*{Construction}\quad
As Table~\ref{tab:tree construction} shows, the construction
of the parenthesis structure takes roughly $1.5$-times the
amount of time of allocation a pointer structure for the tree.
Constructing the tag sequence is considerable slower, about
ten times as much as building the parenthesis structure.
This is because for each opening and for each closing tag, 
a separate \TAG{sarray} is constructed (see bottom left of 
Figure~\ref{fig:example}).
The last column shows the time for building the four
tag-to-tag tables described in Section~\ref{ssec:relativetag}.
We also show the
XML parsing time in the first column of the table.

\paragraph*{Full Traversals}\quad
\begin{table}[th]
{\small
\begin{center}
\begin{tabular}{|r|c|c|c|c|c|c|}
\hline
&\multicolumn{3}{|c|}{recursive, all nodes}&\multicolumn{3}{|c|}{element nodes, SXSI}\\
file&\#nod&pointer&SXSI&\#nod&rec.&//*\\
\hline
XMark116M&6&33&109&1.7&71&153\\
XMark223M&12&63&209&3.3&137&296\\
XMark559M&30&164&535&8.4&345&756\\
Treebank83M&7&57&184&2.4&136&292\\
medline122M&9&48&164&2.9&112&244\\\hline
\end{tabular}
\end{center}
}
\caption{Traversal times (in ms), \#nodes (in millions)}
\label{tab:tree}
\end{table}
The left part of Table~\ref{tab:tree} shows that
a full tree traversal through all nodes is between
$3.2$ and $3.4$ times slower with SXSI, than with
a pointer tree data structure.
Note that the pointers are allocated in pre-order too
giving optimal performance for pre-order traversal.
As a comparison, if the pointers are allocated in post-order,
then traversal time for the pr-order traversal is almost
twice as slow as the numbers reported, and if pointers are
allocated in in-order, then the times are a bit over twice
as slow; see~\cite{DBLP:conf/alenex/ArroyueloCNS10} for a discussion of the phenomenon.
It should also be noted that for other access patterns, such as
random root-to-leaf traversals, the time difference between
pointer and succinct trees is much larger, factors of
up to $100$ are measure in~\cite{DBLP:conf/alenex/ArroyueloCNS10}.

In the right part of Table~\ref{tab:tree} we see the number
of element nodes in these trees, and the time it takes for
SXSI to recurse through those node: either using a small
recursive C-function (column ``rec.''), or
using the automaton for the XPath query //*,
and executing in counting mode.

\paragraph*{Tagged Traversals}\quad
\begin{table}[t]
{\small
\begin{center}
\begin{tabular}{|r|c|c|c|c|}
\hline
tag&\#nodes&jump(C++)&//(cou)&//(mat)\\\hline
category&1040&1.2&1.6&1.7\\
price &10141&2.3&2.9&3.1\\
listitem&63179&16&22&24\\
keyword&73070&11&12&14\\\hline
\end{tabular}
\end{center}
}
\caption{Iteration through all tag-labeled nodes over XMark116M}
\label{tab:tag_Xmark}
\end{table}
Here the speed of the TaggedDesc and TaggedFoll functions
is investigated. Using these two functions,
three different traversal through all nodes
with a given label are considered: first, by a small C++ function,
and second and third by
our automata through a //label query in counting and materializing
modes, respectively.
For instance, Table~\ref{tab:tag_Xmark} shows that
iterating through all keyword-nodes of the 116MB Xmark document
takes essentially the same time for all three methods
(11--14ms). This is in contrast to some other labels:
for listitem for instance, the count-automaton traversal
is $1.5$-times slower than the C++ traversal.
This can be explained by the fact that listitem is a
recursive tag: there are in fact 23298 listitem nodes that appear
as descendants of listitem nodes. Hence, at each listitem node
the automaton issues a taggedDescendant to
search for further nodes. The other labels such as keyword and
category do not appear recursively. Since this information
is part of our tree index (cf. Section~\ref{ssec:relativetag}), 
the automaton run function
avoids all these taggedDesc calls, which brings
the speed almost up to the one of the C++ function.

\subsection{XPath Tree Queries}
We benchmark tree queries using the queries given in
Figure~\ref{fig:treequeries}.
\begin{figure}
\begin{framed}{\small
\begin{tabular}{l@{\hspace{3pt}}l}
Q01 & /site/regions\\
Q02 & /site/regions/*/item\\
Q03 & /site/closed\_auctions/closed\_auction\\
&\hspace{2cm} /annotation/description/text/keyword\\
Q04 & //listitem//keyword\\
Q05 & /site/closed\_auctions/closed\_auction[ \\
& \hspace{1.5cm} annotation/description/text/keyword]/date\\
Q06 & /site/closed\_auctions/closed\_auction[ .//keyword]/date\\
Q07 & /site/people/person[ profile/gender and profile/age]/name\\
Q08 & /site/people/person[ phone or homepage]/name \\
Q09 & /site/people/person[ address and (phone or homepage) and \\
&  (creditcard or profile)]/name\\
Q10 & //listitem[not(.//keyword/emph)]//parlist \\
Q11 & //listitem[ (.//keyword or .//emph) and \\
& (.//emph or .//bold)]/parlist\\
Q12 & //people[ .//person[not(address)] and\\ &.//person[not(watches)]]/person[watches]\\
Q13 & /*[ .//* ]\\
Q14 & //* \\
Q15 & //*//* \\
Q16 & //*//*//* \\
Q17 & //*//*//*//* \\
\end{tabular}}
\caption{Tree oriented queries}
\label{fig:treequeries}
\end{framed}
\end{figure}
Queries Q01 to Q12 are taken from the XPathMark benchmark
\cite{fra07}, derived from the XMark XQuery benchmark suite. 
Q13 to Q17 are ``crash tests'' that are either simple (Q13 selects only the
root since it always has at least one descendant in our files) or
generate roughly the same amount of results but with various
intermediate result sizes.
\paragraph{Query answering time}
For this experiment we use XMark
documents of size 116MB and 1GB. In the cases of MonetDB and Qizx, the
files were indexed using the default settings. Let us describe in
detail Figure~\ref{fig:xmarkrun}.  Each of the six graphs should
be read as follows. For each query (Q1 to Q17), the graph reports as
vertical bars the relative running time of the three engines with
respect to SXSI's running time (therefore SXSI's score is always
100\%). In these graphs a higher bar means that the engine was
slower.  We also give at the top of each bar the average running time
for the query in millisecond (or seconds, if the number is suffixed
with an ``s''). For instance, in the first graph ---labelled ``116 MB
(counting)''--- we can see that for query Q1, SXSI evaluates the query
in 1.3ms, MonetDB 6.8ms (or roughly 500\% of SXSI's speed) and QizX
3.5 ms (or roughly 275\% of SXSI's speed). For count queries, the
timing for all three engines are given side by side (SXSI, MonetDB and
QizX in that order).  For full reporting queries however, we want to
gauge precisely the amount of time spent during materialization and
during serialization. The definition of materialization seems to fit
both MonetDB and SXSI: create a data-structure in memory which holds
the resulting nodes \emph{in order} and \emph{without duplicates} such
that access of the first result in pre-order can be done in
constant time, and accessing the next resulting node in pre-order
can also be done in constant time. The timing for both SXSI and
MonetDB are given in the graphs labelled ``(materialization)''. As we
explained earlier, QizX interleaves evaluation of the query and
serialization, therefore we only compared it to SXSI and MonetDB in
the ``(materialization+serialization)'' series. We also checked that
all three engines generated in the end the same amount of data while in
serialization mode and that they generated valid XML documents (in
particular, characters such as ``\texttt{\&}'' were escaped
correctly).


\newcommand{\B}[1]{\textbf{#1}}
\newcommand{\I}[1]{\underline{#1}}
\definecolor{sxsi1}{HTML}{AAFF22}
\definecolor{monet1}{HTML}{FF8844}
\definecolor{qizx1}{HTML}{0088FF}
\definecolor{sxsi2}{HTML}{77CC00}
\definecolor{monet2}{HTML}{DD6622}
\definecolor{qizx2}{HTML}{0066DD}

\begin{figure*}[th!]
\begin{framed}
{ \scriptsize \sf
\def\svgwidth{\textwidth}
\vspace{1mm}
\begin{minipage}{5pt}
\rotatebox{90}{116MB (counting)}
\end{minipage}
\begin{minipage}{\textwidth}
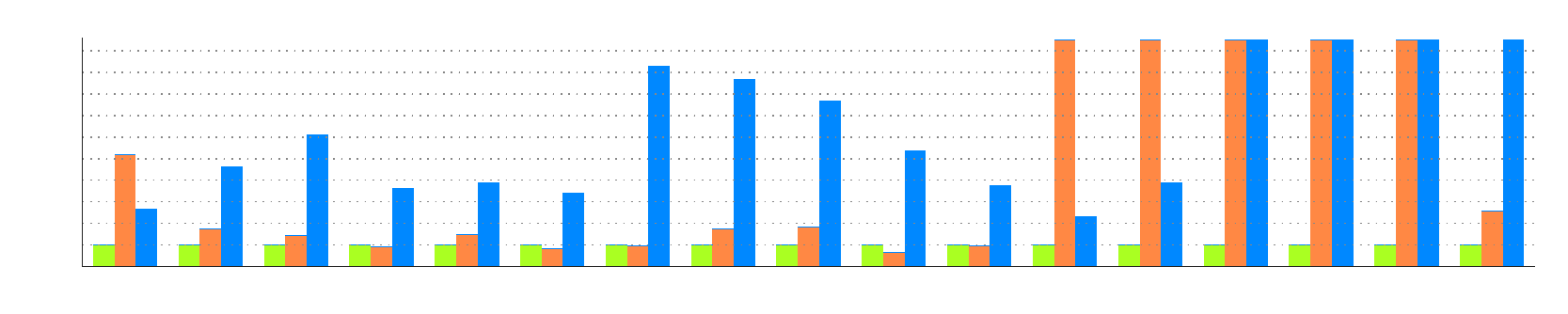
\end{minipage}

\vspace{1mm}

\begin{minipage}{5pt}
\rotatebox{90}{116MB (materialization)}
\end{minipage}
\begin{minipage}{\textwidth}
\def\svgwidth{\textwidth}
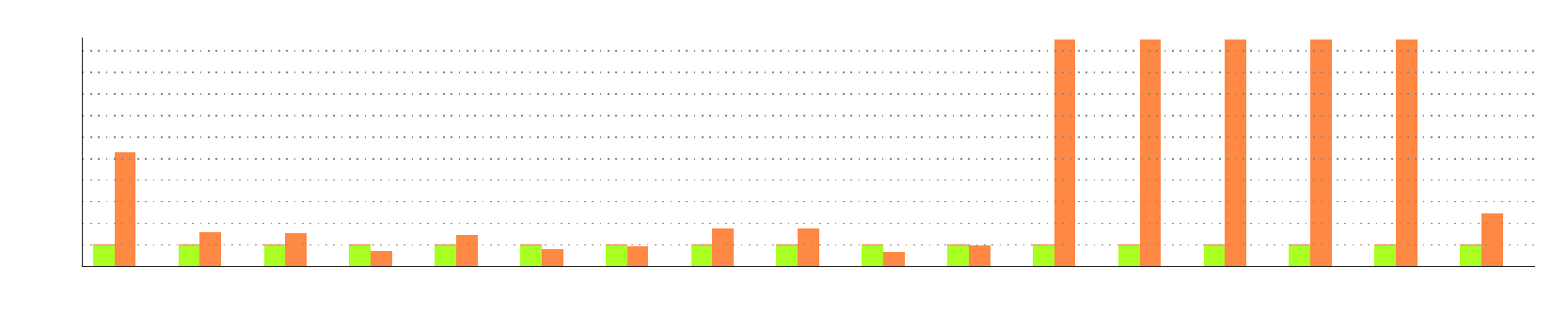
\end{minipage}

\vspace{1mm}

\begin{minipage}{5pt}
\rotatebox{90}{116MB (mat. + serialization)}
\end{minipage}
\begin{minipage}{\textwidth}
\def\svgwidth{\textwidth}
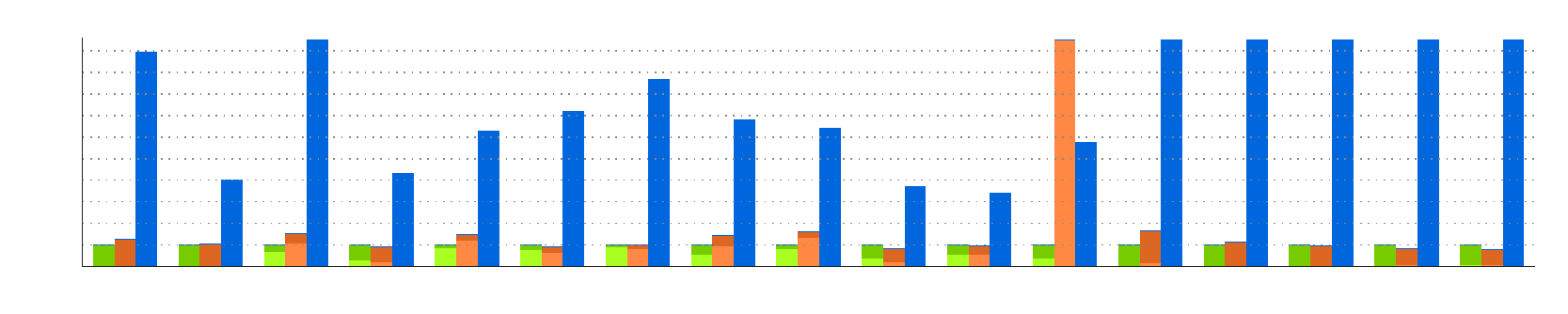
\end{minipage}

\vspace{1mm}

\begin{minipage}{5pt}
\rotatebox{90}{1GB (counting)}
\end{minipage}
\begin{minipage}{\textwidth}
\def\svgwidth{\textwidth}
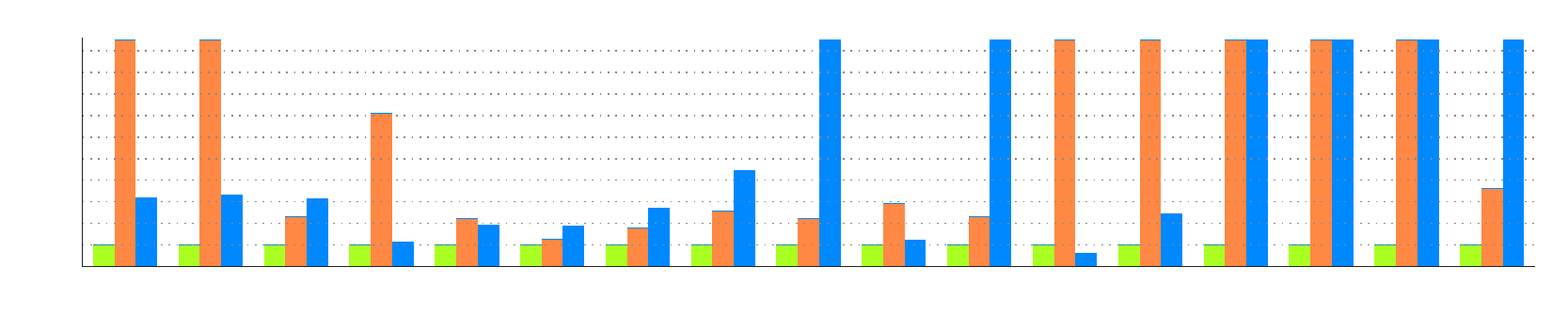
\end{minipage}

\vspace{1mm}

\begin{minipage}{5pt}
\rotatebox{90}{1GB (materialization)}
\end{minipage}
\begin{minipage}{\textwidth}
\def\svgwidth{\textwidth}
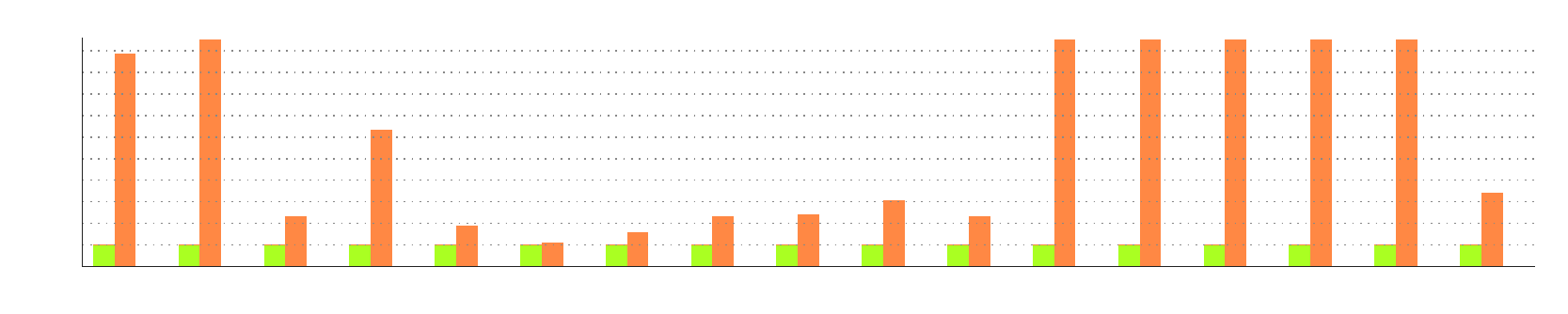
\end{minipage}

\vspace{2mm}

\begin{minipage}{5pt}
\rotatebox{90}{1GB (mat. + serialization)}
\end{minipage}
\begin{minipage}{\textwidth}
\def\svgwidth{\textwidth}
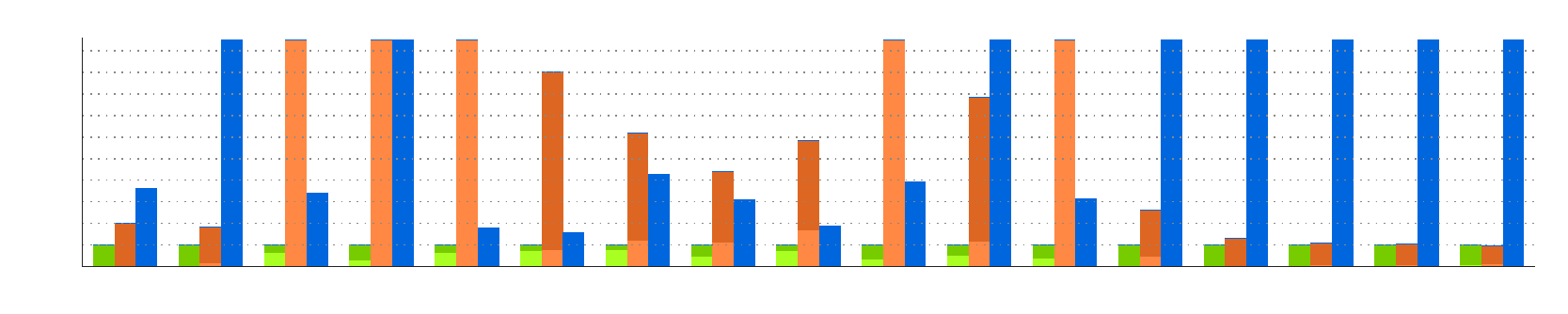
\end{minipage}

{\color{sxsi1}\rule{8pt}{6pt}}SXSI query time\hspace{1mm}%
{\color{sxsi2}\rule{8pt}{6pt}}SXSI serialization time\hspace{1mm}%
{\color{monet1}\rule{8pt}{6pt}}MonetDB query time\hspace{1mm}%
{\color{monet2}\rule{8pt}{6pt}}MonetDB serialization time\hspace{1mm}%
{\color{qizx1}\rule{8pt}{6pt}}Qizx query time\hspace{1mm}%
{\color{qizx2}\rule{8pt}{6pt}}Qizx query + serialization time\\
+++ : query could not be run or took more than 15 minutes
}

\caption{Running times for the tree based queries (in milliseconds or
  seconds and as percent of SXSI's speed. Lower bars are better.)}
\label{fig:xmarkrun}
\end{framed}
\end{figure*}

From the results of Figure~\ref{fig:xmarkrun}, we see how the different
components of SXSI contribute to the efficient evaluation model.
Fully qualified paths, such as queries Q1--3 and Q5 illustrate the
sheer speed of the tree structure and in particular the efficiency of
its basic operations (such as FirstChild and NextSibling, which are
used for the \texttt{child} axis), as well as the efficient execution
scheme provided by the automaton. The descendant axes (used e.g. in
Q4, Q6, Q10--12) show the impact of the jumping primitives and the
computation of relevant nodes. Complex filters (Q6--12) show how the
alternating automata can efficiently evaluate complex Boolean formulas
corresponding to structural conditions over subtrees of a given node,
including negations of paths.

Finally, Q12 to Q16 illustrate the robustness of our automata
model. Indeed while such queries might seem unrealistic, the good
performances that we obtain are the combination of \textit{(i)} using an automata
based evaluator (which factors in the states of the automaton
 all the necessary computation and thus do not materialize
unneeded intermediate results) and \textit{(ii)} our implementation of
lazy result sets, which shifts the burden of walking through the
document as much as possible to the serialization process.
\paragraph{Memory use and precision}
While it is straightforward to predict the memory consumption
of our engine with respect to the index part (the full index is mapped
in memory excluding the Auxiliary Text, see
Figure~\ref{fig:index}), the behaviour of the automaton evaluation
function is unclear. Indeed, to speed-up the computation we
create memoization tables, we handle partial result sets, and we
perform recursive procedures which might be as deep as the
\emph{binary encoding} of the XML document (since we recurse on
FirstChild and NextSibling move) thus increasing the size of the call stack.
\begin{figure}
\begin{framed}
\def\svgwidth{\columnwidth}
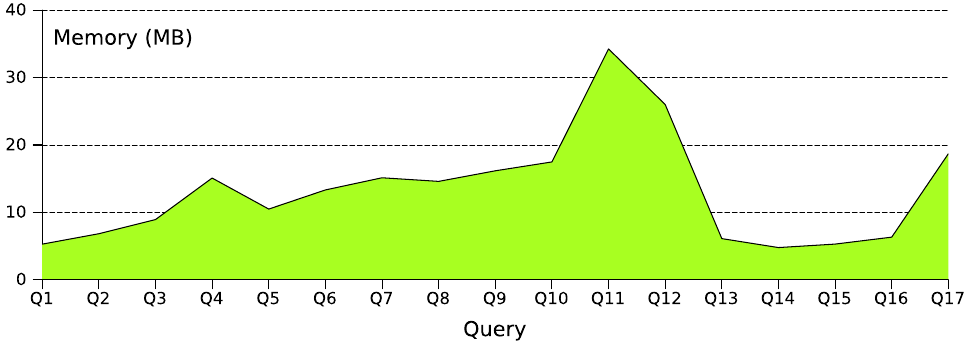
\caption{Memory use in MB for XMark 116MB documents (excluding the index)}
\label{fig:memrun}
\end{framed}
\end{figure}
We report in Figure~\ref{fig:memrun} the memory consumption for the
automata evalutation of materialization queries. This includes the
size of the recursive call stack, the size of OCaml's heap (which is
grown dynamically by OCaml's garbage collector to accomadate the
memory need). On the heap are allocated the memoization tables,
intermediary structures and final result sets. As one can see, the
memory use is very modest, peaking at 32 MB for query Q11.  While we
do not compare directly with MonetDB or QizX for memory consumption
(since these engines try to max out the memory use to achieve better
speed) we see that we can reach comparable (if not greater) speed
while being very conservative memory wise.

\begin{figure}
\begin{framed}
\newcommand{\mytexta}[1]{\scriptsize #1}
{\sf \tiny
\def\svgwidth{\columnwidth}
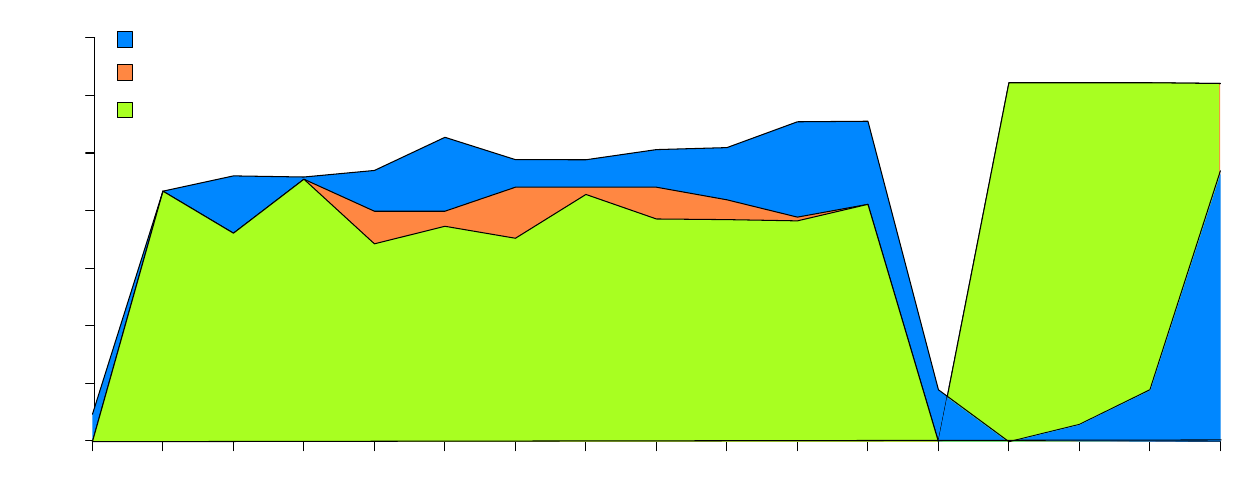 }
\caption{Comparison of visited, marked and result nodes for each query
  (logarithmic scale)}
\label{fig:precision}
\end{framed}
\end{figure}

To gauge the precision of our automata based approach, we report in
Figure~\ref{fig:precision} for each query:
\begin{itemize}
\item the number of visited nodes (that is, the number of nodes that
  are given as argument to the \textbf{top\_down\_run} function);
\item the number of marked nodes (that is, the number of nodes that
  considered as potential results at some point during evaluation);
\item and finally the number of result nodes for the query.
\end{itemize}

A first observation is that the number of marked nodes is almost
always the same as the number of result nodes (save for query Q5--10).
This shows that using automata and early evaluation of Boolean
formulas, we can decide early during query evaluation whether a node
is indeed a result or not. Another point of interest is that for several
queries (Q2, Q4, Q14--17) we \emph{only} visit result nodes. While it
might be expected for queries Q14--17 for which virtually every node
is a result, queries such as Q2 or Q4 are very selective. However,
these queries provide enough information for the runtime analysis of
relevant nodes to be \emph{exact} and therefore only touch result
nodes. In general of course, the number of traversed node is larger
than the number of resulting nodes but always far less than the whole
document. Queries Q14 to Q17 show the impact of lazy result sets
where we mark several nodes (whole subtrees actually) in one function
call, and therefore manage to return more nodes than we have actually
visited.
Lastly, we can see that the shapes of the ``Visited Nodes'' curve in
Figure~\ref{fig:precision} and the memory use (Figure~\ref{fig:memrun})
are quite similar (the former being flattened by the logarithmic
scale). This (quite expectedly) show that the number of visited nodes
(and not the number of result nodes or intermediary results) impacts
directly the memory consumption of our query engine.

\subsection{XPath Text Queries}
We tested the text capabilities of our XPath engine against the most
advanced text oriented features of other query engines.

\paragraph*{Qizx/DB}\quad We use the newly introduced \emph{Full-Text}
extension of XQuery available in Qizx/DB v.~3.0. We try to formulate
the queries as efficiently as possible 
while preserving the semantics of our
original queries. The query used always gave better results than
its pure XPath counterpart. In particular, we use the
\texttt{ftcontains} text predicate~\cite{xqueryfulltext}
implemented by Qizx/DB. The \texttt{ftcontains} predicate
allows to express not only \emph{contains}-like queries but also
Boolean operations on text predicates, regular expression matching and
so on. It is more efficient than the standard \texttt{contains}.

\paragraph*{MonetDB}\quad MonetDB supports some full-text capabilities through
the use of the PF/Tijah text index~\cite{pftija}. However, this
index only supports a complex \texttt{about} operator, which returns
approximate matches and \emph{ranks} results by order of
relevance. Since \texttt{about} queries returned very different
results than standard \texttt{contains} one, we did not include them
but rather used the standard (and un-optimized) \texttt{fn:contains}
functions from the XPath standard library. Interestingly, while these
text functions are unoptimized, they sometimes outperform QizX/DB's
full-text implementation. However, the reader should keep in mind that
MonetDB's timing are given only for reference and are not used here as
a direct comparison (since they implement the full semantics of
\texttt{fn:contains} itself relying on \texttt{fn:string} conversions
as defined in the XPath specification, they are bound to be slower
than the full-text alternative).

Experiments are made on a 122MB Medline file. This file contains
bibliographic information about  life sciences and biomedical
publications. This test file features 5,732,159 text elements, for a
total amount of 95MB of text content. Figure~\ref{fig:textqueries}
shows the text queries that we test.
\begin{figure*}
\begin{framed}\small
\begin{tabular}{|@{}ll|@{\,}c@{\,}|*{5}{@{\,}c@{\,}|}}
\hline
     &   &   & \multicolumn{5}{c|}{Query running time in ms}\\
&  & \# results & S Text & S Auto & S Total & Q & M\\
\hline
T1  & //Article[ .//AbstractText[ contains ( . , "plus") ] ] & 358 
& 3.1  & 2.8 & 5.9 & 41.2 & 224\\

T2  & //Article[ .//AbstractText[ contains ( . , "brain")  ] ] & 1104
& 5.2 & 9.0 & 14.2 & 69.5 & 420\\

T3  & //Article[ .//Year[ . = "1997"  ] ] & 1237
& 2.2 & 8.7 & 10.9 & 48 & 230\\

T4  & //MedlineCitation/Article/AuthorList/Author[ LastName[starts-with( . , "Bar")]] & 630
& 2.24 & 0.25 & 2.5 & 111.25 & 167.85 \\

T5  & //*[ .//LastName[ contains( . , "Nguyen") ] ] & 154 
& 1.9 & 0.4 &  2.3 & 765 & 1433\\

T6  & //*//*[ contains( . , "epididymis") ] & 88 
& 2.0 &0.2 & 2.2 & 1221 & 5218 \\
T7  & //*[ .//PublicationType[ ends-with( . , "Article") ]] & 81187 
&29.3 & 170.7 & 200 & 438  & 1531 \\

T8  & //MedlineCitation[ .//Country[ . = "AUSTRALIA" ]] & 326
& 2.0 &2.6 & 4.6     &  38   &  54   \\
\hline
\end{tabular}\\
S Text: SXSI's text collection~~~S Auto: SXSI's Automata query engine~~~S Total: Total running time for SXSI\\
Q: total running time for Qizx/DB~~~M: total running time for MonetDB/XQuery

\caption{Text oriented queries}
\label{fig:textqueries}
\label{fig:restextq}
\end{framed}
\end{figure*}
We use count queries for all three engines.
The table in Figure~\ref{fig:restextq} summarizes the running times for
each query. As we target very selective text queries, we also give,
for each query, the number of results it returned. Since for these
queries our automata worked in ``bottom-up'' mode, we
detail the two following operations:
\begin{itemize}
\item Calling the text predicate \emph{globally} on the
  text collection, thus retrieving all the probable matches of the
  query (\emph{S Text} column in the table of Figure~\ref{fig:restextq})
\item Running the automaton bottom up from the set of probable matches
  to keep those satisfying the path expression (\emph{S Auto} 
  in the table of Figure~\ref{fig:restextq})
\end{itemize}
As it is clear from the experiments, the bottom-up strategy pays
off. The only down-side of this approach is that the automaton uses
Parent moves, which are less efficient than FirstChild and
NextSibling. This can be seen in query T7 where the increase in the
number of results makes the relative slowness of the automata more
visible. However our evaluator still outperforms the other engines
even in this case.

\subsection{Biological Sequence Queries}
As a last experiment, we demonstrate the versatility of SXSI by showing that
it can be used as a very efficient Biological database, answering queries which
make use of both the tree structure and a tailored text index.
More precisely, we
create XML files that combine gene annotations with their DNA sequences.
A sample DTD for these files is given in Figure~\ref{fig:bioxmldtd}.
\begin{figure}
\begin{framed}
\begin{verbatim}
<!ELEMENT  chromosome  (name, gene*) >
<!ELEMENT  name        #PCDATA >
<!ELEMENT  gene        (name, strand, biotype, status, 
      description?, promoter, sequence, transcript*) >
<!ELEMENT  strand      #PCDATA >
<!ELEMENT  biotype     #PCDATA >
<!ELEMENT  status      #PCDATA >
<!ELEMENT  description #PCDATA >
<!ELEMENT  promoter    #PCDATA >
<!ELEMENT  sequence    #PCDATA >
<!ELEMENT  transcript  (name, start, end, exon*, 
                        sequence, protein?) >
<!ELEMENT  start       #PCDATA >
<!ELEMENT  end         #PCDATA >
<!ELEMENT  exon        (name, start, end, sequence) >
\end{verbatim}
\caption{DTD for bio-genetic data}
\label{fig:bioxmldtd}
\end{framed}
\end{figure}
In this DTD, the elements \texttt{promoter} and \texttt{sequence} are of
particular interest: they store the DNA represented as long sequences of A,
T, C, G characters. The other
\texttt{\#PCDATA} elements store the gene annotation data such as 
positions, names and so on.

Our experiment data is composed from human chromosome
five\footnote{Ensemble Human genome release 59, August 2010.} which
contains 2719 genes having in total 8330 different transcripts.
For each gene, we include 1000 base pairs of its upstream promoter sequence,
the gene sequence itself (all exons and introns included), and
annotation information such as
gene's 
biotype and description. 
Additionally, we include all known transcripts of each gene, that is,
sequences of the exons they contain as well as the concatenation of
these exons. 
The resulting textual content is highly repetitive
since each one of the exon sequences can appear in many transcripts.
Highly repetitive data has been shown to compress well using certain
run-length encoded text-indexes \cite{Maekinen2010}, thus,
here the text-index implementation is switched to use 
RLCSA~\cite{Maekinen2010} instead of the FM-index.
In this example, the final XML file\footnote{\url{http://www.cs.helsinki.fi/group/suds/sxsi/data/}} is 132 MB
while the text-index requires only 63 MB of memory plus 59 MB for the samples array.
The full index,
including tree and text, is around 135 MB, that is only as big as
the original document.
The resulting XML document 
contains 323318 elements of which
65286 are either \texttt{promoter} or \texttt{sequence} nodes containing genetic
data.

To do biologically relevant XML queries, we extend our engine to
support PSSM queries (Position Specific Scoring Matrix) which allows
to search for transcription factor binding sites from genes' promoter regions.
Input for this query is an Position Frequency Matrix (PFM) and
a minimum threshold for a valid match. 
The matrices can be found
from e.g. the Jaspar
database \cite{jaspar}.

In a nutshell, PFM's have one row for each symbol of
the alphabet (in our case 4 rows A, T, C, G) and one column for each
position in the pattern to search. For instance, the PFM:
\[
\left[\begin{array}{ccccc}
      A & 0 & 20 & 10 & 1\\
      T & 30 & 10 & 0 & 0\\
      C & 0  & 0  & 10 & 20\\
      G & 18 & 6 & 6 & 6 \\
    \end{array}\right]
\]
\noindent denotes patterns of length four, and the substring AGCT would
get the score 0 + 6 + 10 + 10 = 26. 
To form the PSSM query, the PFM matrix is first converted into log-odds form 
to take into account the uneven background distribution of nucleotide frequences.
Then the PSSM query takes such a matrix as
well as a threshold and returns all text elements whose content scores
more than the given threshold. Table~\ref{tab:bio} gives the running 
times for XPath queries using the
PSSM predicates and RLCSA, with block size 128 and sample rate 16, as the text index. 
The table summarizes also the number of results, 
the length of the search pattern and
the value of the threshold.
\begin{table}
\begin{framed}
\begin{tabular}{*{5}{|@{~}l@{~}}|}
\hline
query & \# results & Text & Auto & Total\\
\hline
//promoter[ PSSM( ., M1)] & 134 & 85.1 & 7.4 & 92.5\\
//promoter[ PSSM( ., M2)] & 4 & 4.35 & 1.15 & 5.51\\
//promoter[ PSSM( ., M3)] & 1 & 6.5 & 0.38 & 6.97 \\
//exon[ .//sequence[ PSSM( ., M1) ] ] & 434 & 85.5 & 7.5 & 92.6\\
//exon[ .//sequence[ PSSM( ., M2) ] ] & 25 & 4.3  & 1.28 & 5.6\\
//exon[ .//sequence[ PSSM( ., M3)  ] ] & 9 & 6.4 & 0.62 & 7.0\\
//*[ PSSM(., M1) ] & 1875 & 85.04 & 7.6 & 92.6\\
//*[ PSSM(., M2) ] & 184  & 4.3 & 1.19 & 5.5\\
//*[ PSSM(., M3) ] & 51 & 6.4 & 0.58 & 6.9\\
\hline
\end{tabular}
M1 : Jaspar ID = MA0031.1, length = 8, threshold = 5000\\
M2 : Jaspar ID = MA0050.1, length = 12, threshold = 100000\\
M3 : Jaspar ID = MA0017.1, length = 14, threshold = 300000
\caption{Running time for PSSM queries in ms}
\label{tab:bio}
\end{framed}
\end{table}
It is interesting to remark that since the document is a very flat
and shallow structure, the automaton/tree part of the query evaluates
always very quickly (7ms or under). The PSSM scheme also allows to
write biologically meaningful queries that would otherwise be
impossible or very hard to write with regular expressions or a
regular full-text extension. Yet, we did not have to modify our core
engine, only the text index was modified in isolation to add PSSM
capabilities, the automata and tree machinery remaining unchanged.

\no{
\subsection{Remarks}
We also compared with Tauro~\cite{tauro}.  
Yet, as it uses a tailored query language, we could not
produce comparable results.

We limited the experiments to natural language XML, although our engine
(unlike the inverted file based engines) supports
as well queries on XML databases of continuous sequences such as DNA and proteins.
Realistic queries on such biosequence XMLs require
approximate / regular expression search functionalities, that we
already support but whose experimental study is out of
the scope of this paper.
}

%% file: count_xmark_01_04.pdf_tex

\begingroup
  \makeatletter
  \providecommand\color[2][]{%
    \errmessage{(Inkscape) Color is used for the text in Inkscape, but the package 'color.sty' is not loaded}
    \renewcommand\color[2][]{}%
  }
  \providecommand\transparent[1]{%
    \errmessage{(Inkscape) Transparency is used (non-zero) for the text in Inkscape, but the package 'transparent.sty' is not loaded}
    \renewcommand\transparent[1]{}%
  }
  \providecommand\rotatebox[2]{#2}
  \ifx\svgwidth\undefined
    \setlength{\unitlength}{800pt}
  \else
    \setlength{\unitlength}{\svgwidth}
  \fi
  \global\let\svgwidth\undefined
  \makeatother
  \begin{picture}(1,0.2)%
    \put(0,0){\includegraphics[width=\unitlength]{count_xmark_01_04.pdf}}%
    \put(0.0455,0.0263){\makebox(0,0)[rb]{\smash{0}}}%
    \put(0.0455,0.0401){\makebox(0,0)[rb]{\smash{100}}}%
    \put(0.0455,0.0538){\makebox(0,0)[rb]{\smash{200}}}%
    \put(0.0455,0.0676){\makebox(0,0)[rb]{\smash{300}}}%
    \put(0.0455,0.0814){\makebox(0,0)[rb]{\smash{400}}}%
    \put(0.0455,0.0951){\makebox(0,0)[rb]{\smash{500}}}%
    \put(0.0455,0.1089){\makebox(0,0)[rb]{\smash{600}}}%
    \put(0.0455,0.1226){\makebox(0,0)[rb]{\smash{700}}}%
    \put(0.0455,0.1364){\makebox(0,0)[rb]{\smash{800}}}%
    \put(0.0455,0.1502){\makebox(0,0)[rb]{\smash{900}}}%
    \put(0.0455,0.1639){\makebox(0,0)[rb]{\smash{1000}}}%
    \put(0.0797,0.0113){\makebox(0,0)[b]{\smash{Q1}}}%
    \put(0.1342,0.0113){\makebox(0,0)[b]{\smash{Q2}}}%
    \put(0.1887,0.0113){\makebox(0,0)[b]{\smash{Q3}}}%
    \put(0.2432,0.0113){\makebox(0,0)[b]{\smash{Q4}}}%
    \put(0.2977,0.0113){\makebox(0,0)[b]{\smash{Q5}}}%
    \put(0.3522,0.0113){\makebox(0,0)[b]{\smash{Q6}}}%
    \put(0.4067,0.0113){\makebox(0,0)[b]{\smash{Q7}}}%
    \put(0.4612,0.0113){\makebox(0,0)[b]{\smash{Q8}}}%
    \put(0.5157,0.0113){\makebox(0,0)[b]{\smash{Q9}}}%
    \put(0.5702,0.0113){\makebox(0,0)[b]{\smash{Q10}}}%
    \put(0.6247,0.0113){\makebox(0,0)[b]{\smash{Q11}}}%
    \put(0.6792,0.0113){\makebox(0,0)[b]{\smash{Q12}}}%
    \put(0.7337,0.0113){\makebox(0,0)[b]{\smash{Q13}}}%
    \put(0.7882,0.0113){\makebox(0,0)[b]{\smash{Q14}}}%
    \put(0.8427,0.0113){\makebox(0,0)[b]{\smash{Q15}}}%
    \put(0.8972,0.0113){\makebox(0,0)[b]{\smash{Q16}}}%
    \put(0.9517,0.0113){\makebox(0,0)[b]{\smash{Q17}}}%
    \put(0.0175,0.1942){\makebox(0,0)[lb]{\smash{\% of}}}%
    \put(0.0175,0.1792){\makebox(0,0)[lb]{\smash{SXSI}}}%
    \put(0.0455,0.0263){\makebox(0,0)[rb]{\smash{0}}}%
    \put(0.0455,0.0401){\makebox(0,0)[rb]{\smash{100}}}%
    \put(0.0455,0.0538){\makebox(0,0)[rb]{\smash{200}}}%
    \put(0.0455,0.0676){\makebox(0,0)[rb]{\smash{300}}}%
    \put(0.0455,0.0814){\makebox(0,0)[rb]{\smash{400}}}%
    \put(0.0455,0.0951){\makebox(0,0)[rb]{\smash{500}}}%
    \put(0.0455,0.1089){\makebox(0,0)[rb]{\smash{600}}}%
    \put(0.0455,0.1226){\makebox(0,0)[rb]{\smash{700}}}%
    \put(0.0455,0.1364){\makebox(0,0)[rb]{\smash{800}}}%
    \put(0.0455,0.1502){\makebox(0,0)[rb]{\smash{900}}}%
    \put(0.0455,0.1639){\makebox(0,0)[rb]{\smash{1000}}}%
    \put(0.0797,0.0113){\makebox(0,0)[b]{\smash{Q1}}}%
    \put(0.1342,0.0113){\makebox(0,0)[b]{\smash{Q2}}}%
    \put(0.1887,0.0113){\makebox(0,0)[b]{\smash{Q3}}}%
    \put(0.2432,0.0113){\makebox(0,0)[b]{\smash{Q4}}}%
    \put(0.2977,0.0113){\makebox(0,0)[b]{\smash{Q5}}}%
    \put(0.3522,0.0113){\makebox(0,0)[b]{\smash{Q6}}}%
    \put(0.4067,0.0113){\makebox(0,0)[b]{\smash{Q7}}}%
    \put(0.4612,0.0113){\makebox(0,0)[b]{\smash{Q8}}}%
    \put(0.5157,0.0113){\makebox(0,0)[b]{\smash{Q9}}}%
    \put(0.5702,0.0113){\makebox(0,0)[b]{\smash{Q10}}}%
    \put(0.6247,0.0113){\makebox(0,0)[b]{\smash{Q11}}}%
    \put(0.6792,0.0113){\makebox(0,0)[b]{\smash{Q12}}}%
    \put(0.7337,0.0113){\makebox(0,0)[b]{\smash{Q13}}}%
    \put(0.7882,0.0113){\makebox(0,0)[b]{\smash{Q14}}}%
    \put(0.8427,0.0113){\makebox(0,0)[b]{\smash{Q15}}}%
    \put(0.8972,0.0113){\makebox(0,0)[b]{\smash{Q16}}}%
    \put(0.9517,0.0113){\makebox(0,0)[b]{\smash{Q17}}}%
    \put(0.0175,0.1942){\makebox(0,0)[lb]{\smash{\% of}}}%
    \put(0.0175,0.1792){\makebox(0,0)[lb]{\smash{SXSI}}}%
    \put(0.0698,0.1759){\rotatebox{90}{\makebox(0,0)[lb]{\smash{1.3}}}}%
    \put(0.1243,0.1759){\rotatebox{90}{\makebox(0,0)[lb]{\smash{7.1}}}}%
    \put(0.1788,0.1759){\rotatebox{90}{\makebox(0,0)[lb]{\smash{11.8}}}}%
    \put(0.2333,0.1759){\rotatebox{90}{\makebox(0,0)[lb]{\smash{18.2}}}}%
    \put(0.2878,0.1759){\rotatebox{90}{\makebox(0,0)[lb]{\smash{14.4}}}}%
    \put(0.3423,0.1759){\rotatebox{90}{\makebox(0,0)[lb]{\smash{20.9}}}}%
    \put(0.3968,0.1759){\rotatebox{90}{\makebox(0,0)[lb]{\smash{24.8}}}}%
    \put(0.4513,0.1759){\rotatebox{90}{\makebox(0,0)[lb]{\smash{21.8}}}}%
    \put(0.5058,0.1759){\rotatebox{90}{\makebox(0,0)[lb]{\smash{27.2}}}}%
    \put(0.5603,0.1759){\rotatebox{90}{\makebox(0,0)[lb]{\smash{53.5}}}}%
    \put(0.6148,0.1759){\rotatebox{90}{\makebox(0,0)[lb]{\smash{136.8}}}}%
    \put(0.6693,0.1759){\rotatebox{90}{\makebox(0,0)[lb]{\smash{33.5}}}}%
    \put(0.7238,0.1759){\rotatebox{90}{\makebox(0,0)[lb]{\smash{1.9}}}}%
    \put(0.7782,0.1759){\rotatebox{90}{\makebox(0,0)[lb]{\smash{0.7}}}}%
    \put(0.8327,0.1759){\rotatebox{90}{\makebox(0,0)[lb]{\smash{1.3}}}}%
    \put(0.8872,0.1759){\rotatebox{90}{\makebox(0,0)[lb]{\smash{2.4}}}}%
    \put(0.9417,0.1759){\rotatebox{90}{\makebox(0,0)[lb]{\smash{134.4}}}}%
    \put(0.0455,0.0263){\makebox(0,0)[rb]{\smash{0}}}%
    \put(0.0455,0.0401){\makebox(0,0)[rb]{\smash{100}}}%
    \put(0.0455,0.0538){\makebox(0,0)[rb]{\smash{200}}}%
    \put(0.0455,0.0676){\makebox(0,0)[rb]{\smash{300}}}%
    \put(0.0455,0.0814){\makebox(0,0)[rb]{\smash{400}}}%
    \put(0.0455,0.0951){\makebox(0,0)[rb]{\smash{500}}}%
    \put(0.0455,0.1089){\makebox(0,0)[rb]{\smash{600}}}%
    \put(0.0455,0.1226){\makebox(0,0)[rb]{\smash{700}}}%
    \put(0.0455,0.1364){\makebox(0,0)[rb]{\smash{800}}}%
    \put(0.0455,0.1502){\makebox(0,0)[rb]{\smash{900}}}%
    \put(0.0455,0.1639){\makebox(0,0)[rb]{\smash{1000}}}%
    \put(0.0797,0.0113){\makebox(0,0)[b]{\smash{Q1}}}%
    \put(0.1342,0.0113){\makebox(0,0)[b]{\smash{Q2}}}%
    \put(0.1887,0.0113){\makebox(0,0)[b]{\smash{Q3}}}%
    \put(0.2432,0.0113){\makebox(0,0)[b]{\smash{Q4}}}%
    \put(0.2977,0.0113){\makebox(0,0)[b]{\smash{Q5}}}%
    \put(0.3522,0.0113){\makebox(0,0)[b]{\smash{Q6}}}%
    \put(0.4067,0.0113){\makebox(0,0)[b]{\smash{Q7}}}%
    \put(0.4612,0.0113){\makebox(0,0)[b]{\smash{Q8}}}%
    \put(0.5157,0.0113){\makebox(0,0)[b]{\smash{Q9}}}%
    \put(0.5702,0.0113){\makebox(0,0)[b]{\smash{Q10}}}%
    \put(0.6247,0.0113){\makebox(0,0)[b]{\smash{Q11}}}%
    \put(0.6792,0.0113){\makebox(0,0)[b]{\smash{Q12}}}%
    \put(0.7337,0.0113){\makebox(0,0)[b]{\smash{Q13}}}%
    \put(0.7882,0.0113){\makebox(0,0)[b]{\smash{Q14}}}%
    \put(0.8427,0.0113){\makebox(0,0)[b]{\smash{Q15}}}%
    \put(0.8972,0.0113){\makebox(0,0)[b]{\smash{Q16}}}%
    \put(0.9517,0.0113){\makebox(0,0)[b]{\smash{Q17}}}%
    \put(0.0175,0.1942){\makebox(0,0)[lb]{\smash{\% of}}}%
    \put(0.0175,0.1792){\makebox(0,0)[lb]{\smash{SXSI}}}%
    \put(0.0834,0.1759){\rotatebox{90}{\makebox(0,0)[lb]{\smash{6.8}}}}%
    \put(0.1379,0.1759){\rotatebox{90}{\makebox(0,0)[lb]{\smash{12.4}}}}%
    \put(0.1924,0.1759){\rotatebox{90}{\makebox(0,0)[lb]{\smash{17.5}}}}%
    \put(0.2469,0.1759){\rotatebox{90}{\makebox(0,0)[lb]{\smash{17.0}}}}%
    \put(0.3014,0.1759){\rotatebox{90}{\makebox(0,0)[lb]{\smash{21.6}}}}%
    \put(0.3559,0.1759){\rotatebox{90}{\makebox(0,0)[lb]{\smash{18.2}}}}%
    \put(0.4104,0.1759){\rotatebox{90}{\makebox(0,0)[lb]{\smash{24.0}}}}%
    \put(0.4649,0.1759){\rotatebox{90}{\makebox(0,0)[lb]{\smash{37.8}}}}%
    \put(0.5194,0.1759){\rotatebox{90}{\makebox(0,0)[lb]{\smash{49.9}}}}%
    \put(0.5739,0.1759){\rotatebox{90}{\makebox(0,0)[lb]{\smash{34.4}}}}%
    \put(0.6284,0.1759){\rotatebox{90}{\makebox(0,0)[lb]{\smash{135.3}}}}%
    \put(0.6829,0.1759){\rotatebox{90}{\makebox(0,0)[lb]{\smash{+++}}}}%
    \put(0.7374,0.1759){\rotatebox{90}{\makebox(0,0)[lb]{\smash{130.7}}}}%
    \put(0.7919,0.1759){\rotatebox{90}{\makebox(0,0)[lb]{\smash{47.5}}}}%
    \put(0.8464,0.1759){\rotatebox{90}{\makebox(0,0)[lb]{\smash{146.7}}}}%
    \put(0.9009,0.1759){\rotatebox{90}{\makebox(0,0)[lb]{\smash{245.7}}}}%
    \put(0.9554,0.1759){\rotatebox{90}{\makebox(0,0)[lb]{\smash{349.8}}}}%
    \put(0.0455,0.0263){\makebox(0,0)[rb]{\smash{0}}}%
    \put(0.0455,0.0401){\makebox(0,0)[rb]{\smash{100}}}%
    \put(0.0455,0.0538){\makebox(0,0)[rb]{\smash{200}}}%
    \put(0.0455,0.0676){\makebox(0,0)[rb]{\smash{300}}}%
    \put(0.0455,0.0814){\makebox(0,0)[rb]{\smash{400}}}%
    \put(0.0455,0.0951){\makebox(0,0)[rb]{\smash{500}}}%
    \put(0.0455,0.1089){\makebox(0,0)[rb]{\smash{600}}}%
    \put(0.0455,0.1226){\makebox(0,0)[rb]{\smash{700}}}%
    \put(0.0455,0.1364){\makebox(0,0)[rb]{\smash{800}}}%
    \put(0.0455,0.1502){\makebox(0,0)[rb]{\smash{900}}}%
    \put(0.0455,0.1639){\makebox(0,0)[rb]{\smash{1000}}}%
    \put(0.0797,0.0113){\makebox(0,0)[b]{\smash{Q1}}}%
    \put(0.1342,0.0113){\makebox(0,0)[b]{\smash{Q2}}}%
    \put(0.1887,0.0113){\makebox(0,0)[b]{\smash{Q3}}}%
    \put(0.2432,0.0113){\makebox(0,0)[b]{\smash{Q4}}}%
    \put(0.2977,0.0113){\makebox(0,0)[b]{\smash{Q5}}}%
    \put(0.3522,0.0113){\makebox(0,0)[b]{\smash{Q6}}}%
    \put(0.4067,0.0113){\makebox(0,0)[b]{\smash{Q7}}}%
    \put(0.4612,0.0113){\makebox(0,0)[b]{\smash{Q8}}}%
    \put(0.5157,0.0113){\makebox(0,0)[b]{\smash{Q9}}}%
    \put(0.5702,0.0113){\makebox(0,0)[b]{\smash{Q10}}}%
    \put(0.6247,0.0113){\makebox(0,0)[b]{\smash{Q11}}}%
    \put(0.6792,0.0113){\makebox(0,0)[b]{\smash{Q12}}}%
    \put(0.7337,0.0113){\makebox(0,0)[b]{\smash{Q13}}}%
    \put(0.7882,0.0113){\makebox(0,0)[b]{\smash{Q14}}}%
    \put(0.8427,0.0113){\makebox(0,0)[b]{\smash{Q15}}}%
    \put(0.8972,0.0113){\makebox(0,0)[b]{\smash{Q16}}}%
    \put(0.9517,0.0113){\makebox(0,0)[b]{\smash{Q17}}}%
    \put(0.0175,0.1942){\makebox(0,0)[lb]{\smash{\% of}}}%
    \put(0.0175,0.1792){\makebox(0,0)[lb]{\smash{SXSI}}}%
    \put(0.0971,0.1759){\rotatebox{90}{\makebox(0,0)[lb]{\smash{3.5}}}}%
    \put(0.1516,0.1759){\rotatebox{90}{\makebox(0,0)[lb]{\smash{33.0}}}}%
    \put(0.2061,0.1759){\rotatebox{90}{\makebox(0,0)[lb]{\smash{72.2}}}}%
    \put(0.2606,0.1759){\rotatebox{90}{\makebox(0,0)[lb]{\smash{65.8}}}}%
    \put(0.315,0.1759){\rotatebox{90}{\makebox(0,0)[lb]{\smash{56.2}}}}%
    \put(0.3695,0.1759){\rotatebox{90}{\makebox(0,0)[lb]{\smash{71.8}}}}%
    \put(0.424,0.1759){\rotatebox{90}{\makebox(0,0)[lb]{\smash{230.5}}}}%
    \put(0.4785,0.1759){\rotatebox{90}{\makebox(0,0)[lb]{\smash{188.8}}}}%
    \put(0.533,0.1759){\rotatebox{90}{\makebox(0,0)[lb]{\smash{208.2}}}}%
    \put(0.5875,0.1759){\rotatebox{90}{\makebox(0,0)[lb]{\smash{287.0}}}}%
    \put(0.642,0.1759){\rotatebox{90}{\makebox(0,0)[lb]{\smash{515.2}}}}%
    \put(0.6965,0.1759){\rotatebox{90}{\makebox(0,0)[lb]{\smash{77.0}}}}%
    \put(0.751,0.1759){\rotatebox{90}{\makebox(0,0)[lb]{\smash{7.2}}}}%
    \put(0.8055,0.1759){\rotatebox{90}{\makebox(0,0)[lb]{\smash{127.9s}}}}%
    \put(0.86,0.1759){\rotatebox{90}{\makebox(0,0)[lb]{\smash{177.1s}}}}%
    \put(0.9145,0.1759){\rotatebox{90}{\makebox(0,0)[lb]{\smash{312.9s}}}}%
    \put(0.969,0.1759){\rotatebox{90}{\makebox(0,0)[lb]{\smash{371.1s}}}}%
  \end{picture}%
\endgroup

%% file: mat_xmark_01_04.pdf_tex

\begingroup
  \makeatletter
  \providecommand\color[2][]{%
    \errmessage{(Inkscape) Color is used for the text in Inkscape, but the package 'color.sty' is not loaded}
    \renewcommand\color[2][]{}%
  }
  \providecommand\transparent[1]{%
    \errmessage{(Inkscape) Transparency is used (non-zero) for the text in Inkscape, but the package 'transparent.sty' is not loaded}
    \renewcommand\transparent[1]{}%
  }
  \providecommand\rotatebox[2]{#2}
  \ifx\svgwidth\undefined
    \setlength{\unitlength}{800pt}
  \else
    \setlength{\unitlength}{\svgwidth}
  \fi
  \global\let\svgwidth\undefined
  \makeatother
  \begin{picture}(1,0.2)%
    \put(0,0){\includegraphics[width=\unitlength]{mat_xmark_01_04.pdf}}%
    \put(0.0455,0.0263){\makebox(0,0)[rb]{\smash{0}}}%
    \put(0.0455,0.0401){\makebox(0,0)[rb]{\smash{100}}}%
    \put(0.0455,0.0538){\makebox(0,0)[rb]{\smash{200}}}%
    \put(0.0455,0.0676){\makebox(0,0)[rb]{\smash{300}}}%
    \put(0.0455,0.0814){\makebox(0,0)[rb]{\smash{400}}}%
    \put(0.0455,0.0951){\makebox(0,0)[rb]{\smash{500}}}%
    \put(0.0455,0.1089){\makebox(0,0)[rb]{\smash{600}}}%
    \put(0.0455,0.1226){\makebox(0,0)[rb]{\smash{700}}}%
    \put(0.0455,0.1364){\makebox(0,0)[rb]{\smash{800}}}%
    \put(0.0455,0.1502){\makebox(0,0)[rb]{\smash{900}}}%
    \put(0.0455,0.1639){\makebox(0,0)[rb]{\smash{1000}}}%
    \put(0.0797,0.0113){\makebox(0,0)[b]{\smash{Q1}}}%
    \put(0.1342,0.0113){\makebox(0,0)[b]{\smash{Q2}}}%
    \put(0.1887,0.0113){\makebox(0,0)[b]{\smash{Q3}}}%
    \put(0.2432,0.0113){\makebox(0,0)[b]{\smash{Q4}}}%
    \put(0.2977,0.0113){\makebox(0,0)[b]{\smash{Q5}}}%
    \put(0.3522,0.0113){\makebox(0,0)[b]{\smash{Q6}}}%
    \put(0.4067,0.0113){\makebox(0,0)[b]{\smash{Q7}}}%
    \put(0.4612,0.0113){\makebox(0,0)[b]{\smash{Q8}}}%
    \put(0.5157,0.0113){\makebox(0,0)[b]{\smash{Q9}}}%
    \put(0.5702,0.0113){\makebox(0,0)[b]{\smash{Q10}}}%
    \put(0.6247,0.0113){\makebox(0,0)[b]{\smash{Q11}}}%
    \put(0.6792,0.0113){\makebox(0,0)[b]{\smash{Q12}}}%
    \put(0.7337,0.0113){\makebox(0,0)[b]{\smash{Q13}}}%
    \put(0.7882,0.0113){\makebox(0,0)[b]{\smash{Q14}}}%
    \put(0.8427,0.0113){\makebox(0,0)[b]{\smash{Q15}}}%
    \put(0.8972,0.0113){\makebox(0,0)[b]{\smash{Q16}}}%
    \put(0.9517,0.0113){\makebox(0,0)[b]{\smash{Q17}}}%
    \put(0.0175,0.1942){\makebox(0,0)[lb]{\smash{\% of}}}%
    \put(0.0175,0.1792){\makebox(0,0)[lb]{\smash{SXSI}}}%
    \put(0.0455,0.0263){\makebox(0,0)[rb]{\smash{0}}}%
    \put(0.0455,0.0401){\makebox(0,0)[rb]{\smash{100}}}%
    \put(0.0455,0.0538){\makebox(0,0)[rb]{\smash{200}}}%
    \put(0.0455,0.0676){\makebox(0,0)[rb]{\smash{300}}}%
    \put(0.0455,0.0814){\makebox(0,0)[rb]{\smash{400}}}%
    \put(0.0455,0.0951){\makebox(0,0)[rb]{\smash{500}}}%
    \put(0.0455,0.1089){\makebox(0,0)[rb]{\smash{600}}}%
    \put(0.0455,0.1226){\makebox(0,0)[rb]{\smash{700}}}%
    \put(0.0455,0.1364){\makebox(0,0)[rb]{\smash{800}}}%
    \put(0.0455,0.1502){\makebox(0,0)[rb]{\smash{900}}}%
    \put(0.0455,0.1639){\makebox(0,0)[rb]{\smash{1000}}}%
    \put(0.0797,0.0113){\makebox(0,0)[b]{\smash{Q1}}}%
    \put(0.1342,0.0113){\makebox(0,0)[b]{\smash{Q2}}}%
    \put(0.1887,0.0113){\makebox(0,0)[b]{\smash{Q3}}}%
    \put(0.2432,0.0113){\makebox(0,0)[b]{\smash{Q4}}}%
    \put(0.2977,0.0113){\makebox(0,0)[b]{\smash{Q5}}}%
    \put(0.3522,0.0113){\makebox(0,0)[b]{\smash{Q6}}}%
    \put(0.4067,0.0113){\makebox(0,0)[b]{\smash{Q7}}}%
    \put(0.4612,0.0113){\makebox(0,0)[b]{\smash{Q8}}}%
    \put(0.5157,0.0113){\makebox(0,0)[b]{\smash{Q9}}}%
    \put(0.5702,0.0113){\makebox(0,0)[b]{\smash{Q10}}}%
    \put(0.6247,0.0113){\makebox(0,0)[b]{\smash{Q11}}}%
    \put(0.6792,0.0113){\makebox(0,0)[b]{\smash{Q12}}}%
    \put(0.7337,0.0113){\makebox(0,0)[b]{\smash{Q13}}}%
    \put(0.7882,0.0113){\makebox(0,0)[b]{\smash{Q14}}}%
    \put(0.8427,0.0113){\makebox(0,0)[b]{\smash{Q15}}}%
    \put(0.8972,0.0113){\makebox(0,0)[b]{\smash{Q16}}}%
    \put(0.9517,0.0113){\makebox(0,0)[b]{\smash{Q17}}}%
    \put(0.0175,0.1942){\makebox(0,0)[lb]{\smash{\% of}}}%
    \put(0.0175,0.1792){\makebox(0,0)[lb]{\smash{SXSI}}}%
    \put(0.0698,0.1759){\rotatebox{90}{\makebox(0,0)[lb]{\smash{1.1}}}}%
    \put(0.1243,0.1759){\rotatebox{90}{\makebox(0,0)[lb]{\smash{7.3}}}}%
    \put(0.1788,0.1759){\rotatebox{90}{\makebox(0,0)[lb]{\smash{11.9}}}}%
    \put(0.2333,0.1759){\rotatebox{90}{\makebox(0,0)[lb]{\smash{22.7}}}}%
    \put(0.2878,0.1759){\rotatebox{90}{\makebox(0,0)[lb]{\smash{15.0}}}}%
    \put(0.3423,0.1759){\rotatebox{90}{\makebox(0,0)[lb]{\smash{20.6}}}}%
    \put(0.3968,0.1759){\rotatebox{90}{\makebox(0,0)[lb]{\smash{25.8}}}}%
    \put(0.4513,0.1759){\rotatebox{90}{\makebox(0,0)[lb]{\smash{21.8}}}}%
    \put(0.5058,0.1759){\rotatebox{90}{\makebox(0,0)[lb]{\smash{28.7}}}}%
    \put(0.5603,0.1759){\rotatebox{90}{\makebox(0,0)[lb]{\smash{54.2}}}}%
    \put(0.6148,0.1759){\rotatebox{90}{\makebox(0,0)[lb]{\smash{135.8}}}}%
    \put(0.6693,0.1759){\rotatebox{90}{\makebox(0,0)[lb]{\smash{34.3}}}}%
    \put(0.7238,0.1759){\rotatebox{90}{\makebox(0,0)[lb]{\smash{1.8}}}}%
    \put(0.7782,0.1759){\rotatebox{90}{\makebox(0,0)[lb]{\smash{0.7}}}}%
    \put(0.8327,0.1759){\rotatebox{90}{\makebox(0,0)[lb]{\smash{1.3}}}}%
    \put(0.8872,0.1759){\rotatebox{90}{\makebox(0,0)[lb]{\smash{2.4}}}}%
    \put(0.9417,0.1759){\rotatebox{90}{\makebox(0,0)[lb]{\smash{144.3}}}}%
    \put(0.0455,0.0263){\makebox(0,0)[rb]{\smash{0}}}%
    \put(0.0455,0.0401){\makebox(0,0)[rb]{\smash{100}}}%
    \put(0.0455,0.0538){\makebox(0,0)[rb]{\smash{200}}}%
    \put(0.0455,0.0676){\makebox(0,0)[rb]{\smash{300}}}%
    \put(0.0455,0.0814){\makebox(0,0)[rb]{\smash{400}}}%
    \put(0.0455,0.0951){\makebox(0,0)[rb]{\smash{500}}}%
    \put(0.0455,0.1089){\makebox(0,0)[rb]{\smash{600}}}%
    \put(0.0455,0.1226){\makebox(0,0)[rb]{\smash{700}}}%
    \put(0.0455,0.1364){\makebox(0,0)[rb]{\smash{800}}}%
    \put(0.0455,0.1502){\makebox(0,0)[rb]{\smash{900}}}%
    \put(0.0455,0.1639){\makebox(0,0)[rb]{\smash{1000}}}%
    \put(0.0797,0.0113){\makebox(0,0)[b]{\smash{Q1}}}%
    \put(0.1342,0.0113){\makebox(0,0)[b]{\smash{Q2}}}%
    \put(0.1887,0.0113){\makebox(0,0)[b]{\smash{Q3}}}%
    \put(0.2432,0.0113){\makebox(0,0)[b]{\smash{Q4}}}%
    \put(0.2977,0.0113){\makebox(0,0)[b]{\smash{Q5}}}%
    \put(0.3522,0.0113){\makebox(0,0)[b]{\smash{Q6}}}%
    \put(0.4067,0.0113){\makebox(0,0)[b]{\smash{Q7}}}%
    \put(0.4612,0.0113){\makebox(0,0)[b]{\smash{Q8}}}%
    \put(0.5157,0.0113){\makebox(0,0)[b]{\smash{Q9}}}%
    \put(0.5702,0.0113){\makebox(0,0)[b]{\smash{Q10}}}%
    \put(0.6247,0.0113){\makebox(0,0)[b]{\smash{Q11}}}%
    \put(0.6792,0.0113){\makebox(0,0)[b]{\smash{Q12}}}%
    \put(0.7337,0.0113){\makebox(0,0)[b]{\smash{Q13}}}%
    \put(0.7882,0.0113){\makebox(0,0)[b]{\smash{Q14}}}%
    \put(0.8427,0.0113){\makebox(0,0)[b]{\smash{Q15}}}%
    \put(0.8972,0.0113){\makebox(0,0)[b]{\smash{Q16}}}%
    \put(0.9517,0.0113){\makebox(0,0)[b]{\smash{Q17}}}%
    \put(0.0175,0.1942){\makebox(0,0)[lb]{\smash{\% of}}}%
    \put(0.0175,0.1792){\makebox(0,0)[lb]{\smash{SXSI}}}%
    \put(0.0834,0.1759){\rotatebox{90}{\makebox(0,0)[lb]{\smash{5.9}}}}%
    \put(0.1379,0.1759){\rotatebox{90}{\makebox(0,0)[lb]{\smash{11.8}}}}%
    \put(0.1924,0.1759){\rotatebox{90}{\makebox(0,0)[lb]{\smash{18.3}}}}%
    \put(0.2469,0.1759){\rotatebox{90}{\makebox(0,0)[lb]{\smash{16.3}}}}%
    \put(0.3014,0.1759){\rotatebox{90}{\makebox(0,0)[lb]{\smash{21.7}}}}%
    \put(0.3559,0.1759){\rotatebox{90}{\makebox(0,0)[lb]{\smash{16.8}}}}%
    \put(0.4104,0.1759){\rotatebox{90}{\makebox(0,0)[lb]{\smash{24.4}}}}%
    \put(0.4649,0.1759){\rotatebox{90}{\makebox(0,0)[lb]{\smash{37.9}}}}%
    \put(0.5194,0.1759){\rotatebox{90}{\makebox(0,0)[lb]{\smash{49.9}}}}%
    \put(0.5739,0.1759){\rotatebox{90}{\makebox(0,0)[lb]{\smash{34.8}}}}%
    \put(0.6284,0.1759){\rotatebox{90}{\makebox(0,0)[lb]{\smash{134.1}}}}%
    \put(0.6829,0.1759){\rotatebox{90}{\makebox(0,0)[lb]{\smash{+++}}}}%
    \put(0.7374,0.1759){\rotatebox{90}{\makebox(0,0)[lb]{\smash{128.4}}}}%
    \put(0.7919,0.1759){\rotatebox{90}{\makebox(0,0)[lb]{\smash{40.3}}}}%
    \put(0.8464,0.1759){\rotatebox{90}{\makebox(0,0)[lb]{\smash{146.2}}}}%
    \put(0.9009,0.1759){\rotatebox{90}{\makebox(0,0)[lb]{\smash{244.6}}}}%
    \put(0.9554,0.1759){\rotatebox{90}{\makebox(0,0)[lb]{\smash{351.6}}}}%
  \end{picture}%
\endgroup

%% file: matser_xmark_01_04.pdf_tex

\begingroup
  \makeatletter
  \providecommand\color[2][]{%
    \errmessage{(Inkscape) Color is used for the text in Inkscape, but the package 'color.sty' is not loaded}
    \renewcommand\color[2][]{}%
  }
  \providecommand\transparent[1]{%
    \errmessage{(Inkscape) Transparency is used (non-zero) for the text in Inkscape, but the package 'transparent.sty' is not loaded}
    \renewcommand\transparent[1]{}%
  }
  \providecommand\rotatebox[2]{#2}
  \ifx\svgwidth\undefined
    \setlength{\unitlength}{800pt}
  \else
    \setlength{\unitlength}{\svgwidth}
  \fi
  \global\let\svgwidth\undefined
  \makeatother
  \begin{picture}(1,0.2)%
    \put(0,0){\includegraphics[width=\unitlength]{matser_xmark_01_04.pdf}}%
    \put(0.0455,0.0263){\makebox(0,0)[rb]{\smash{0}}}%
    \put(0.0455,0.0401){\makebox(0,0)[rb]{\smash{100}}}%
    \put(0.0455,0.0538){\makebox(0,0)[rb]{\smash{200}}}%
    \put(0.0455,0.0676){\makebox(0,0)[rb]{\smash{300}}}%
    \put(0.0455,0.0814){\makebox(0,0)[rb]{\smash{400}}}%
    \put(0.0455,0.0951){\makebox(0,0)[rb]{\smash{500}}}%
    \put(0.0455,0.1089){\makebox(0,0)[rb]{\smash{600}}}%
    \put(0.0455,0.1226){\makebox(0,0)[rb]{\smash{700}}}%
    \put(0.0455,0.1364){\makebox(0,0)[rb]{\smash{800}}}%
    \put(0.0455,0.1502){\makebox(0,0)[rb]{\smash{900}}}%
    \put(0.0455,0.1639){\makebox(0,0)[rb]{\smash{1000}}}%
    \put(0.0797,0.0113){\makebox(0,0)[b]{\smash{Q1}}}%
    \put(0.1342,0.0113){\makebox(0,0)[b]{\smash{Q2}}}%
    \put(0.1887,0.0113){\makebox(0,0)[b]{\smash{Q3}}}%
    \put(0.2432,0.0113){\makebox(0,0)[b]{\smash{Q4}}}%
    \put(0.2977,0.0113){\makebox(0,0)[b]{\smash{Q5}}}%
    \put(0.3522,0.0113){\makebox(0,0)[b]{\smash{Q6}}}%
    \put(0.4067,0.0113){\makebox(0,0)[b]{\smash{Q7}}}%
    \put(0.4612,0.0113){\makebox(0,0)[b]{\smash{Q8}}}%
    \put(0.5157,0.0113){\makebox(0,0)[b]{\smash{Q9}}}%
    \put(0.5702,0.0113){\makebox(0,0)[b]{\smash{Q10}}}%
    \put(0.6247,0.0113){\makebox(0,0)[b]{\smash{Q11}}}%
    \put(0.6792,0.0113){\makebox(0,0)[b]{\smash{Q12}}}%
    \put(0.7337,0.0113){\makebox(0,0)[b]{\smash{Q13}}}%
    \put(0.7882,0.0113){\makebox(0,0)[b]{\smash{Q14}}}%
    \put(0.8427,0.0113){\makebox(0,0)[b]{\smash{Q15}}}%
    \put(0.8972,0.0113){\makebox(0,0)[b]{\smash{Q16}}}%
    \put(0.9517,0.0113){\makebox(0,0)[b]{\smash{Q17}}}%
    \put(0.0175,0.1942){\makebox(0,0)[lb]{\smash{\% of}}}%
    \put(0.0175,0.1792){\makebox(0,0)[lb]{\smash{SXSI}}}%
    \put(0.0455,0.0263){\makebox(0,0)[rb]{\smash{0}}}%
    \put(0.0455,0.0401){\makebox(0,0)[rb]{\smash{100}}}%
    \put(0.0455,0.0538){\makebox(0,0)[rb]{\smash{200}}}%
    \put(0.0455,0.0676){\makebox(0,0)[rb]{\smash{300}}}%
    \put(0.0455,0.0814){\makebox(0,0)[rb]{\smash{400}}}%
    \put(0.0455,0.0951){\makebox(0,0)[rb]{\smash{500}}}%
    \put(0.0455,0.1089){\makebox(0,0)[rb]{\smash{600}}}%
    \put(0.0455,0.1226){\makebox(0,0)[rb]{\smash{700}}}%
    \put(0.0455,0.1364){\makebox(0,0)[rb]{\smash{800}}}%
    \put(0.0455,0.1502){\makebox(0,0)[rb]{\smash{900}}}%
    \put(0.0455,0.1639){\makebox(0,0)[rb]{\smash{1000}}}%
    \put(0.0797,0.0113){\makebox(0,0)[b]{\smash{Q1}}}%
    \put(0.1342,0.0113){\makebox(0,0)[b]{\smash{Q2}}}%
    \put(0.1887,0.0113){\makebox(0,0)[b]{\smash{Q3}}}%
    \put(0.2432,0.0113){\makebox(0,0)[b]{\smash{Q4}}}%
    \put(0.2977,0.0113){\makebox(0,0)[b]{\smash{Q5}}}%
    \put(0.3522,0.0113){\makebox(0,0)[b]{\smash{Q6}}}%
    \put(0.4067,0.0113){\makebox(0,0)[b]{\smash{Q7}}}%
    \put(0.4612,0.0113){\makebox(0,0)[b]{\smash{Q8}}}%
    \put(0.5157,0.0113){\makebox(0,0)[b]{\smash{Q9}}}%
    \put(0.5702,0.0113){\makebox(0,0)[b]{\smash{Q10}}}%
    \put(0.6247,0.0113){\makebox(0,0)[b]{\smash{Q11}}}%
    \put(0.6792,0.0113){\makebox(0,0)[b]{\smash{Q12}}}%
    \put(0.7337,0.0113){\makebox(0,0)[b]{\smash{Q13}}}%
    \put(0.7882,0.0113){\makebox(0,0)[b]{\smash{Q14}}}%
    \put(0.8427,0.0113){\makebox(0,0)[b]{\smash{Q15}}}%
    \put(0.8972,0.0113){\makebox(0,0)[b]{\smash{Q16}}}%
    \put(0.9517,0.0113){\makebox(0,0)[b]{\smash{Q17}}}%
    \put(0.0175,0.1942){\makebox(0,0)[lb]{\smash{\% of}}}%
    \put(0.0175,0.1792){\makebox(0,0)[lb]{\smash{SXSI}}}%
    \put(0.0698,0.1759){\rotatebox{90}{\makebox(0,0)[lb]{\smash{391.0}}}}%
    \put(0.1243,0.1759){\rotatebox{90}{\makebox(0,0)[lb]{\smash{417.0}}}}%
    \put(0.1788,0.1759){\rotatebox{90}{\makebox(0,0)[lb]{\smash{17.0}}}}%
    \put(0.2333,0.1759){\rotatebox{90}{\makebox(0,0)[lb]{\smash{80.0}}}}%
    \put(0.2878,0.1759){\rotatebox{90}{\makebox(0,0)[lb]{\smash{18.0}}}}%
    \put(0.3423,0.1759){\rotatebox{90}{\makebox(0,0)[lb]{\smash{27.0}}}}%
    \put(0.3968,0.1759){\rotatebox{90}{\makebox(0,0)[lb]{\smash{30.0}}}}%
    \put(0.4513,0.1759){\rotatebox{90}{\makebox(0,0)[lb]{\smash{40.0}}}}%
    \put(0.5058,0.1759){\rotatebox{90}{\makebox(0,0)[lb]{\smash{37.0}}}}%
    \put(0.5603,0.1759){\rotatebox{90}{\makebox(0,0)[lb]{\smash{161.0}}}}%
    \put(0.6148,0.1759){\rotatebox{90}{\makebox(0,0)[lb]{\smash{251.0}}}}%
    \put(0.6693,0.1759){\rotatebox{90}{\makebox(0,0)[lb]{\smash{91.0}}}}%
    \put(0.7238,0.1759){\rotatebox{90}{\makebox(0,0)[lb]{\smash{784.0}}}}%
    \put(0.7782,0.1759){\rotatebox{90}{\makebox(0,0)[lb]{\smash{5.7s}}}}%
    \put(0.8327,0.1759){\rotatebox{90}{\makebox(0,0)[lb]{\smash{5.7s}}}}%
    \put(0.8872,0.1759){\rotatebox{90}{\makebox(0,0)[lb]{\smash{4.9s}}}}%
    \put(0.9417,0.1759){\rotatebox{90}{\makebox(0,0)[lb]{\smash{4.3s}}}}%
    \put(0.0455,0.0263){\makebox(0,0)[rb]{\smash{0}}}%
    \put(0.0455,0.0401){\makebox(0,0)[rb]{\smash{100}}}%
    \put(0.0455,0.0538){\makebox(0,0)[rb]{\smash{200}}}%
    \put(0.0455,0.0676){\makebox(0,0)[rb]{\smash{300}}}%
    \put(0.0455,0.0814){\makebox(0,0)[rb]{\smash{400}}}%
    \put(0.0455,0.0951){\makebox(0,0)[rb]{\smash{500}}}%
    \put(0.0455,0.1089){\makebox(0,0)[rb]{\smash{600}}}%
    \put(0.0455,0.1226){\makebox(0,0)[rb]{\smash{700}}}%
    \put(0.0455,0.1364){\makebox(0,0)[rb]{\smash{800}}}%
    \put(0.0455,0.1502){\makebox(0,0)[rb]{\smash{900}}}%
    \put(0.0455,0.1639){\makebox(0,0)[rb]{\smash{1000}}}%
    \put(0.0797,0.0113){\makebox(0,0)[b]{\smash{Q1}}}%
    \put(0.1342,0.0113){\makebox(0,0)[b]{\smash{Q2}}}%
    \put(0.1887,0.0113){\makebox(0,0)[b]{\smash{Q3}}}%
    \put(0.2432,0.0113){\makebox(0,0)[b]{\smash{Q4}}}%
    \put(0.2977,0.0113){\makebox(0,0)[b]{\smash{Q5}}}%
    \put(0.3522,0.0113){\makebox(0,0)[b]{\smash{Q6}}}%
    \put(0.4067,0.0113){\makebox(0,0)[b]{\smash{Q7}}}%
    \put(0.4612,0.0113){\makebox(0,0)[b]{\smash{Q8}}}%
    \put(0.5157,0.0113){\makebox(0,0)[b]{\smash{Q9}}}%
    \put(0.5702,0.0113){\makebox(0,0)[b]{\smash{Q10}}}%
    \put(0.6247,0.0113){\makebox(0,0)[b]{\smash{Q11}}}%
    \put(0.6792,0.0113){\makebox(0,0)[b]{\smash{Q12}}}%
    \put(0.7337,0.0113){\makebox(0,0)[b]{\smash{Q13}}}%
    \put(0.7882,0.0113){\makebox(0,0)[b]{\smash{Q14}}}%
    \put(0.8427,0.0113){\makebox(0,0)[b]{\smash{Q15}}}%
    \put(0.8972,0.0113){\makebox(0,0)[b]{\smash{Q16}}}%
    \put(0.9517,0.0113){\makebox(0,0)[b]{\smash{Q17}}}%
    \put(0.0175,0.1942){\makebox(0,0)[lb]{\smash{\% of}}}%
    \put(0.0175,0.1792){\makebox(0,0)[lb]{\smash{SXSI}}}%
    \put(0.0834,0.1759){\rotatebox{90}{\makebox(0,0)[lb]{\smash{497.0}}}}%
    \put(0.1379,0.1759){\rotatebox{90}{\makebox(0,0)[lb]{\smash{453.0}}}}%
    \put(0.1924,0.1759){\rotatebox{90}{\makebox(0,0)[lb]{\smash{26.0}}}}%
    \put(0.2469,0.1759){\rotatebox{90}{\makebox(0,0)[lb]{\smash{73.0}}}}%
    \put(0.3014,0.1759){\rotatebox{90}{\makebox(0,0)[lb]{\smash{27.0}}}}%
    \put(0.3559,0.1759){\rotatebox{90}{\makebox(0,0)[lb]{\smash{26.0}}}}%
    \put(0.4104,0.1759){\rotatebox{90}{\makebox(0,0)[lb]{\smash{30.0}}}}%
    \put(0.4649,0.1759){\rotatebox{90}{\makebox(0,0)[lb]{\smash{58.0}}}}%
    \put(0.5194,0.1759){\rotatebox{90}{\makebox(0,0)[lb]{\smash{60.0}}}}%
    \put(0.5739,0.1759){\rotatebox{90}{\makebox(0,0)[lb]{\smash{135.0}}}}%
    \put(0.6284,0.1759){\rotatebox{90}{\makebox(0,0)[lb]{\smash{240.0}}}}%
    \put(0.6829,0.1759){\rotatebox{90}{\makebox(0,0)[lb]{\smash{+++}}}}%
    \put(0.7374,0.1759){\rotatebox{90}{\makebox(0,0)[lb]{\smash{1.3s}}}}%
    \put(0.7919,0.1759){\rotatebox{90}{\makebox(0,0)[lb]{\smash{6.5s}}}}%
    \put(0.8464,0.1759){\rotatebox{90}{\makebox(0,0)[lb]{\smash{5.5s}}}}%
    \put(0.9009,0.1759){\rotatebox{90}{\makebox(0,0)[lb]{\smash{4.2s}}}}%
    \put(0.9554,0.1759){\rotatebox{90}{\makebox(0,0)[lb]{\smash{3.5s}}}}%
    \put(0.0455,0.0263){\makebox(0,0)[rb]{\smash{0}}}%
    \put(0.0455,0.0401){\makebox(0,0)[rb]{\smash{100}}}%
    \put(0.0455,0.0538){\makebox(0,0)[rb]{\smash{200}}}%
    \put(0.0455,0.0676){\makebox(0,0)[rb]{\smash{300}}}%
    \put(0.0455,0.0814){\makebox(0,0)[rb]{\smash{400}}}%
    \put(0.0455,0.0951){\makebox(0,0)[rb]{\smash{500}}}%
    \put(0.0455,0.1089){\makebox(0,0)[rb]{\smash{600}}}%
    \put(0.0455,0.1226){\makebox(0,0)[rb]{\smash{700}}}%
    \put(0.0455,0.1364){\makebox(0,0)[rb]{\smash{800}}}%
    \put(0.0455,0.1502){\makebox(0,0)[rb]{\smash{900}}}%
    \put(0.0455,0.1639){\makebox(0,0)[rb]{\smash{1000}}}%
    \put(0.0797,0.0113){\makebox(0,0)[b]{\smash{Q1}}}%
    \put(0.1342,0.0113){\makebox(0,0)[b]{\smash{Q2}}}%
    \put(0.1887,0.0113){\makebox(0,0)[b]{\smash{Q3}}}%
    \put(0.2432,0.0113){\makebox(0,0)[b]{\smash{Q4}}}%
    \put(0.2977,0.0113){\makebox(0,0)[b]{\smash{Q5}}}%
    \put(0.3522,0.0113){\makebox(0,0)[b]{\smash{Q6}}}%
    \put(0.4067,0.0113){\makebox(0,0)[b]{\smash{Q7}}}%
    \put(0.4612,0.0113){\makebox(0,0)[b]{\smash{Q8}}}%
    \put(0.5157,0.0113){\makebox(0,0)[b]{\smash{Q9}}}%
    \put(0.5702,0.0113){\makebox(0,0)[b]{\smash{Q10}}}%
    \put(0.6247,0.0113){\makebox(0,0)[b]{\smash{Q11}}}%
    \put(0.6792,0.0113){\makebox(0,0)[b]{\smash{Q12}}}%
    \put(0.7337,0.0113){\makebox(0,0)[b]{\smash{Q13}}}%
    \put(0.7882,0.0113){\makebox(0,0)[b]{\smash{Q14}}}%
    \put(0.8427,0.0113){\makebox(0,0)[b]{\smash{Q15}}}%
    \put(0.8972,0.0113){\makebox(0,0)[b]{\smash{Q16}}}%
    \put(0.9517,0.0113){\makebox(0,0)[b]{\smash{Q17}}}%
    \put(0.0175,0.1942){\makebox(0,0)[lb]{\smash{\% of}}}%
    \put(0.0175,0.1792){\makebox(0,0)[lb]{\smash{SXSI}}}%
    \put(0.0971,0.1759){\rotatebox{90}{\makebox(0,0)[lb]{\smash{3.9s}}}}%
    \put(0.1516,0.1759){\rotatebox{90}{\makebox(0,0)[lb]{\smash{1.7s}}}}%
    \put(0.2061,0.1759){\rotatebox{90}{\makebox(0,0)[lb]{\smash{217.0}}}}%
    \put(0.2606,0.1759){\rotatebox{90}{\makebox(0,0)[lb]{\smash{344.0}}}}%
    \put(0.315,0.1759){\rotatebox{90}{\makebox(0,0)[lb]{\smash{113.0}}}}%
    \put(0.3695,0.1759){\rotatebox{90}{\makebox(0,0)[lb]{\smash{194.0}}}}%
    \put(0.424,0.1759){\rotatebox{90}{\makebox(0,0)[lb]{\smash{260.0}}}}%
    \put(0.4785,0.1759){\rotatebox{90}{\makebox(0,0)[lb]{\smash{273.0}}}}%
    \put(0.533,0.1759){\rotatebox{90}{\makebox(0,0)[lb]{\smash{238.0}}}}%
    \put(0.5875,0.1759){\rotatebox{90}{\makebox(0,0)[lb]{\smash{602.0}}}}%
    \put(0.642,0.1759){\rotatebox{90}{\makebox(0,0)[lb]{\smash{851.0}}}}%
    \put(0.6965,0.1759){\rotatebox{90}{\makebox(0,0)[lb]{\smash{523.0}}}}%
    \put(0.751,0.1759){\rotatebox{90}{\makebox(0,0)[lb]{\smash{96.0s}}}}%
    \put(0.8055,0.1759){\rotatebox{90}{\makebox(0,0)[lb]{\smash{209.2s}}}}%
    \put(0.86,0.1759){\rotatebox{90}{\makebox(0,0)[lb]{\smash{82.4s}}}}%
    \put(0.9145,0.1759){\rotatebox{90}{\makebox(0,0)[lb]{\smash{817.4s}}}}%
    \put(0.969,0.1759){\rotatebox{90}{\makebox(0,0)[lb]{\smash{112.9s}}}}%
  \end{picture}%
\endgroup

%% file: count_xmark_10.pdf_tex

\begingroup
  \makeatletter
  \providecommand\color[2][]{%
    \errmessage{(Inkscape) Color is used for the text in Inkscape, but the package 'color.sty' is not loaded}
    \renewcommand\color[2][]{}%
  }
  \providecommand\transparent[1]{%
    \errmessage{(Inkscape) Transparency is used (non-zero) for the text in Inkscape, but the package 'transparent.sty' is not loaded}
    \renewcommand\transparent[1]{}%
  }
  \providecommand\rotatebox[2]{#2}
  \ifx\svgwidth\undefined
    \setlength{\unitlength}{800pt}
  \else
    \setlength{\unitlength}{\svgwidth}
  \fi
  \global\let\svgwidth\undefined
  \makeatother
  \begin{picture}(1,0.2)%
    \put(0,0){\includegraphics[width=\unitlength]{count_xmark_10.pdf}}%
    \put(0.0455,0.0263){\makebox(0,0)[rb]{\smash{0}}}%
    \put(0.0455,0.0401){\makebox(0,0)[rb]{\smash{100}}}%
    \put(0.0455,0.0538){\makebox(0,0)[rb]{\smash{200}}}%
    \put(0.0455,0.0676){\makebox(0,0)[rb]{\smash{300}}}%
    \put(0.0455,0.0814){\makebox(0,0)[rb]{\smash{400}}}%
    \put(0.0455,0.0951){\makebox(0,0)[rb]{\smash{500}}}%
    \put(0.0455,0.1089){\makebox(0,0)[rb]{\smash{600}}}%
    \put(0.0455,0.1226){\makebox(0,0)[rb]{\smash{700}}}%
    \put(0.0455,0.1364){\makebox(0,0)[rb]{\smash{800}}}%
    \put(0.0455,0.1502){\makebox(0,0)[rb]{\smash{900}}}%
    \put(0.0455,0.1639){\makebox(0,0)[rb]{\smash{1000}}}%
    \put(0.0797,0.0113){\makebox(0,0)[b]{\smash{Q1}}}%
    \put(0.1342,0.0113){\makebox(0,0)[b]{\smash{Q2}}}%
    \put(0.1887,0.0113){\makebox(0,0)[b]{\smash{Q3}}}%
    \put(0.2432,0.0113){\makebox(0,0)[b]{\smash{Q4}}}%
    \put(0.2977,0.0113){\makebox(0,0)[b]{\smash{Q5}}}%
    \put(0.3522,0.0113){\makebox(0,0)[b]{\smash{Q6}}}%
    \put(0.4067,0.0113){\makebox(0,0)[b]{\smash{Q7}}}%
    \put(0.4612,0.0113){\makebox(0,0)[b]{\smash{Q8}}}%
    \put(0.5157,0.0113){\makebox(0,0)[b]{\smash{Q9}}}%
    \put(0.5702,0.0113){\makebox(0,0)[b]{\smash{Q10}}}%
    \put(0.6247,0.0113){\makebox(0,0)[b]{\smash{Q11}}}%
    \put(0.6792,0.0113){\makebox(0,0)[b]{\smash{Q12}}}%
    \put(0.7337,0.0113){\makebox(0,0)[b]{\smash{Q13}}}%
    \put(0.7882,0.0113){\makebox(0,0)[b]{\smash{Q14}}}%
    \put(0.8427,0.0113){\makebox(0,0)[b]{\smash{Q15}}}%
    \put(0.8972,0.0113){\makebox(0,0)[b]{\smash{Q16}}}%
    \put(0.9517,0.0113){\makebox(0,0)[b]{\smash{Q17}}}%
    \put(0.0175,0.1942){\makebox(0,0)[lb]{\smash{\% of}}}%
    \put(0.0175,0.1792){\makebox(0,0)[lb]{\smash{SXSI}}}%
    \put(0.0455,0.0263){\makebox(0,0)[rb]{\smash{0}}}%
    \put(0.0455,0.0401){\makebox(0,0)[rb]{\smash{100}}}%
    \put(0.0455,0.0538){\makebox(0,0)[rb]{\smash{200}}}%
    \put(0.0455,0.0676){\makebox(0,0)[rb]{\smash{300}}}%
    \put(0.0455,0.0814){\makebox(0,0)[rb]{\smash{400}}}%
    \put(0.0455,0.0951){\makebox(0,0)[rb]{\smash{500}}}%
    \put(0.0455,0.1089){\makebox(0,0)[rb]{\smash{600}}}%
    \put(0.0455,0.1226){\makebox(0,0)[rb]{\smash{700}}}%
    \put(0.0455,0.1364){\makebox(0,0)[rb]{\smash{800}}}%
    \put(0.0455,0.1502){\makebox(0,0)[rb]{\smash{900}}}%
    \put(0.0455,0.1639){\makebox(0,0)[rb]{\smash{1000}}}%
    \put(0.0797,0.0113){\makebox(0,0)[b]{\smash{Q1}}}%
    \put(0.1342,0.0113){\makebox(0,0)[b]{\smash{Q2}}}%
    \put(0.1887,0.0113){\makebox(0,0)[b]{\smash{Q3}}}%
    \put(0.2432,0.0113){\makebox(0,0)[b]{\smash{Q4}}}%
    \put(0.2977,0.0113){\makebox(0,0)[b]{\smash{Q5}}}%
    \put(0.3522,0.0113){\makebox(0,0)[b]{\smash{Q6}}}%
    \put(0.4067,0.0113){\makebox(0,0)[b]{\smash{Q7}}}%
    \put(0.4612,0.0113){\makebox(0,0)[b]{\smash{Q8}}}%
    \put(0.5157,0.0113){\makebox(0,0)[b]{\smash{Q9}}}%
    \put(0.5702,0.0113){\makebox(0,0)[b]{\smash{Q10}}}%
    \put(0.6247,0.0113){\makebox(0,0)[b]{\smash{Q11}}}%
    \put(0.6792,0.0113){\makebox(0,0)[b]{\smash{Q12}}}%
    \put(0.7337,0.0113){\makebox(0,0)[b]{\smash{Q13}}}%
    \put(0.7882,0.0113){\makebox(0,0)[b]{\smash{Q14}}}%
    \put(0.8427,0.0113){\makebox(0,0)[b]{\smash{Q15}}}%
    \put(0.8972,0.0113){\makebox(0,0)[b]{\smash{Q16}}}%
    \put(0.9517,0.0113){\makebox(0,0)[b]{\smash{Q17}}}%
    \put(0.0175,0.1942){\makebox(0,0)[lb]{\smash{\% of}}}%
    \put(0.0175,0.1792){\makebox(0,0)[lb]{\smash{SXSI}}}%
    \put(0.0698,0.1759){\rotatebox{90}{\makebox(0,0)[lb]{\smash{1.1}}}}%
    \put(0.1243,0.1759){\rotatebox{90}{\makebox(0,0)[lb]{\smash{55.0}}}}%
    \put(0.1788,0.1759){\rotatebox{90}{\makebox(0,0)[lb]{\smash{97.9}}}}%
    \put(0.2333,0.1759){\rotatebox{90}{\makebox(0,0)[lb]{\smash{178.8}}}}%
    \put(0.2878,0.1759){\rotatebox{90}{\makebox(0,0)[lb]{\smash{114.8}}}}%
    \put(0.3423,0.1759){\rotatebox{90}{\makebox(0,0)[lb]{\smash{179.2}}}}%
    \put(0.3968,0.1759){\rotatebox{90}{\makebox(0,0)[lb]{\smash{218.2}}}}%
    \put(0.4513,0.1759){\rotatebox{90}{\makebox(0,0)[lb]{\smash{199.3}}}}%
    \put(0.5058,0.1759){\rotatebox{90}{\makebox(0,0)[lb]{\smash{245.6}}}}%
    \put(0.5603,0.1759){\rotatebox{90}{\makebox(0,0)[lb]{\smash{489.7}}}}%
    \put(0.6148,0.1759){\rotatebox{90}{\makebox(0,0)[lb]{\smash{1.1s}}}}%
    \put(0.6693,0.1759){\rotatebox{90}{\makebox(0,0)[lb]{\smash{263.7}}}}%
    \put(0.7238,0.1759){\rotatebox{90}{\makebox(0,0)[lb]{\smash{5.2}}}}%
    \put(0.7782,0.1759){\rotatebox{90}{\makebox(0,0)[lb]{\smash{13.5}}}}%
    \put(0.8327,0.1759){\rotatebox{90}{\makebox(0,0)[lb]{\smash{1.6}}}}%
    \put(0.8872,0.1759){\rotatebox{90}{\makebox(0,0)[lb]{\smash{2.4}}}}%
    \put(0.9417,0.1759){\rotatebox{90}{\makebox(0,0)[lb]{\smash{1.3s}}}}%
    \put(0.0455,0.0263){\makebox(0,0)[rb]{\smash{0}}}%
    \put(0.0455,0.0401){\makebox(0,0)[rb]{\smash{100}}}%
    \put(0.0455,0.0538){\makebox(0,0)[rb]{\smash{200}}}%
    \put(0.0455,0.0676){\makebox(0,0)[rb]{\smash{300}}}%
    \put(0.0455,0.0814){\makebox(0,0)[rb]{\smash{400}}}%
    \put(0.0455,0.0951){\makebox(0,0)[rb]{\smash{500}}}%
    \put(0.0455,0.1089){\makebox(0,0)[rb]{\smash{600}}}%
    \put(0.0455,0.1226){\makebox(0,0)[rb]{\smash{700}}}%
    \put(0.0455,0.1364){\makebox(0,0)[rb]{\smash{800}}}%
    \put(0.0455,0.1502){\makebox(0,0)[rb]{\smash{900}}}%
    \put(0.0455,0.1639){\makebox(0,0)[rb]{\smash{1000}}}%
    \put(0.0797,0.0113){\makebox(0,0)[b]{\smash{Q1}}}%
    \put(0.1342,0.0113){\makebox(0,0)[b]{\smash{Q2}}}%
    \put(0.1887,0.0113){\makebox(0,0)[b]{\smash{Q3}}}%
    \put(0.2432,0.0113){\makebox(0,0)[b]{\smash{Q4}}}%
    \put(0.2977,0.0113){\makebox(0,0)[b]{\smash{Q5}}}%
    \put(0.3522,0.0113){\makebox(0,0)[b]{\smash{Q6}}}%
    \put(0.4067,0.0113){\makebox(0,0)[b]{\smash{Q7}}}%
    \put(0.4612,0.0113){\makebox(0,0)[b]{\smash{Q8}}}%
    \put(0.5157,0.0113){\makebox(0,0)[b]{\smash{Q9}}}%
    \put(0.5702,0.0113){\makebox(0,0)[b]{\smash{Q10}}}%
    \put(0.6247,0.0113){\makebox(0,0)[b]{\smash{Q11}}}%
    \put(0.6792,0.0113){\makebox(0,0)[b]{\smash{Q12}}}%
    \put(0.7337,0.0113){\makebox(0,0)[b]{\smash{Q13}}}%
    \put(0.7882,0.0113){\makebox(0,0)[b]{\smash{Q14}}}%
    \put(0.8427,0.0113){\makebox(0,0)[b]{\smash{Q15}}}%
    \put(0.8972,0.0113){\makebox(0,0)[b]{\smash{Q16}}}%
    \put(0.9517,0.0113){\makebox(0,0)[b]{\smash{Q17}}}%
    \put(0.0175,0.1942){\makebox(0,0)[lb]{\smash{\% of}}}%
    \put(0.0175,0.1792){\makebox(0,0)[lb]{\smash{SXSI}}}%
    \put(0.0834,0.1759){\rotatebox{90}{\makebox(0,0)[lb]{\smash{12.9}}}}%
    \put(0.1379,0.1759){\rotatebox{90}{\makebox(0,0)[lb]{\smash{640.0}}}}%
    \put(0.1924,0.1759){\rotatebox{90}{\makebox(0,0)[lb]{\smash{225.3}}}}%
    \put(0.2469,0.1759){\rotatebox{90}{\makebox(0,0)[lb]{\smash{1.3s}}}}%
    \put(0.3014,0.1759){\rotatebox{90}{\makebox(0,0)[lb]{\smash{257.1}}}}%
    \put(0.3559,0.1759){\rotatebox{90}{\makebox(0,0)[lb]{\smash{229.8}}}}%
    \put(0.4104,0.1759){\rotatebox{90}{\makebox(0,0)[lb]{\smash{386.4}}}}%
    \put(0.4649,0.1759){\rotatebox{90}{\makebox(0,0)[lb]{\smash{514.4}}}}%
    \put(0.5194,0.1759){\rotatebox{90}{\makebox(0,0)[lb]{\smash{551.0}}}}%
    \put(0.5739,0.1759){\rotatebox{90}{\makebox(0,0)[lb]{\smash{1.4s}}}}%
    \put(0.6284,0.1759){\rotatebox{90}{\makebox(0,0)[lb]{\smash{2.5s}}}}%
    \put(0.6829,0.1759){\rotatebox{90}{\makebox(0,0)[lb]{\smash{+++}}}}%
    \put(0.7374,0.1759){\rotatebox{90}{\makebox(0,0)[lb]{\smash{3.5s}}}}%
    \put(0.7919,0.1759){\rotatebox{90}{\makebox(0,0)[lb]{\smash{2.8s}}}}%
    \put(0.8464,0.1759){\rotatebox{90}{\makebox(0,0)[lb]{\smash{2.9s}}}}%
    \put(0.9009,0.1759){\rotatebox{90}{\makebox(0,0)[lb]{\smash{3.9s}}}}%
    \put(0.9554,0.1759){\rotatebox{90}{\makebox(0,0)[lb]{\smash{4.8s}}}}%
    \put(0.0455,0.0263){\makebox(0,0)[rb]{\smash{0}}}%
    \put(0.0455,0.0401){\makebox(0,0)[rb]{\smash{100}}}%
    \put(0.0455,0.0538){\makebox(0,0)[rb]{\smash{200}}}%
    \put(0.0455,0.0676){\makebox(0,0)[rb]{\smash{300}}}%
    \put(0.0455,0.0814){\makebox(0,0)[rb]{\smash{400}}}%
    \put(0.0455,0.0951){\makebox(0,0)[rb]{\smash{500}}}%
    \put(0.0455,0.1089){\makebox(0,0)[rb]{\smash{600}}}%
    \put(0.0455,0.1226){\makebox(0,0)[rb]{\smash{700}}}%
    \put(0.0455,0.1364){\makebox(0,0)[rb]{\smash{800}}}%
    \put(0.0455,0.1502){\makebox(0,0)[rb]{\smash{900}}}%
    \put(0.0455,0.1639){\makebox(0,0)[rb]{\smash{1000}}}%
    \put(0.0797,0.0113){\makebox(0,0)[b]{\smash{Q1}}}%
    \put(0.1342,0.0113){\makebox(0,0)[b]{\smash{Q2}}}%
    \put(0.1887,0.0113){\makebox(0,0)[b]{\smash{Q3}}}%
    \put(0.2432,0.0113){\makebox(0,0)[b]{\smash{Q4}}}%
    \put(0.2977,0.0113){\makebox(0,0)[b]{\smash{Q5}}}%
    \put(0.3522,0.0113){\makebox(0,0)[b]{\smash{Q6}}}%
    \put(0.4067,0.0113){\makebox(0,0)[b]{\smash{Q7}}}%
    \put(0.4612,0.0113){\makebox(0,0)[b]{\smash{Q8}}}%
    \put(0.5157,0.0113){\makebox(0,0)[b]{\smash{Q9}}}%
    \put(0.5702,0.0113){\makebox(0,0)[b]{\smash{Q10}}}%
    \put(0.6247,0.0113){\makebox(0,0)[b]{\smash{Q11}}}%
    \put(0.6792,0.0113){\makebox(0,0)[b]{\smash{Q12}}}%
    \put(0.7337,0.0113){\makebox(0,0)[b]{\smash{Q13}}}%
    \put(0.7882,0.0113){\makebox(0,0)[b]{\smash{Q14}}}%
    \put(0.8427,0.0113){\makebox(0,0)[b]{\smash{Q15}}}%
    \put(0.8972,0.0113){\makebox(0,0)[b]{\smash{Q16}}}%
    \put(0.9517,0.0113){\makebox(0,0)[b]{\smash{Q17}}}%
    \put(0.0175,0.1942){\makebox(0,0)[lb]{\smash{\% of}}}%
    \put(0.0175,0.1792){\makebox(0,0)[lb]{\smash{SXSI}}}%
    \put(0.0971,0.1759){\rotatebox{90}{\makebox(0,0)[lb]{\smash{3.5}}}}%
    \put(0.1516,0.1759){\rotatebox{90}{\makebox(0,0)[lb]{\smash{184.0}}}}%
    \put(0.2061,0.1759){\rotatebox{90}{\makebox(0,0)[lb]{\smash{309.8}}}}%
    \put(0.2606,0.1759){\rotatebox{90}{\makebox(0,0)[lb]{\smash{202.8}}}}%
    \put(0.315,0.1759){\rotatebox{90}{\makebox(0,0)[lb]{\smash{223.0}}}}%
    \put(0.3695,0.1759){\rotatebox{90}{\makebox(0,0)[lb]{\smash{335.5}}}}%
    \put(0.424,0.1759){\rotatebox{90}{\makebox(0,0)[lb]{\smash{592.8}}}}%
    \put(0.4785,0.1759){\rotatebox{90}{\makebox(0,0)[lb]{\smash{890.5}}}}%
    \put(0.533,0.1759){\rotatebox{90}{\makebox(0,0)[lb]{\smash{89.7s}}}}%
    \put(0.5875,0.1759){\rotatebox{90}{\makebox(0,0)[lb]{\smash{599.2}}}}%
    \put(0.642,0.1759){\rotatebox{90}{\makebox(0,0)[lb]{\smash{459.0s}}}}%
    \put(0.6965,0.1759){\rotatebox{90}{\makebox(0,0)[lb]{\smash{160.8}}}}%
    \put(0.751,0.1759){\rotatebox{90}{\makebox(0,0)[lb]{\smash{12.8}}}}%
    \put(0.8055,0.1759){\rotatebox{90}{\makebox(0,0)[lb]{\smash{479.0s}}}}%
    \put(0.86,0.1759){\rotatebox{90}{\makebox(0,0)[lb]{\smash{+++}}}}%
    \put(0.9145,0.1759){\rotatebox{90}{\makebox(0,0)[lb]{\smash{+++}}}}%
    \put(0.969,0.1759){\rotatebox{90}{\makebox(0,0)[lb]{\smash{+++}}}}%
  \end{picture}%
\endgroup

%% file: mat_xmark_10.pdf_tex

\begingroup
  \makeatletter
  \providecommand\color[2][]{%
    \errmessage{(Inkscape) Color is used for the text in Inkscape, but the package 'color.sty' is not loaded}
    \renewcommand\color[2][]{}%
  }
  \providecommand\transparent[1]{%
    \errmessage{(Inkscape) Transparency is used (non-zero) for the text in Inkscape, but the package 'transparent.sty' is not loaded}
    \renewcommand\transparent[1]{}%
  }
  \providecommand\rotatebox[2]{#2}
  \ifx\svgwidth\undefined
    \setlength{\unitlength}{800pt}
  \else
    \setlength{\unitlength}{\svgwidth}
  \fi
  \global\let\svgwidth\undefined
  \makeatother
  \begin{picture}(1,0.2)%
    \put(0,0){\includegraphics[width=\unitlength]{mat_xmark_10.pdf}}%
    \put(0.0455,0.0263){\makebox(0,0)[rb]{\smash{0}}}%
    \put(0.0455,0.0401){\makebox(0,0)[rb]{\smash{100}}}%
    \put(0.0455,0.0538){\makebox(0,0)[rb]{\smash{200}}}%
    \put(0.0455,0.0676){\makebox(0,0)[rb]{\smash{300}}}%
    \put(0.0455,0.0814){\makebox(0,0)[rb]{\smash{400}}}%
    \put(0.0455,0.0951){\makebox(0,0)[rb]{\smash{500}}}%
    \put(0.0455,0.1089){\makebox(0,0)[rb]{\smash{600}}}%
    \put(0.0455,0.1226){\makebox(0,0)[rb]{\smash{700}}}%
    \put(0.0455,0.1364){\makebox(0,0)[rb]{\smash{800}}}%
    \put(0.0455,0.1502){\makebox(0,0)[rb]{\smash{900}}}%
    \put(0.0455,0.1639){\makebox(0,0)[rb]{\smash{1000}}}%
    \put(0.0797,0.0113){\makebox(0,0)[b]{\smash{Q1}}}%
    \put(0.1342,0.0113){\makebox(0,0)[b]{\smash{Q2}}}%
    \put(0.1887,0.0113){\makebox(0,0)[b]{\smash{Q3}}}%
    \put(0.2432,0.0113){\makebox(0,0)[b]{\smash{Q4}}}%
    \put(0.2977,0.0113){\makebox(0,0)[b]{\smash{Q5}}}%
    \put(0.3522,0.0113){\makebox(0,0)[b]{\smash{Q6}}}%
    \put(0.4067,0.0113){\makebox(0,0)[b]{\smash{Q7}}}%
    \put(0.4612,0.0113){\makebox(0,0)[b]{\smash{Q8}}}%
    \put(0.5157,0.0113){\makebox(0,0)[b]{\smash{Q9}}}%
    \put(0.5702,0.0113){\makebox(0,0)[b]{\smash{Q10}}}%
    \put(0.6247,0.0113){\makebox(0,0)[b]{\smash{Q11}}}%
    \put(0.6792,0.0113){\makebox(0,0)[b]{\smash{Q12}}}%
    \put(0.7337,0.0113){\makebox(0,0)[b]{\smash{Q13}}}%
    \put(0.7882,0.0113){\makebox(0,0)[b]{\smash{Q14}}}%
    \put(0.8427,0.0113){\makebox(0,0)[b]{\smash{Q15}}}%
    \put(0.8972,0.0113){\makebox(0,0)[b]{\smash{Q16}}}%
    \put(0.9517,0.0113){\makebox(0,0)[b]{\smash{Q17}}}%
    \put(0.0175,0.1942){\makebox(0,0)[lb]{\smash{\% of}}}%
    \put(0.0175,0.1792){\makebox(0,0)[lb]{\smash{SXSI}}}%
    \put(0.0455,0.0263){\makebox(0,0)[rb]{\smash{0}}}%
    \put(0.0455,0.0401){\makebox(0,0)[rb]{\smash{100}}}%
    \put(0.0455,0.0538){\makebox(0,0)[rb]{\smash{200}}}%
    \put(0.0455,0.0676){\makebox(0,0)[rb]{\smash{300}}}%
    \put(0.0455,0.0814){\makebox(0,0)[rb]{\smash{400}}}%
    \put(0.0455,0.0951){\makebox(0,0)[rb]{\smash{500}}}%
    \put(0.0455,0.1089){\makebox(0,0)[rb]{\smash{600}}}%
    \put(0.0455,0.1226){\makebox(0,0)[rb]{\smash{700}}}%
    \put(0.0455,0.1364){\makebox(0,0)[rb]{\smash{800}}}%
    \put(0.0455,0.1502){\makebox(0,0)[rb]{\smash{900}}}%
    \put(0.0455,0.1639){\makebox(0,0)[rb]{\smash{1000}}}%
    \put(0.0797,0.0113){\makebox(0,0)[b]{\smash{Q1}}}%
    \put(0.1342,0.0113){\makebox(0,0)[b]{\smash{Q2}}}%
    \put(0.1887,0.0113){\makebox(0,0)[b]{\smash{Q3}}}%
    \put(0.2432,0.0113){\makebox(0,0)[b]{\smash{Q4}}}%
    \put(0.2977,0.0113){\makebox(0,0)[b]{\smash{Q5}}}%
    \put(0.3522,0.0113){\makebox(0,0)[b]{\smash{Q6}}}%
    \put(0.4067,0.0113){\makebox(0,0)[b]{\smash{Q7}}}%
    \put(0.4612,0.0113){\makebox(0,0)[b]{\smash{Q8}}}%
    \put(0.5157,0.0113){\makebox(0,0)[b]{\smash{Q9}}}%
    \put(0.5702,0.0113){\makebox(0,0)[b]{\smash{Q10}}}%
    \put(0.6247,0.0113){\makebox(0,0)[b]{\smash{Q11}}}%
    \put(0.6792,0.0113){\makebox(0,0)[b]{\smash{Q12}}}%
    \put(0.7337,0.0113){\makebox(0,0)[b]{\smash{Q13}}}%
    \put(0.7882,0.0113){\makebox(0,0)[b]{\smash{Q14}}}%
    \put(0.8427,0.0113){\makebox(0,0)[b]{\smash{Q15}}}%
    \put(0.8972,0.0113){\makebox(0,0)[b]{\smash{Q16}}}%
    \put(0.9517,0.0113){\makebox(0,0)[b]{\smash{Q17}}}%
    \put(0.0175,0.1942){\makebox(0,0)[lb]{\smash{\% of}}}%
    \put(0.0175,0.1792){\makebox(0,0)[lb]{\smash{SXSI}}}%
    \put(0.0698,0.1759){\rotatebox{90}{\makebox(0,0)[lb]{\smash{1.5}}}}%
    \put(0.1243,0.1759){\rotatebox{90}{\makebox(0,0)[lb]{\smash{62.3}}}}%
    \put(0.1788,0.1759){\rotatebox{90}{\makebox(0,0)[lb]{\smash{97.0}}}}%
    \put(0.2333,0.1759){\rotatebox{90}{\makebox(0,0)[lb]{\smash{203.4}}}}%
    \put(0.2878,0.1759){\rotatebox{90}{\makebox(0,0)[lb]{\smash{135.0}}}}%
    \put(0.3423,0.1759){\rotatebox{90}{\makebox(0,0)[lb]{\smash{198.4}}}}%
    \put(0.3968,0.1759){\rotatebox{90}{\makebox(0,0)[lb]{\smash{231.0}}}}%
    \put(0.4513,0.1759){\rotatebox{90}{\makebox(0,0)[lb]{\smash{212.6}}}}%
    \put(0.5058,0.1759){\rotatebox{90}{\makebox(0,0)[lb]{\smash{261.8}}}}%
    \put(0.5603,0.1759){\rotatebox{90}{\makebox(0,0)[lb]{\smash{486.3}}}}%
    \put(0.6148,0.1759){\rotatebox{90}{\makebox(0,0)[lb]{\smash{1.1s}}}}%
    \put(0.6693,0.1759){\rotatebox{90}{\makebox(0,0)[lb]{\smash{296.3}}}}%
    \put(0.7238,0.1759){\rotatebox{90}{\makebox(0,0)[lb]{\smash{2.3}}}}%
    \put(0.7782,0.1759){\rotatebox{90}{\makebox(0,0)[lb]{\smash{21.9}}}}%
    \put(0.8327,0.1759){\rotatebox{90}{\makebox(0,0)[lb]{\smash{1.3}}}}%
    \put(0.8872,0.1759){\rotatebox{90}{\makebox(0,0)[lb]{\smash{4.3}}}}%
    \put(0.9417,0.1759){\rotatebox{90}{\makebox(0,0)[lb]{\smash{1.3s}}}}%
    \put(0.0455,0.0263){\makebox(0,0)[rb]{\smash{0}}}%
    \put(0.0455,0.0401){\makebox(0,0)[rb]{\smash{100}}}%
    \put(0.0455,0.0538){\makebox(0,0)[rb]{\smash{200}}}%
    \put(0.0455,0.0676){\makebox(0,0)[rb]{\smash{300}}}%
    \put(0.0455,0.0814){\makebox(0,0)[rb]{\smash{400}}}%
    \put(0.0455,0.0951){\makebox(0,0)[rb]{\smash{500}}}%
    \put(0.0455,0.1089){\makebox(0,0)[rb]{\smash{600}}}%
    \put(0.0455,0.1226){\makebox(0,0)[rb]{\smash{700}}}%
    \put(0.0455,0.1364){\makebox(0,0)[rb]{\smash{800}}}%
    \put(0.0455,0.1502){\makebox(0,0)[rb]{\smash{900}}}%
    \put(0.0455,0.1639){\makebox(0,0)[rb]{\smash{1000}}}%
    \put(0.0797,0.0113){\makebox(0,0)[b]{\smash{Q1}}}%
    \put(0.1342,0.0113){\makebox(0,0)[b]{\smash{Q2}}}%
    \put(0.1887,0.0113){\makebox(0,0)[b]{\smash{Q3}}}%
    \put(0.2432,0.0113){\makebox(0,0)[b]{\smash{Q4}}}%
    \put(0.2977,0.0113){\makebox(0,0)[b]{\smash{Q5}}}%
    \put(0.3522,0.0113){\makebox(0,0)[b]{\smash{Q6}}}%
    \put(0.4067,0.0113){\makebox(0,0)[b]{\smash{Q7}}}%
    \put(0.4612,0.0113){\makebox(0,0)[b]{\smash{Q8}}}%
    \put(0.5157,0.0113){\makebox(0,0)[b]{\smash{Q9}}}%
    \put(0.5702,0.0113){\makebox(0,0)[b]{\smash{Q10}}}%
    \put(0.6247,0.0113){\makebox(0,0)[b]{\smash{Q11}}}%
    \put(0.6792,0.0113){\makebox(0,0)[b]{\smash{Q12}}}%
    \put(0.7337,0.0113){\makebox(0,0)[b]{\smash{Q13}}}%
    \put(0.7882,0.0113){\makebox(0,0)[b]{\smash{Q14}}}%
    \put(0.8427,0.0113){\makebox(0,0)[b]{\smash{Q15}}}%
    \put(0.8972,0.0113){\makebox(0,0)[b]{\smash{Q16}}}%
    \put(0.9517,0.0113){\makebox(0,0)[b]{\smash{Q17}}}%
    \put(0.0175,0.1942){\makebox(0,0)[lb]{\smash{\% of}}}%
    \put(0.0175,0.1792){\makebox(0,0)[lb]{\smash{SXSI}}}%
    \put(0.0834,0.1759){\rotatebox{90}{\makebox(0,0)[lb]{\smash{15.0}}}}%
    \put(0.1379,0.1759){\rotatebox{90}{\makebox(0,0)[lb]{\smash{662.0}}}}%
    \put(0.1924,0.1759){\rotatebox{90}{\makebox(0,0)[lb]{\smash{223.9}}}}%
    \put(0.2469,0.1759){\rotatebox{90}{\makebox(0,0)[lb]{\smash{1.3s}}}}%
    \put(0.3014,0.1759){\rotatebox{90}{\makebox(0,0)[lb]{\smash{255.3}}}}%
    \put(0.3559,0.1759){\rotatebox{90}{\makebox(0,0)[lb]{\smash{217.9}}}}%
    \put(0.4104,0.1759){\rotatebox{90}{\makebox(0,0)[lb]{\smash{362.0}}}}%
    \put(0.4649,0.1759){\rotatebox{90}{\makebox(0,0)[lb]{\smash{497.3}}}}%
    \put(0.5194,0.1759){\rotatebox{90}{\makebox(0,0)[lb]{\smash{634.5}}}}%
    \put(0.5739,0.1759){\rotatebox{90}{\makebox(0,0)[lb]{\smash{1.5s}}}}%
    \put(0.6284,0.1759){\rotatebox{90}{\makebox(0,0)[lb]{\smash{2.6s}}}}%
    \put(0.6829,0.1759){\rotatebox{90}{\makebox(0,0)[lb]{\smash{+++}}}}%
    \put(0.7374,0.1759){\rotatebox{90}{\makebox(0,0)[lb]{\smash{3.4s}}}}%
    \put(0.7919,0.1759){\rotatebox{90}{\makebox(0,0)[lb]{\smash{1.8s}}}}%
    \put(0.8464,0.1759){\rotatebox{90}{\makebox(0,0)[lb]{\smash{3.0s}}}}%
    \put(0.9009,0.1759){\rotatebox{90}{\makebox(0,0)[lb]{\smash{3.9s}}}}%
    \put(0.9554,0.1759){\rotatebox{90}{\makebox(0,0)[lb]{\smash{4.6s}}}}%
  \end{picture}%
\endgroup

%% file: matser_xmark_10.pdf_tex

\begingroup
  \makeatletter
  \providecommand\color[2][]{%
    \errmessage{(Inkscape) Color is used for the text in Inkscape, but the package 'color.sty' is not loaded}
    \renewcommand\color[2][]{}%
  }
  \providecommand\transparent[1]{%
    \errmessage{(Inkscape) Transparency is used (non-zero) for the text in Inkscape, but the package 'transparent.sty' is not loaded}
    \renewcommand\transparent[1]{}%
  }
  \providecommand\rotatebox[2]{#2}
  \ifx\svgwidth\undefined
    \setlength{\unitlength}{800pt}
  \else
    \setlength{\unitlength}{\svgwidth}
  \fi
  \global\let\svgwidth\undefined
  \makeatother
  \begin{picture}(1,0.2)%
    \put(0,0){\includegraphics[width=\unitlength]{matser_xmark_10.pdf}}%
    \put(0.0455,0.0263){\makebox(0,0)[rb]{\smash{0}}}%
    \put(0.0455,0.0401){\makebox(0,0)[rb]{\smash{100}}}%
    \put(0.0455,0.0538){\makebox(0,0)[rb]{\smash{200}}}%
    \put(0.0455,0.0676){\makebox(0,0)[rb]{\smash{300}}}%
    \put(0.0455,0.0814){\makebox(0,0)[rb]{\smash{400}}}%
    \put(0.0455,0.0951){\makebox(0,0)[rb]{\smash{500}}}%
    \put(0.0455,0.1089){\makebox(0,0)[rb]{\smash{600}}}%
    \put(0.0455,0.1226){\makebox(0,0)[rb]{\smash{700}}}%
    \put(0.0455,0.1364){\makebox(0,0)[rb]{\smash{800}}}%
    \put(0.0455,0.1502){\makebox(0,0)[rb]{\smash{900}}}%
    \put(0.0455,0.1639){\makebox(0,0)[rb]{\smash{1000}}}%
    \put(0.0797,0.0113){\makebox(0,0)[b]{\smash{Q1}}}%
    \put(0.1342,0.0113){\makebox(0,0)[b]{\smash{Q2}}}%
    \put(0.1887,0.0113){\makebox(0,0)[b]{\smash{Q3}}}%
    \put(0.2432,0.0113){\makebox(0,0)[b]{\smash{Q4}}}%
    \put(0.2977,0.0113){\makebox(0,0)[b]{\smash{Q5}}}%
    \put(0.3522,0.0113){\makebox(0,0)[b]{\smash{Q6}}}%
    \put(0.4067,0.0113){\makebox(0,0)[b]{\smash{Q7}}}%
    \put(0.4612,0.0113){\makebox(0,0)[b]{\smash{Q8}}}%
    \put(0.5157,0.0113){\makebox(0,0)[b]{\smash{Q9}}}%
    \put(0.5702,0.0113){\makebox(0,0)[b]{\smash{Q10}}}%
    \put(0.6247,0.0113){\makebox(0,0)[b]{\smash{Q11}}}%
    \put(0.6792,0.0113){\makebox(0,0)[b]{\smash{Q12}}}%
    \put(0.7337,0.0113){\makebox(0,0)[b]{\smash{Q13}}}%
    \put(0.7882,0.0113){\makebox(0,0)[b]{\smash{Q14}}}%
    \put(0.8427,0.0113){\makebox(0,0)[b]{\smash{Q15}}}%
    \put(0.8972,0.0113){\makebox(0,0)[b]{\smash{Q16}}}%
    \put(0.9517,0.0113){\makebox(0,0)[b]{\smash{Q17}}}%
    \put(0.0175,0.1942){\makebox(0,0)[lb]{\smash{\% of}}}%
    \put(0.0175,0.1792){\makebox(0,0)[lb]{\smash{SXSI}}}%
    \put(0.0455,0.0263){\makebox(0,0)[rb]{\smash{0}}}%
    \put(0.0455,0.0401){\makebox(0,0)[rb]{\smash{100}}}%
    \put(0.0455,0.0538){\makebox(0,0)[rb]{\smash{200}}}%
    \put(0.0455,0.0676){\makebox(0,0)[rb]{\smash{300}}}%
    \put(0.0455,0.0814){\makebox(0,0)[rb]{\smash{400}}}%
    \put(0.0455,0.0951){\makebox(0,0)[rb]{\smash{500}}}%
    \put(0.0455,0.1089){\makebox(0,0)[rb]{\smash{600}}}%
    \put(0.0455,0.1226){\makebox(0,0)[rb]{\smash{700}}}%
    \put(0.0455,0.1364){\makebox(0,0)[rb]{\smash{800}}}%
    \put(0.0455,0.1502){\makebox(0,0)[rb]{\smash{900}}}%
    \put(0.0455,0.1639){\makebox(0,0)[rb]{\smash{1000}}}%
    \put(0.0797,0.0113){\makebox(0,0)[b]{\smash{Q1}}}%
    \put(0.1342,0.0113){\makebox(0,0)[b]{\smash{Q2}}}%
    \put(0.1887,0.0113){\makebox(0,0)[b]{\smash{Q3}}}%
    \put(0.2432,0.0113){\makebox(0,0)[b]{\smash{Q4}}}%
    \put(0.2977,0.0113){\makebox(0,0)[b]{\smash{Q5}}}%
    \put(0.3522,0.0113){\makebox(0,0)[b]{\smash{Q6}}}%
    \put(0.4067,0.0113){\makebox(0,0)[b]{\smash{Q7}}}%
    \put(0.4612,0.0113){\makebox(0,0)[b]{\smash{Q8}}}%
    \put(0.5157,0.0113){\makebox(0,0)[b]{\smash{Q9}}}%
    \put(0.5702,0.0113){\makebox(0,0)[b]{\smash{Q10}}}%
    \put(0.6247,0.0113){\makebox(0,0)[b]{\smash{Q11}}}%
    \put(0.6792,0.0113){\makebox(0,0)[b]{\smash{Q12}}}%
    \put(0.7337,0.0113){\makebox(0,0)[b]{\smash{Q13}}}%
    \put(0.7882,0.0113){\makebox(0,0)[b]{\smash{Q14}}}%
    \put(0.8427,0.0113){\makebox(0,0)[b]{\smash{Q15}}}%
    \put(0.8972,0.0113){\makebox(0,0)[b]{\smash{Q16}}}%
    \put(0.9517,0.0113){\makebox(0,0)[b]{\smash{Q17}}}%
    \put(0.0175,0.1942){\makebox(0,0)[lb]{\smash{\% of}}}%
    \put(0.0175,0.1792){\makebox(0,0)[lb]{\smash{SXSI}}}%
    \put(0.0698,0.1759){\rotatebox{90}{\makebox(0,0)[lb]{\smash{3.8s}}}}%
    \put(0.1243,0.1759){\rotatebox{90}{\makebox(0,0)[lb]{\smash{4.0s}}}}%
    \put(0.1788,0.1759){\rotatebox{90}{\makebox(0,0)[lb]{\smash{153.0}}}}%
    \put(0.2333,0.1759){\rotatebox{90}{\makebox(0,0)[lb]{\smash{766.0}}}}%
    \put(0.2878,0.1759){\rotatebox{90}{\makebox(0,0)[lb]{\smash{212.0}}}}%
    \put(0.3423,0.1759){\rotatebox{90}{\makebox(0,0)[lb]{\smash{287.0}}}}%
    \put(0.3968,0.1759){\rotatebox{90}{\makebox(0,0)[lb]{\smash{305.0}}}}%
    \put(0.4513,0.1759){\rotatebox{90}{\makebox(0,0)[lb]{\smash{460.0}}}}%
    \put(0.5058,0.1759){\rotatebox{90}{\makebox(0,0)[lb]{\smash{377.0}}}}%
    \put(0.5603,0.1759){\rotatebox{90}{\makebox(0,0)[lb]{\smash{1.5s}}}}%
    \put(0.6148,0.1759){\rotatebox{90}{\makebox(0,0)[lb]{\smash{2.2s}}}}%
    \put(0.6693,0.1759){\rotatebox{90}{\makebox(0,0)[lb]{\smash{895.0}}}}%
    \put(0.7238,0.1759){\rotatebox{90}{\makebox(0,0)[lb]{\smash{7.6s}}}}%
    \put(0.7782,0.1759){\rotatebox{90}{\makebox(0,0)[lb]{\smash{55.5s}}}}%
    \put(0.8327,0.1759){\rotatebox{90}{\makebox(0,0)[lb]{\smash{55.5s}}}}%
    \put(0.8872,0.1759){\rotatebox{90}{\makebox(0,0)[lb]{\smash{47.9s}}}}%
    \put(0.9417,0.1759){\rotatebox{90}{\makebox(0,0)[lb]{\smash{41.6s}}}}%
    \put(0.0455,0.0263){\makebox(0,0)[rb]{\smash{0}}}%
    \put(0.0455,0.0401){\makebox(0,0)[rb]{\smash{100}}}%
    \put(0.0455,0.0538){\makebox(0,0)[rb]{\smash{200}}}%
    \put(0.0455,0.0676){\makebox(0,0)[rb]{\smash{300}}}%
    \put(0.0455,0.0814){\makebox(0,0)[rb]{\smash{400}}}%
    \put(0.0455,0.0951){\makebox(0,0)[rb]{\smash{500}}}%
    \put(0.0455,0.1089){\makebox(0,0)[rb]{\smash{600}}}%
    \put(0.0455,0.1226){\makebox(0,0)[rb]{\smash{700}}}%
    \put(0.0455,0.1364){\makebox(0,0)[rb]{\smash{800}}}%
    \put(0.0455,0.1502){\makebox(0,0)[rb]{\smash{900}}}%
    \put(0.0455,0.1639){\makebox(0,0)[rb]{\smash{1000}}}%
    \put(0.0797,0.0113){\makebox(0,0)[b]{\smash{Q1}}}%
    \put(0.1342,0.0113){\makebox(0,0)[b]{\smash{Q2}}}%
    \put(0.1887,0.0113){\makebox(0,0)[b]{\smash{Q3}}}%
    \put(0.2432,0.0113){\makebox(0,0)[b]{\smash{Q4}}}%
    \put(0.2977,0.0113){\makebox(0,0)[b]{\smash{Q5}}}%
    \put(0.3522,0.0113){\makebox(0,0)[b]{\smash{Q6}}}%
    \put(0.4067,0.0113){\makebox(0,0)[b]{\smash{Q7}}}%
    \put(0.4612,0.0113){\makebox(0,0)[b]{\smash{Q8}}}%
    \put(0.5157,0.0113){\makebox(0,0)[b]{\smash{Q9}}}%
    \put(0.5702,0.0113){\makebox(0,0)[b]{\smash{Q10}}}%
    \put(0.6247,0.0113){\makebox(0,0)[b]{\smash{Q11}}}%
    \put(0.6792,0.0113){\makebox(0,0)[b]{\smash{Q12}}}%
    \put(0.7337,0.0113){\makebox(0,0)[b]{\smash{Q13}}}%
    \put(0.7882,0.0113){\makebox(0,0)[b]{\smash{Q14}}}%
    \put(0.8427,0.0113){\makebox(0,0)[b]{\smash{Q15}}}%
    \put(0.8972,0.0113){\makebox(0,0)[b]{\smash{Q16}}}%
    \put(0.9517,0.0113){\makebox(0,0)[b]{\smash{Q17}}}%
    \put(0.0175,0.1942){\makebox(0,0)[lb]{\smash{\% of}}}%
    \put(0.0175,0.1792){\makebox(0,0)[lb]{\smash{SXSI}}}%
    \put(0.0834,0.1759){\rotatebox{90}{\makebox(0,0)[lb]{\smash{7.7s}}}}%
    \put(0.1379,0.1759){\rotatebox{90}{\makebox(0,0)[lb]{\smash{7.5s}}}}%
    \put(0.1924,0.1759){\rotatebox{90}{\makebox(0,0)[lb]{\smash{2.6s}}}}%
    \put(0.2469,0.1759){\rotatebox{90}{\makebox(0,0)[lb]{\smash{16.2s}}}}%
    \put(0.3014,0.1759){\rotatebox{90}{\makebox(0,0)[lb]{\smash{2.6s}}}}%
    \put(0.3559,0.1759){\rotatebox{90}{\makebox(0,0)[lb]{\smash{2.6s}}}}%
    \put(0.4104,0.1759){\rotatebox{90}{\makebox(0,0)[lb]{\smash{1.9s}}}}%
    \put(0.4649,0.1759){\rotatebox{90}{\makebox(0,0)[lb]{\smash{2.0s}}}}%
    \put(0.5194,0.1759){\rotatebox{90}{\makebox(0,0)[lb]{\smash{2.2s}}}}%
    \put(0.5739,0.1759){\rotatebox{90}{\makebox(0,0)[lb]{\smash{16.5s}}}}%
    \put(0.6284,0.1759){\rotatebox{90}{\makebox(0,0)[lb]{\smash{17.7s}}}}%
    \put(0.6829,0.1759){\rotatebox{90}{\makebox(0,0)[lb]{\smash{+++}}}}%
    \put(0.7374,0.1759){\rotatebox{90}{\makebox(0,0)[lb]{\smash{20.2s}}}}%
    \put(0.7919,0.1759){\rotatebox{90}{\makebox(0,0)[lb]{\smash{73.9s}}}}%
    \put(0.8464,0.1759){\rotatebox{90}{\makebox(0,0)[lb]{\smash{60.5s}}}}%
    \put(0.9009,0.1759){\rotatebox{90}{\makebox(0,0)[lb]{\smash{50.2s}}}}%
    \put(0.9554,0.1759){\rotatebox{90}{\makebox(0,0)[lb]{\smash{40.0s}}}}%
    \put(0.0455,0.0263){\makebox(0,0)[rb]{\smash{0}}}%
    \put(0.0455,0.0401){\makebox(0,0)[rb]{\smash{100}}}%
    \put(0.0455,0.0538){\makebox(0,0)[rb]{\smash{200}}}%
    \put(0.0455,0.0676){\makebox(0,0)[rb]{\smash{300}}}%
    \put(0.0455,0.0814){\makebox(0,0)[rb]{\smash{400}}}%
    \put(0.0455,0.0951){\makebox(0,0)[rb]{\smash{500}}}%
    \put(0.0455,0.1089){\makebox(0,0)[rb]{\smash{600}}}%
    \put(0.0455,0.1226){\makebox(0,0)[rb]{\smash{700}}}%
    \put(0.0455,0.1364){\makebox(0,0)[rb]{\smash{800}}}%
    \put(0.0455,0.1502){\makebox(0,0)[rb]{\smash{900}}}%
    \put(0.0455,0.1639){\makebox(0,0)[rb]{\smash{1000}}}%
    \put(0.0797,0.0113){\makebox(0,0)[b]{\smash{Q1}}}%
    \put(0.1342,0.0113){\makebox(0,0)[b]{\smash{Q2}}}%
    \put(0.1887,0.0113){\makebox(0,0)[b]{\smash{Q3}}}%
    \put(0.2432,0.0113){\makebox(0,0)[b]{\smash{Q4}}}%
    \put(0.2977,0.0113){\makebox(0,0)[b]{\smash{Q5}}}%
    \put(0.3522,0.0113){\makebox(0,0)[b]{\smash{Q6}}}%
    \put(0.4067,0.0113){\makebox(0,0)[b]{\smash{Q7}}}%
    \put(0.4612,0.0113){\makebox(0,0)[b]{\smash{Q8}}}%
    \put(0.5157,0.0113){\makebox(0,0)[b]{\smash{Q9}}}%
    \put(0.5702,0.0113){\makebox(0,0)[b]{\smash{Q10}}}%
    \put(0.6247,0.0113){\makebox(0,0)[b]{\smash{Q11}}}%
    \put(0.6792,0.0113){\makebox(0,0)[b]{\smash{Q12}}}%
    \put(0.7337,0.0113){\makebox(0,0)[b]{\smash{Q13}}}%
    \put(0.7882,0.0113){\makebox(0,0)[b]{\smash{Q14}}}%
    \put(0.8427,0.0113){\makebox(0,0)[b]{\smash{Q15}}}%
    \put(0.8972,0.0113){\makebox(0,0)[b]{\smash{Q16}}}%
    \put(0.9517,0.0113){\makebox(0,0)[b]{\smash{Q17}}}%
    \put(0.0175,0.1942){\makebox(0,0)[lb]{\smash{\% of}}}%
    \put(0.0175,0.1792){\makebox(0,0)[lb]{\smash{SXSI}}}%
    \put(0.0971,0.1759){\rotatebox{90}{\makebox(0,0)[lb]{\smash{13.8s}}}}%
    \put(0.1516,0.1759){\rotatebox{90}{\makebox(0,0)[lb]{\smash{323.7s}}}}%
    \put(0.2061,0.1759){\rotatebox{90}{\makebox(0,0)[lb]{\smash{519.0}}}}%
    \put(0.2606,0.1759){\rotatebox{90}{\makebox(0,0)[lb]{\smash{40.8s}}}}%
    \put(0.315,0.1759){\rotatebox{90}{\makebox(0,0)[lb]{\smash{380.0}}}}%
    \put(0.3695,0.1759){\rotatebox{90}{\makebox(0,0)[lb]{\smash{460.0}}}}%
    \put(0.424,0.1759){\rotatebox{90}{\makebox(0,0)[lb]{\smash{1.3s}}}}%
    \put(0.4785,0.1759){\rotatebox{90}{\makebox(0,0)[lb]{\smash{1.4s}}}}%
    \put(0.533,0.1759){\rotatebox{90}{\makebox(0,0)[lb]{\smash{707.0}}}}%
    \put(0.5875,0.1759){\rotatebox{90}{\makebox(0,0)[lb]{\smash{6.1s}}}}%
    \put(0.642,0.1759){\rotatebox{90}{\makebox(0,0)[lb]{\smash{101.5s}}}}%
    \put(0.6965,0.1759){\rotatebox{90}{\makebox(0,0)[lb]{\smash{2.8s}}}}%
    \put(0.751,0.1759){\rotatebox{90}{\makebox(0,0)[lb]{\smash{514.7s}}}}%
    \put(0.8055,0.1759){\rotatebox{90}{\makebox(0,0)[lb]{\smash{+++}}}}%
    \put(0.86,0.1759){\rotatebox{90}{\makebox(0,0)[lb]{\smash{+++}}}}%
    \put(0.9145,0.1759){\rotatebox{90}{\makebox(0,0)[lb]{\smash{+++}}}}%
    \put(0.969,0.1759){\rotatebox{90}{\makebox(0,0)[lb]{\smash{+++}}}}%
  \end{picture}%
\endgroup

%% file: runtime_mem.pdf_tex

\begingroup
  \makeatletter
  \providecommand\color[2][]{%
    \errmessage{(Inkscape) Color is used for the text in Inkscape, but the package 'color.sty' is not loaded}
    \renewcommand\color[2][]{}%
  }
  \providecommand\transparent[1]{%
    \errmessage{(Inkscape) Transparency is used (non-zero) for the text in Inkscape, but the package 'transparent.sty' is not loaded}
    \renewcommand\transparent[1]{}%
  }
  \providecommand\rotatebox[2]{#2}
  \ifx\svgwidth\undefined
    \setlength{\unitlength}{466.5pt}
  \else
    \setlength{\unitlength}{\svgwidth}
  \fi
  \global\let\svgwidth\undefined
  \makeatother
  \begin{picture}(1,0.35209003)%
    \put(0,0){\includegraphics[width=\unitlength]{runtime_mem.pdf}}%
  \end{picture}%
\endgroup

%% file: precision_edited.pdf_tex

\begingroup
  \makeatletter
  \providecommand\color[2][]{%
    \errmessage{(Inkscape) Color is used for the text in Inkscape, but the package 'color.sty' is not loaded}
    \renewcommand\color[2][]{}%
  }
  \providecommand\transparent[1]{%
    \errmessage{(Inkscape) Transparency is used (non-zero) for the text in Inkscape, but the package 'transparent.sty' is not loaded}
    \renewcommand\transparent[1]{}%
  }
  \providecommand\rotatebox[2]{#2}
  \ifx\svgwidth\undefined
    \setlength{\unitlength}{594.1246582pt}
  \else
    \setlength{\unitlength}{\svgwidth}
  \fi
  \global\let\svgwidth\undefined
  \makeatother
  \begin{picture}(1,0.40657104)%
    \put(0,0){\includegraphics[width=\unitlength]{precision_edited.pdf}}%
    \put(-0.00002366,0.39400529){\color[rgb]{0,0,0}\makebox(0,0)[lb]{\smash{\mytexta{Number of Nodes}}}}%
    \put(0.04039207,0.36864698){\color[rgb]{0,0,0}\makebox(0,0)[lb]{\smash{10$^7$}}}%
    \put(0.04039207,0.32270864){\color[rgb]{0,0,0}\makebox(0,0)[lb]{\smash{10$^6$}}}%
    \put(0.04039207,0.27706366){\color[rgb]{0,0,0}\makebox(0,0)[lb]{\smash{10$^5$}}}%
    \put(0.04039207,0.23057346){\color[rgb]{0,0,0}\makebox(0,0)[lb]{\smash{10$^4$}}}%
    \put(0.04039207,0.1849285){\color[rgb]{0,0,0}\makebox(0,0)[lb]{\smash{10$^3$}}}%
    \put(0.04039207,0.13674768){\color[rgb]{0,0,0}\makebox(0,0)[lb]{\smash{10$^2$}}}%
    \put(0.04678063,0.09194798){\color[rgb]{0,0,0}\makebox(0,0)[lb]{\smash{10}}}%
    \put(0.05625623,0.04545774){\color[rgb]{0,0,0}\makebox(0,0)[lb]{\smash{1}}}%
    \put(0.11072781,0.36900725){\color[rgb]{0,0,0}\makebox(0,0)[lb]{\smash{\mytexta{Visited Nodes}}}}%
    \put(0.11060157,0.3423362){\color[rgb]{0,0,0}\makebox(0,0)[lb]{\smash{\mytexta{Marked Nodes}}}}%
    \put(0.1105937,0.31229586){\color[rgb]{0,0,0}\makebox(0,0)[lb]{\smash{\mytexta{Result Nodes}}}}%
    \put(0.5105439,0.00348726){\color[rgb]{0,0,0}\makebox(0,0)[lb]{\smash{\mytexta{Query}}}}%
    \put(0.06486051,0.0291007){\color[rgb]{0,0,0}\makebox(0,0)[lb]{\smash{Q1}}}%
    \put(0.12110791,0.0291007){\color[rgb]{0,0,0}\makebox(0,0)[lb]{\smash{Q2}}}%
    \put(0.17853219,0.0291007){\color[rgb]{0,0,0}\makebox(0,0)[lb]{\smash{Q3}}}%
    \put(0.23525693,0.0291007){\color[rgb]{0,0,0}\makebox(0,0)[lb]{\smash{Q4}}}%
    \put(0.29225648,0.0291007){\color[rgb]{0,0,0}\makebox(0,0)[lb]{\smash{Q5}}}%
    \put(0.34916401,0.0291007){\color[rgb]{0,0,0}\makebox(0,0)[lb]{\smash{Q6}}}%
    \put(0.40601102,0.0291007){\color[rgb]{0,0,0}\makebox(0,0)[lb]{\smash{Q7}}}%
    \put(0.46319338,0.0291007){\color[rgb]{0,0,0}\makebox(0,0)[lb]{\smash{Q8}}}%
    \put(0.52001936,0.0291007){\color[rgb]{0,0,0}\makebox(0,0)[lb]{\smash{Q9}}}%
    \put(0.57290108,0.0291007){\color[rgb]{0,0,0}\makebox(0,0)[lb]{\smash{Q10}}}%
    \put(0.6306857,0.0291007){\color[rgb]{0,0,0}\makebox(0,0)[lb]{\smash{Q11}}}%
    \put(0.68663986,0.0291007){\color[rgb]{0,0,0}\makebox(0,0)[lb]{\smash{Q12}}}%
    \put(0.74406415,0.0291007){\color[rgb]{0,0,0}\makebox(0,0)[lb]{\smash{Q13}}}%
    \put(0.80078889,0.0291007){\color[rgb]{0,0,0}\makebox(0,0)[lb]{\smash{Q14}}}%
    \put(0.85778843,0.0291007){\color[rgb]{0,0,0}\makebox(0,0)[lb]{\smash{Q15}}}%
    \put(0.91469599,0.0291007){\color[rgb]{0,0,0}\makebox(0,0)[lb]{\smash{Q16}}}%
    \put(0.97153635,0.0291007){\color[rgb]{0,0,0}\makebox(0,0)[lb]{\smash{Q17}}}%
  \end{picture}%
\endgroup

%% file: conclusions.tex
\section{Conclusions and Future Work}
\label{sec:concl}

We have presented SXSI, a system for representing an XML collection in compact
form so that fast indexed XPath queries can be carried out on it. 
Even in its current prototype stage, SXSI is already competitive
with well-known efficient systems such as MonetDB and Qizx. As such, a number
of avenues for future work are open. We mention the broadest ones here.

Handling updates to the collections is possible in
principle, as there are dynamic data structures for sequences, trees, and
text collections \cite{CHLS07,MN08,SN08}. What remains to be verified is how 
practical those theoretical solutions really are.

As seen, the compact data structures support several fancy operations beyond
those actually used by our XPath evaluator. A matter of future work is to
explore other evaluation strategies that take advantage of those nonstandard
capabilities. 
As an example, the current XPath evaluator does not use
the range search capabilities of structure $Doc$ of
Section~\ref{sec:text}.
Another interesting challenge is to support XPath
string-value semantics, where strings spanning more than one text node can
be searched for. This, at least at a rough level, is not hard to achieve with 
our FM-index, by removing the \$-terminators and marking them
on a separate bitmap instead.
Beyond that, we would like to extend our implementation to
full XPath 1.0, and add core functionalities of XQuery.